\documentclass[%
reprint,
 amsmath,amssymb,
 aps,
]{revtex4-2}

\usepackage{graphicx}
\usepackage{dcolumn}
\usepackage{bm}

\usepackage{graphicx}
\usepackage{amsmath}
\usepackage{pifont}
\usepackage{amssymb}
\usepackage{float}
\usepackage{xcolor}
\usepackage{mathtools}

\usepackage{xr}
\usepackage[makeroom]{cancel}
\usepackage{tikz}
\usetikzlibrary{spy}
\usepackage{pgfplots}
\pgfplotsset{compat = newest}
\usepackage{pgfplotstable}
\usetikzlibrary{positioning, decorations.pathreplacing, calc, intersections, pgfplots.groupplots,fadings}
\usetikzlibrary{arrows,intersections}
\usepackage{tikz-cd}
\usepackage{xcolor}
\usetikzlibrary{fadings}
\usepackage{physics}
\usepackage{verbatim}
\usepackage{siunitx}
\usetikzlibrary{matrix}
\usepackage{dcolumn}
\usepackage{bm}
\usetikzlibrary{decorations.markings}
\pgfkeys{
  /pgf/arrow keys/.cd,
  pitch/.code={%
    \pgfmathsetmacro\pgfarrowpitch{#1}
    \pgfmathsetmacro\pgfarrowsinpitch{abs(sin(\pgfarrowpitch))}
    \pgfmathsetmacro\pgfarrowcospitch{abs(cos(\pgfarrowpitch))}
  },
}

\usetikzlibrary{decorations.pathreplacing}
\usepackage{xifthen}

\newcommand{\mycolorbar}[3]
{   
\begin{scope}[xshift=-10]
        \foreach \x [count=\c] in {#3}{ \xdef\numcolo{\c}}
        \pgfmathsetmacro{\pieceheight}{#1/(\numcolo-1)}
        \xdef\lowcolo{}
        \foreach \x [count=\c] in {#3}
        {   \ifthenelse{\c = 1}
            {}
            {   \fill[bottom color=\lowcolo,top color=\x] (0,{(\c-2)*\pieceheight}) rectangle (#2,{(\c-1)*\pieceheight});
            }
            \xdef\lowcolo{\x}
        }
        \end{scope}
}

\definecolor{darkblue}{RGB}{8, 66, 102}
\definecolor{darkblue2}{RGB}{42, 102, 143}
\definecolor{darkblue3}{RGB}{186, 199, 216}
\definecolor{darkred}{RGB}{231, 117, 0}
\definecolor{bluegrey}{RGB}{42, 102, 143}

\pgfdeclarearrow{
  name = Cone,
  defaults = {       
    length     = +2.3pt +3.4,
    width'     = +0pt +0.2,
    line width = +0pt 1 1,
    pitch      = +0, 
  },
  cache = false,     
  setup code = {},   
  drawing code = {   
    \pgfmathsetmacro\pgfarrowhalfwidth{.38\pgfarrowwidth}
    \pgfmathsetmacro\pgfarrowhalfwidthsin{\pgfarrowhalfwidth*\pgfarrowsinpitch}
    \pgfpathellipse{\pgfpointorigin}{\pgfqpoint{\pgfarrowhalfwidthsin pt}{0pt}}{\pgfqpoint{0pt}{\pgfarrowhalfwidth pt}}
    \pgfusepath{fill}
    \pgfmathsetmacro\pgfarrowlengthcos{\pgfarrowlength*\pgfarrowcospitch}
    \pgfmathparse{\pgfarrowlengthcos>\pgfarrowhalfwidthsin}
    \ifnum\pgfmathresult=1
      \pgfmathsetmacro\pgfarrowlengthtemp{\pgfarrowhalfwidthsin*\pgfarrowhalfwidthsin/\pgfarrowlengthcos}
      \pgfmathsetmacro\pgfarrowwidthtemp{\pgfarrowhalfwidth/\pgfarrowlengthcos*sqrt(\pgfarrowlengthcos*\pgfarrowlengthcos-\pgfarrowhalfwidthsin*\pgfarrowhalfwidthsin)}
      \pgfpathmoveto{\pgfqpoint{\pgfarrowlengthcos pt}{0pt}}
      \pgfpathlineto{\pgfqpoint{\pgfarrowlengthtemp pt}{ \pgfarrowwidthtemp pt}}
      \pgfpathlineto{\pgfqpoint{\pgfarrowlengthtemp pt}{-\pgfarrowwidthtemp pt}}
      \pgfpathclose
      \pgfusepath{fill}
    \fi
    \pgfpathmoveto{\pgfpointorigin}
  }
}
\usepackage{xr}

\newcommand{\mycomment}[1]{}

\begin{document}

\newcommand\gauss[2]{1/(#2*sqrt(2*pi))*exp(-((x-#1)^2)/(2*#2^2))} 

\preprint{APS/123-QED}

\title{Stability and quasi-Periodicity of Many-Body Localized Dynamics}

\author{Peyman Azodi}
\email{pazodi@princeton.edu}
\author{Herschel A.Rabitz}%

\affiliation{%
 Department of Chemistry, Princeton University, Princeton, New Jersey 08544, USA 
}%

\date{\today}

\begin{abstract}
The connection between entanglement dynamics and non-equilibrium statistics in isolated many-body quantum systems has been established both theoretically and experimentally. Many-Body Localization (MBL), a phenomenon where interacting particles in disordered (i.e., random) chains fail to thermalize, exemplifies this connection. However, the systematic proof of critical phenomena such as MBL remains challenging due to the lack of robust methods for analyzing many-body entanglement dynamics. In this paper, we identify MBL through quasi-periodic dynamics in the entanglement evolution of subsystems in a disordered Heisenberg chain. This new form of characterizing MBL, through stable quasi-periodic dynamics of entanglement-where ``stable" means they persist in the thermodynamic limit—concretely distinguishes between two competing scenarios: fully localized behavior of subsystems or slowly, exponentially slow in disorder, thermalizing subsystems—a heated controversy in the literature. Utilizing perturbation theory, we derive the entanglement dynamics of single spins through an infinite perturbative series, while also modeling rare Griffiths regions (locally thermal inclusions). Our results prove that in regimes of sufficiently strong disorder, the entanglement evolution of individual subsystems remains quasi-periodic in the thermodynamic limit, thereby providing concrete evidence for the stability of MBL dynamics in disordered Heisenberg chains. This behavior contrasts with the widely reported logarithmic growth of subsystem entanglement in the MBL phase. We show that the logarithmic growth observed in prior studies arises from statistical ensemble averaging, which is prohibited due to the intrinsic non-ergodic dynamics characteristic of MBL systems, rooted in their quasi-periodic features.

\end{abstract}

\maketitle

\section{Introduction} 
Many-Body Localized (MBL) systems avoid thermalization under their own unitary dynamics and have been the subject of extensive study for over a decade \cite{anderson1958absence,fleishman1980interactions,pal2010many,basko2006metal,oganesyan2007localization,nandkishore2015many, abanin2019colloquium,gornyi2005interacting}. Research into MBL has primarily relied on numerical simulations of 1D chains, revealing an emergent phase transition between the MBL and thermal phases, mediated by the so-called critical phase in finite-size systems \cite{SCHOLLWOCK201196, PhysRevB.94.214206, PhysRevB.91.081103, PhysRevB.94.241104, khemani2017critical, PhysRevX.5.041047, PhysRevX.7.021013,PhysRevX.5.031033,PhysRevLett.122.040601,PhysRevB.102.125134, PhysRevB.107.115132}. In addition to numerical analyses, the phenomenological proposal of l-bits \cite{huse2014phenomenology, serbyn2013local}, strong disorder Renormalization Group (RG) techniques \cite{igloi2018strong, PhysRevB.99.224205, PhysRevLett.122.040601, PhysRevX.5.031032}, and Imbrie's systematic approach to diagonalizing a particular random Hamiltonian \cite{imbrie2016many} represent the primary analytical results on the existence of MBL.

Capturing the MBL dynamics through numerical simulations requires access to the asymptotic limits of infinite time and infinite length in 1D disordered chains, which has proven to be even more challenging than originally anticipated \cite{sierant2022challenges, morningstar2022avalanches, Sierant_2025}. This difficulty has sparked ongoing debates about the stability and existence \cite{PhysRevE.104.054105, PhysRevE.102.062144} of the MBL phase, suggesting that MBL is limited to finite 1D chains, and breaks down in the thermodynamic limit. These debates are primarily focused on many-body resonances and the emergence of local Markovian baths in the chain, which can lead to thermal avalanches in the system \cite{Panda_2020, sels2022bath, crowley2022constructive, PhysRevLett.124.243601,PhysRevB.100.104204, PhysRevLett.124.186601, abanin2021distinguishing, PhysRevB.102.100202}.

Prior to these debates, other studies had pointed to destabilizing effects from rare, weakly-disordered regions (known as Griffiths regions) on MBL dynamics \cite{luitz2017small, de2017stability}. More recently, numerical studies of these rare regions via open-system equations (Lindblad formulation) have suggested that the MBL phase transition may occur at significantly stronger disorder, i.e., higher randomness in the chain, compared to previous understandings \cite{morningstar2022avalanches}. Experimental studies have confirmed that the destabilizing effects of rare regions diminish as their length shrinks \cite{rubio2019many, leonard2023probing}. Additionally, numerical investigations of 1D chains have shown unbounded growth of subsystems entanglement entropy \cite{PhysRevLett.124.243601, bardarson2012unbounded}, challenging the phenomenology of the MBL phase, and have suggested that MBL could break down even in the absence of rare Griffiths regions \cite{PhysRevB.103.024203}.

\par Despite extensive research in the field, as discussed above, fundamental controversies persist regarding the existence, stability, and dynamical properties of the Many-Body Localization (MBL) phase. In this paper, we introduce a new approach to studying MBL by analyzing entanglement dynamics of subsystems in the frequency domain. Specifically, we analytically identify stable quasi-periodic behavior in the entanglement evolution of subsystems in the thermodynamic limit. These quasi-periodic features manifest as singularities in the frequency spectrum of entanglement dynamics, appearing as delta functions at non-zero frequencies in the Fourier domain or as poles on the imaginary axis in the Laplace domain. While such singularities are expected in finite systems, their persistence in infinite chains and at long time scales is incompatible with thermal behavior, where entanglement is expected to saturate over time \cite{rigol2008thermalization, kaufman2016quantum}. The presence of singularities indicates non-saturating, quasi-periodic dynamics. Hence, by proving the persistence of these singularities in thermodynamic limit, we rule out the possibility of exponentially slow thermalization and establish the robustness of MBL dynamics in disordered chains, which has been a point of controversy in the literature \cite{Panda_2020}.

\par Many-body localization (MBL) can be understood through the analogy of a music track with distinct beats and instruments. Imagine a song where each instrument plays a well-defined frequency mode, forming the track's rhythm. These beats remain clear throughout the song, just as MBL preserves localized dynamics over time. Now, suppose we gradually increase the number of instrument players—in a thermalizing system, where the structure is lost over time, the overlapping contributions from an infinitely large ensemble of instruments would create a chaotic, unstructured sound, obscuring the original beat until no clear rhythm remains. This mirrors thermalization, where information spreads, and subsystems entanglement grows without bound in a spin chain. However, if the effects of these additional instruments remain significantly weaker than the dominant beats, the track maintains its recognizable initial rhythm indefinitely, even as more instruments join, resembling MBL. In our frequency-domain analysis, these persistent beats appear mathematically as delta functions in the Fourier domain and poles in the Laplace domain, marking stable, quasi-periodic (rhythmic) dynamics. Our results show that, in the presence of disorder, these delta functions (or poles) persist throughout the system's evolution, proving that MBL remains stable—just as a song with a dominant rhythm can still be recognized despite added background noise.

\par The reported quasi-periodicity of subsystems' dynamics in the MBL phase contrasts sharply with the widely established logarithmic growth of subsystems' entanglement and other dynamical indicators of thermal behavior, such as the commonly used imbalance, e.g., \cite{serbyn2013universal,bardarson2012unbounded,lukin2019probing,leonard2023probing}. We demonstrate that this observed logarithmic growth can naturally arise from the employed ensemble-averaging approach, which is usually, and frequently, employed in MBL studies e.g., \cite{serbyn2013universal, bardarson2012unbounded, lukin2019probing, leonard2023probing, PhysRevE.104.054105, sierant2022challenges, vsuntajs2020quantum, khemani2017critical}. However, due to the inherent non-ergodicity of the MBL phase—manifested through its quasi-periodic dynamics—such averaging techniques are fundamentally prohibited, making MBL invisible in the averaged dynamics. Consequently, we show that MBL should instead be investigated within individual realizations of the disordered dynamics. In support of this claim, we present additional evidence from existing numerical and experimental studies in the literature (Section \ref{nonerg}). Using the song-track analogy, simultaneously playing (equivalent to averaging) various symphonies of Johannes Brahms—each a coherent Romantic masterpiece (analogous to MBL behavior)—would result in incoherent, chaotic noise (analogous to thermal behavior), characteristically devoid of emotion. In fact, the statistical properties of the resultant noise might well match those obtained from overlapping works by Richard Wagner, Brahms' rival. However, even if their averages yield the same statistical characteristics, this does not imply that Brahms' and Wagner's works share the same artistic character. Analogously, if the ensemble-averaged dynamics of individually MBL spin chains exhibit thermal characteristics (e.g., unbounded entanglement growth), it does not necessarily imply that each individual chain is thermal.

\par To facilitate the analysis of the frequency spectrum of individual spins, we first define the \textit{$\epsilon$-ideally disordered region}s as a mathematical framework to model and quantify disorder within the chain. This concept is later employed to prove the convergence of the perturbative series describing the entanglement dynamics of subsystems. These regions are contrasted with weakly disordered areas in the chain, commonly referred to as rare Griffiths regions in the literature \cite{PhysRevX.5.031032}. Our analysis considers the effects of both $\epsilon$-ideally disordered regions and rare Griffiths regions. 
\par As the central result of this paper, we provide an analytical proof for the existence of a disorder strength interval where singularities in the frequency spectrum of individual spins persist in the thermodynamic limit. This result proves that, in sufficiently strong disordered Heisenberg chains, the entanglement dynamics of subsystems remain quasi-periodic, demonstrating the stability of MBL dynamics, even in the presence of rare Griffiths regions. Furthermore, we demonstrate that the proven quasi-periodicity of subsystem entanglement fundamentally prohibit the use of numerical ensemble averaging—a method commonly employed in the literature to study MBL- due to the breakdown of ergodicity. This methodological limitation can lead to the observed logarithmic, unbounded growth of entanglement in this regime, which fails to capture the quasi-periodic nature of the MBL phase. Our work contributes to the foundational understanding of MBL by establishing its stability through a new analytical approach. By introducing a concrete criterion for MBL—the presence of stable singularities in the frequency domain—our findings contribute to resolving a key controversy in the literature regarding the nature and stability of the MBL phase.

 {This paper is organized as follows. In Section \ref{prel}, we introduce the model, specifically the disordered Heisenberg chain (subsection \ref{Hchain}), define $\epsilon$-ideally disordered regions (subsection \ref{epsiloni}), and discuss how MBL can be identified through quasi-periodic features, which manifest as singularities in the frequency domain (subsection \ref{spectrum}). In Section \ref{sec3}, we derive the random evolution of entanglement for individual spins by analyzing both low-order ($4^{\text{th}}$-order) perturbations of the Hamiltonian (subsection \ref{lowoo}) and higher-order contributions (subsection \ref{higho}). Building on these results and our earlier discussion on the connection between MBL and singularities in the frequency domain, we demonstrate in Section \ref{smoothness} that MBL is directly reflected in the smoothness of the Fourier spectrum of subsystem entanglement dynamics. In this section, we establish a concrete criterion for proving the stability of MBL by analyzing the energy distribution across frequency modes in the entanglement evolution. Our main proof of the stability of MBL is presented in Section \ref{stabil}, where we systematically account for contributions from both highly disordered ($\epsilon$-ideally disordered) regions (subsection \ref{stabil-eps}) and Griffiths regions (subsection \ref{gr}). As a key implication of our findings, Section \ref{nonerg} discusses why MBL cannot be reliably captured using traditional numerical methods, particularly ensemble averaging. To support this claim, we first provide a formal mathematical argument (subsection \ref{nonegr1}) demonstrating the fundamental incompatibility between ensemble averaging and the non-ergodic nature of MBL. We then explain how this limitation has led to misinterpretations of logarithmic entanglement growth in MBL systems (subsection \ref{nonerg2}). Furthermore, we highlight numerical and experimental studies from the literature that illustrate the failure of averaging techniques in detecting MBL (subsection \ref{numrr}). Finally, Section \ref{conc} provides concluding remarks based on our findings. For the reader’s convenience, the Appendices include detailed mathematical proofs and numerical demonstrations supporting our results.}

\section{Preliminary definitions}\label{prel}

\subsection{Disordered Heisenberg Chains}\label{Hchain}

\par The following Hamiltonian describes a one-dimensional lattice of \( N \) spin-\(\frac{1}{2}\) particles with nearest-neighbor spin-spin exchange interactions and additional disorder induced by random fields at each lattice site. The chain is initially in the anti-ferromagnetic order.

\begin{equation}\label{Hamil}
    \mathbf{H}={{J\sum_{k=1}^{N-1}{\mathbf{S}_k\mathbf{S}_{k+1}}}}
    +{\sum_{k=1}^{N}{h_k{S}_{k}^{z}}},
\end{equation}
where \( \mathbf{S}_k = \{S_k^x, S_k^y, S_k^z\} \) represents the spin operator for the \( k \)-th particle, \( J \) is the nearest-neighbor coupling strength, and \( h_k \) are independent random fields. The first term in the Hamiltonian represents the spin-spin interaction, while the second term accounts for the disordered local fields. The random fields $h_k$, which introduce disorder into the system, are drawn from a probability density \( P_h(x) \). In this work, as in most studies of the many-body localization (MBL) phenomenon, \( P_h(x) \) is assumed to be the uniform distribution over the interval \([-W, W]\).

\subsection{$\epsilon$-Ideally Disordered Regions}\label{epsiloni}

\begin{figure}
    \centering

    \begin{tikzpicture}[line width=2]
    \node at (-3.2,0.3) {$\cdots$};
    \node at (3.55,0.3) {$\cdots$};
    \begin{scope}[xshift=-3.7cm, yshift=-0.72cm]
 \mycolorbar{2}{0.3}{darkred!90!white,darkblue!90!white} 
 \node at (-0.6,1.8) {$\mathsf{W}$};
 \node at (-0.7,0.15) {$\mathsf{-W}$};
    \end{scope}

          \begin{scope}[shift={(-0.06,0)}]
              \draw[blue!50!black,-{Cone[width'=0 1,pitch=40]}](0,0)(0,0)(0,0)--({.5*cos(50)},0.5);
        \shade[ball color=red] (0.175,.24) circle(0.15);
        \begin{scope}[shift={(0.2,0.3)}]
        \fill[darkred!20!white, path fading=south,opacity=0.9] (-0.25,1) -- (0.25,1) -- (0.08,0)--(-0.08,0) -- cycle;
        \fill[darkred!20!white,path fading=north]  (0.08,0)--(0.25,-1)--(-0.25,-1)--(-0.08,0) -- cycle;
        \end{scope}
        \end{scope}
        \begin{scope}[shift={(1.05,0)}]
              \draw[blue!50!black,-{Cone[width'=0 1,pitch=-40]}](0,0)(0,0)(0,0)--({-.55*cos(-40)},0.55);
        \shade[ball color=red] (-0.22,.29) circle(0.15);  
                \begin{scope}[shift={(-0.21,0.3)}]
        \fill[darkred!40!white, path fading=south,opacity=0.9] (-0.25,1) -- (0.25,1) -- (0.08,0)--(-0.08,0) -- cycle;
        \fill[darkred!40!white,path fading=north]  (0.08,0)--(0.25,-1)--(-0.25,-1)--(-0.08,0) -- cycle;
        \end{scope}
        \end{scope}
        \begin{scope}[shift={(1.5,0)}]
              \draw[blue!50!black,-{Cone[width'=0 1,pitch=150]}](0,0)(0,0)(0,0)--({-.55*cos(100)},0.55);
        \shade[ball color=red] (0.05,.29) circle(0.15);
        \begin{scope}[shift={(0.05,0.3)}]
        \fill[darkred!18!white, path fading=south,opacity=0.9] (-0.25,1) -- (0.25,1) -- (0.08,0)--(-0.08,0) -- cycle;
        \fill[darkred!18!white,path fading=north]  (0.08,0)--(0.25,-1)--(-0.25,-1)--(-0.08,0) -- cycle;
        \end{scope}
        \end{scope}
        \begin{scope}[shift={(2,0.4)}]
              \draw[blue!50!black,-{Cone[width'=0 1,pitch=45]}](0,0)(0,0)(0.48,0)--(0.1,{-.35*cos(45)});
        \shade[ball color=red] (0.25,-.12) circle(0.15);
         \begin{scope}[shift={(0.25,-0.1)}]
        \fill[darkblue!38!white, path fading=south,opacity=0.9] (-0.25,1) -- (0.25,1) -- (0.08,0)--(-0.08,0) -- cycle;
        \fill[darkblue!38!white,path fading=north]  (0.08,0)--(0.25,-1)--(-0.25,-1)--(-0.08,0) -- cycle;
        \end{scope}
        \end{scope}
        \begin{scope}[shift={(-0.8,0.14)}]
              \draw[blue!50!black,-{Cone[width'=0 1,pitch=45]}](0,0)(0,0)(0.48,0)--(0,{.35*cos(45)});
        \shade[ball color=red] (0.25,0.12) circle(0.15); 
        \begin{scope}[shift={(0.25,0.15)}]
        \fill[darkblue!68!white, path fading=south,opacity=0.9] (-0.25,1) -- (0.25,1) -- (0.08,0)--(-0.08,0) -- cycle;
        \fill[darkblue!68!white,path fading=north]  (0.08,0)--(0.25,-1)--(-0.25,-1)--(-0.08,0) -- cycle;
        \end{scope}
        \end{scope}
        \begin{scope}[shift={(-1.3,0)}]
              \draw[blue!50!black,-{Cone[width'=0 1,pitch=130]}](0,0)(0,0)(0,0)--({-.55*cos(100)},0.55);
        \shade[ball color=red] (0.05,.29) circle(0.15); 
        \begin{scope}[shift={(0.05,0.29)}]
        \fill[darkred!68!white, path fading=south,opacity=0.9] (-0.25,1) -- (0.25,1) -- (0.08,0)--(-0.08,0) -- cycle;
        \fill[darkred!68!white,path fading=north]  (0.08,0)--(0.25,-1)--(-0.25,-1)--(-0.08,0) -- cycle;
        \end{scope}
        \end{scope}
        \begin{scope}[shift={(2.75,0.05)}]
              \draw[blue!50!black,-{Cone[width'=0 1,pitch=40]}](0,0)(0,0)({.5*cos(50)},0)--(0,0.5);
        \shade[ball color=red] (0.175,.24) circle(0.15); 
        \begin{scope}[shift={(0.18,0.25)}]
        \fill[darkred!68!white, path fading=south,opacity=0.9] (-0.25,1) -- (0.25,1) -- (0.08,0)--(-0.08,0) -- cycle;
        \fill[darkred!68!white,path fading=north]  (0.08,0)--(0.25,-1)--(-0.25,-1)--(-0.08,0) -- cycle;
        
        \end{scope}
        \end{scope}
 \begin{scope}[shift={(-2.2,0.4)}]
              \draw[blue!50!black,-{Cone[width'=0 1,pitch=55]}](0,0)(0,0)(0,0)--(.48,{-.35*cos(45)});
        \shade[ball color=red] (0.25,-.12) circle(0.15); 

                \begin{scope}[shift={(0.26,-0.11)}]
        \fill[darkblue!78!white, path fading=south,opacity=0.9] (-0.25,1) -- (0.25,1) -- (0.08,0)--(-0.08,0) -- cycle;
        \fill[darkblue!78!white,path fading=north]  (0.08,0)--(0.25,-1)--(-0.25,-1)--(-0.08,0) -- cycle;
        \end{scope}
        \end{scope}
        \begin{scope}[shift={(-2.45,0.05)}]
              \draw[blue!50!black,-{Cone[width'=0 1,pitch=-30]}](0,0)(0,0)(0,0)--({-.35*cos(30)},0.4);
        \shade[ball color=red] (-0.16,0.2) circle(0.15);           \begin{scope}[shift={(-0.15,0.24)}]
        \fill[darkblue!20!white, path fading=south,opacity=0.9] (-0.25,1) -- (0.25,1) -- (0.08,0)--(-0.08,0) -- cycle;
        \fill[darkblue!20!white,path fading=north]  (0.08,0)--(0.25,-1)--(-0.25,-1)--(-0.08,0) -- cycle;
        \end{scope}  
        \end{scope}
    \begin{scope}[fill opacity = 0.1,thin]
    \draw[magenta!25,fill=magenta!40] (-0.185,-0.8) rectangle +(2.03,2.2);
    \end{scope}
    \begin{scope}[fill opacity = 0.1,thin]
    \draw[teal!25,fill=teal!40] (-2.95,-0.8) rectangle +(2.7,2.2);
    \end{scope}
     \begin{scope}[fill opacity = 0.1,thin]
    \draw[teal!25,fill=teal!40] (1.91,-0.8) rectangle +(1.38,2.2);
    \end{scope}
    \end{tikzpicture}

    \caption{The figure illustrates a disordered (random) spin chain, where disorder is introduced through random fields ($h_i$ in (\ref{Hamil})), drawn uniformly from the interval $[-W,W]$. These random fields are represented by shaded highlights for individual spins. Regions exhibiting significant variations in random field values (characterized by diverse shadings) are referred to as $\epsilon$-ideally disordered regions. The level of disorder in these regions—formally defined in Section \ref{epsiloni}—is quantified by a parameter $\epsilon$, which is inversely related to disorder strength. Separating these highly disordered regions are segments with relatively uniform disorder (illustrated by the middle box), termed rare Griffiths regions.}
    \label{fig:121}
\end{figure}
The random structure of the chain (through $h_i$ in (\ref{Hamil})) naturally creates regions of varying disorder strength. In regions of ``low" (as will be defined) disorder, adjacent sites experience random fields with low variance, forming what are known as Griffiths regions. In these regions, most eigenstates of the system are expected to be non-local and extend across the region. These regions can trigger thermalization via avalanche mechanisms \cite{ha2023many, morningstar2022avalanches}. Conversely, regions of strong collective randomness support site-localized (dressed) eigenstates. We define these strongly disordered areas as \(\epsilon\)-\textit{ideally disordered regions}, formally defined in Appendix \ref{epsil} (see Figure \ref{fig:121}).

\subsection{Spins' entanglement dynamics and MBL}\label{spectrum}

\begin{figure*}
\flushleft
    \begin{tikzpicture}[scale=1]

    \draw[dashed, darkblue!60!white, rounded corners, thick] (-1,3.7) -- (17.1,3.7) -- (17.1,-5)-- (11.4,-5)--(11.4,-1) --(-1,-1) -- cycle;

       \draw[dashed, darkblue!60!white, rounded corners, thick] (-1,-1.3) -- (11.2,-1.3) -- (11.2,-5)-- (-1,-5)-- cycle;
  \begin{axis}[xmin=-7,
        xmax=7,
        ymin=0,
        ymax=4.5,
        ytick={0,4},
        yticklabels={,},
        xtick={0},
        xlabel={\textsf{Frequency} $\mathsf{\omega}$},
        ylabel={$|\tilde{\mathcal{Q}}_\mathcal{M}(i\omega)|$},
        width = 0.35\textwidth,
    	height = 0.25\textwidth]
     
     \draw[-stealth,very thick,darkred] (0,0) -- (0,4) ;
\draw[-stealth,very thick,darkred] (2.2,0) -- (2.2,1.5) ;
\draw[-stealth,very thick,darkred] (-2.2,0) -- (-2.2,1.5) ;
\draw[-stealth, thick,darkred] (2.35,0) -- (2.35,0.25) ;
  \draw[-stealth, thick,darkred] (-2.35,0) -- (-2.35,0.25) ;
  \draw[-stealth, thick,darkred] (1.8,0) -- (1.8,0.25) ;
  \draw[-stealth, thick,darkred] (-1.8,0) -- (-1.8,0.25) ;
    \draw[-stealth, thick,darkred] (4,0) -- (4,0.25) ;
  \draw[-stealth, thick,darkred] (-4,0) -- (-4,0.25) ;
      \draw[-stealth,thick,darkred] (0.2,0) -- (0.2,0.25) ;
  \draw[-stealth, thick,darkred] (-0.2,0) -- (-0.2,0.25) ;

     \draw[-stealth,very thick,darkred] (4.5,0) -- (4.5,0.6) ;
\draw[-stealth,very thick,darkred] (-4.5,0) -- (-4.5,0.6) ;
\draw[-stealth, thick,darkred] (5.2,0) -- (5.2,0.15) ;
  \draw[-stealth, thick,darkred] (-5.2,0) -- (-5.2,0.2) ;
  \draw[-stealth, thick,darkred] (4.8,0) -- (4.8,0.15) ;
  \draw[-stealth, thick,darkred] (-4.8,0) -- (-4.8,0.2) ;

      \draw[-stealth,thick,darkred] (0.4,0) -- (0.4,0.2) ;
  \draw[-stealth, thick,darkred] (-0.4,0) -- (-0.4,0.2) ;

     \end{axis}

     \node at (0.4,3.2) {\small{\textsf{(a)}}};
     \node at (0.7,2.5) {$\mathsf{T\rightarrow \infty}$ };
     \node at (0.8,2.1) {\textsf{few-body}};
     \end{tikzpicture}

\centering 
\vspace{-8.5cm}
\begin{tikzpicture}[scale=1]

       \begin{axis}[xmin=-7,
        xmax=7,
        ymin=0,
        ymax=4.5,
        ytick={0,4},
        yticklabels={,},
        xtick={0},
        xlabel={\textsf{Frequency} $\omega$},
        width = 0.35\textwidth,
    	height = 0.25\textwidth]

        \draw[darkblue, thick] plot[smooth] coordinates {(0.06,4) (0.1,1) (0.2,0.5) (0.3,0.33) (0.5, 0.2) (0.8, 0.125) (1, 0.1) (1.4, 0.12) (1.8,0.4) (2.2,1.5) (2.6, 0.4) (3,0.06)( 4,0.2) (4.5, 0.5) (5,0.2) (6,0.05)  (8,0.02)} ;

   \draw[darkblue, thick] plot[smooth] coordinates {(-0.06,4) (-0.1,1) (-0.2,0.5) (-0.3,0.33) (-0.5, 0.2) (-0.8, 0.125) (-1, 0.1) (-1.4, 0.12) (-1.8,0.4) (-2.2,1.5) (-2.6, 0.4) (-3,0.06)( -4,0.2) (-4.5, 0.5) (-5,0.2) (-6,0.05)  (-8,0.02)} ;

    \node at (-4.5,3.9) {$\mathsf{T\sim finite}$ };
   \node at (-3.5,3.2) {\textsf{many/few-body}};
     \end{axis}
   \node at (0.4,3.2) {\small{\textsf{(b)}}};

\end{tikzpicture}

  \flushright
\vspace{-4.7cm}
\begin{tikzpicture}[scale=1]

       \begin{axis}[xmin=-7,
        xmax=7,
        ymin=0,
        ymax=4.5,
        ytick={0,4},
        yticklabels={,},
        xtick={0},
        xlabel={\textsf{Frequency} $\omega$},
        width = 0.35\textwidth,
    	height = 0.25\textwidth]
     
     \draw[-stealth,very thick,darkred] (0,0) -- (0,4) ;

     \draw[-stealth,thick,darkred] (2.2,0) -- (2.2,0.8) ;
\draw[-stealth,thick,darkred] (-2.2,0) -- (-2.2,0.8) ;
  \draw[-stealth, thick,darkred] (4.5,0) -- (4.5,0.4) ;
\draw[-stealth, thick,darkred] (-4.5,0) -- (-4.5,0.4) ;

   \draw[darkred] plot[smooth] coordinates {(0.06,3.5) (0.1,0.8) (0.2,0.35) (0.3,0.2) (0.5, 0.1) (0.8, 0.05) (1, 0.03) (1.4, 0.06) (1.8,0.15) (2.2,0.4) (2.6, 0.1) (3,0.02)( 4,0.03) (4.5, 0.18) (5,0.04) (6,0.04)  (8,0)} ;

   \draw[darkred] plot[smooth] coordinates {(-0.06,3.5) (-0.1,0.8) (-0.2,0.35) (-0.3,0.2) (-0.5, 0.1) (-0.8, 0.05) (-1, 0.03) (-1.4, 0.06) (-1.8,0.15) (-2.2,0.4) (-2.6, 0.1) (-3,0.02)( -4,0.03) (-4.5, 0.18) (-5,0.04) (-6,0.04)  (-8,0)} ;

   \draw[darkblue!50!darkred] plot[smooth] coordinates {(0.06,3.5) (0.1,1.2) (0.2,0.45) (0.3,0.28) (0.5, 0.18) (0.8, 0.12) (1, 0.1) (1.4, 0.13) (1.8,0.35) (2.2,1.2) (2.6, 0.26) (3.1,0.08)(3.5,0.1)( 4,0.3) (4.5, 0.6) (5,0.3) (6,0.05)  (8,0.01)} ;

\draw[darkblue!50!darkred] plot[smooth] coordinates {(-0.06,3.5) (-0.1,1.2) (-0.2,0.45) (-0.3,0.28) (-0.5, 0.18) (-0.8, 0.12) (-1, 0.1) (-1.4, 0.13) (-1.8,0.35) (-2.2,1.2) (-2.6, 0.26) (-3.1,0.08)(-3.5,0.1)( -4,0.3) (-4.5, 0.6) (-5,0.3) (-6,0.05)  (-8,0.01)} ;

    \draw[darkblue, yscale=1.3] plot[smooth] coordinates {(0.06,2.8) (0.1,1.5) (0.2,0.8) (0.3,0.5) (0.5, 0.28) (0.8, 0.2) (1, 0.18) (1.4, 0.2) (1.8,0.5) (2.2,1.6) (2.6, 0.4) (3.1,0.18)(3.5,0.22)( 4,0.44) (4.5, 0.8) (5,0.4) (6,0.1)  (8,0.05)} ;

    \draw[darkblue, yscale=1.3] plot[smooth] coordinates {(-0.06,2.8) (-0.1,1.5) (-0.2,0.8) (-0.3,0.5) (-0.5, 0.28) (-0.8, 0.2) (-1, 0.18) (-1.4, 0.2) (-1.8,0.5) (-2.2,1.6) (-2.6, 0.4) (-3.1,0.18)(-3.5,0.22)( -4,0.44) (-4.5, 0.8) (-5,0.4) (-6,0.1)  (-8,0.05)} ;

        \node at (-3.6,3.95) {$\mathsf{T: {\color{darkblue}{finite}}\rightarrow {\color{darkred}{\mathsf{\infty}}}}$ };
   \node at (-4.2,3.2) {\textsf{many-body}};
     \end{axis}
     \node at (0.7,3.2) {\small{\textsf{(c)-MBL}}};

\end{tikzpicture}

\flushright 
\vspace{-0.3cm}
\begin{tikzpicture}[scale=1]

       \begin{axis}[xmin=-7,
        xmax=7,
        ymin=0,
        ymax=4.5,
        ytick={0,4},
        yticklabels={,},
        xtick={0},
        ylabel={$|\tilde{\mathcal{Q}}_\mathcal{M}(i\omega)|$},
        width = 0.35\textwidth,
    	height = 0.25\textwidth]
     
     \draw[-stealth,very thick,darkred] (0,0) -- (0,4) ;

   \draw[darkred, thin] plot[smooth] coordinates {(0.06,2.5) (0.1,0.5) (0.4, 0.17)  (0.7, 0.11)  (1.5, 0.08)   (8,0.02)} ;

 \draw[darkred] plot[smooth] coordinates {(-0.06,2.5) (-0.1,0.5) (-0.4, 0.17)  (-0.7, 0.11)  (-1.5, 0.08)   (-8,0.02)} ;

   \draw[darkblue!50!darkred, yscale=1.5] plot[smooth] coordinates {(0.06,2.5) (0.1,1.3) (0.2,0.5) (0.3,0.28) (0.5, 0.18) (0.8, 0.12) (1, 0.1) (1.4, 0.13) (1.8,0.3) (2.2,0.6) (2.6, 0.2) (3.1,0.08)(3.5,0.1)( 4,0.2) (4.5, 0.35) (5,0.2) (6,0.05)  (8,0.01)} ;

   \draw[darkblue!50!darkred, yscale=1.5] plot[smooth] coordinates {(-0.06,2.5) (-0.1,1.3) (-0.2,0.5) (-0.3,0.28) (-0.5, 0.18) (-0.8, 0.12) (-1, 0.1) (-1.4, 0.13) (-1.8,0.3) (-2.2,0.6) (-2.6, 0.2) (-3.1,0.08)(-3.5,0.1)( -4,0.2) (-4.5, 0.35) (-5,0.2) (-6,0.05)  (-8,0.01)} ;

    \draw[darkblue, yscale=1.5] plot[smooth] coordinates {(0.06,2.4) (0.1,1.5) (0.2,0.8) (0.3,0.5) (0.5, 0.28) (0.8, 0.2) (1, 0.18) (1.4, 0.2) (1.8,0.5) (2.2,1.6) (2.6, 0.4) (3.1,0.18)(3.5,0.22)( 4,0.44) (4.5, 0.8) (5,0.4) (6,0.1)  (8,0.05)} ;

    \draw[darkblue, yscale=1.5] plot[smooth] coordinates {(-0.06,2.4) (-0.1,1.5) (-0.2,0.8) (-0.3,0.5) (-0.5, 0.28) (-0.8, 0.2) (-1, 0.18) (-1.4, 0.2) (-1.8,0.5) (-2.2,1.6) (-2.6, 0.4) (-3.1,0.18)(-3.5,0.22)( -4,0.44) (-4.5, 0.8) (-5,0.4) (-6,0.1)  (-8,0.05)} ;

        \node at (-3.6,3.95) {$\mathsf{T: {\color{darkblue}{finite}}\rightarrow {\color{darkred}{\mathsf{\infty}}}}$ };
   \node at (-4.2,3.2) {\textsf{many-body}};
     \end{axis}
     \node at (0.9,3.2) {\small{\textsf{(d)-Thermal}}};

\end{tikzpicture}

\flushleft 

\vspace{-3.9cm}  \hspace{-0.1cm}
\begin{tikzpicture}[scale=0.6]

\draw [decorate,decoration={brace,amplitude=6pt},xshift=0pt,yshift=0pt]
(0,-0.15) -- (0,3.1)node [black,midway,xshift=-25pt] {\footnotesize
\textsf{Average}};

\draw [decorate,decoration={brace,amplitude=6pt},xshift=0pt,yshift=0pt]
(11.2,3.1) -- (11.2,-0.15)node [black,midway,xshift=12pt] {$\mathsf{=}$};
\hspace{0 cm} \vspace{-2cm}
      \begin{axis}[xmin=-7,
        xmax=7,
        ymin=0,
        ymax=4.5,
        ytick={0,4},
        yticklabels={,},
        xtick={-5,0,5},
        xticklabels={,0,},
        width = 0.35\textwidth,
    	height = 0.25\textwidth]
     
     \draw[-stealth,very thick,darkred] (0,0) -- (0,4) ;

     \draw[-stealth,thick,darkred] (2.2,0) -- (2.2,1.5) ;
\draw[-stealth,thick,darkred] (-2.2,0) -- (-2.2,1.5) ;
  \draw[-stealth, thick,darkred] (4.5,0) -- (4.5,1) ;
\draw[-stealth, thick,darkred] (-4.5,0) -- (-4.5,1) ;

    \draw[darkred, yscale=1.5] plot[smooth] coordinates {(0.06,2.5) (0.1,0.8) (0.2,0.35) (0.3,0.2) (0.5, 0.1) (0.8, 0.05) (1, 0.03) (1.4, 0.06) (1.8,0.15) (2.2,0.4) (2.6, 0.1) (3,0.02)( 4,0.03) (4.5, 0.18) (5,0.04) (6,0.04)  (8,0)} ;

   \draw[darkred, yscale=1.5] plot[smooth] coordinates {(-0.06,2.5) (-0.1,0.8) (-0.2,0.35) (-0.3,0.2) (-0.5, 0.1) (-0.8, 0.05) (-1, 0.03) (-1.4, 0.06) (-1.8,0.15) (-2.2,0.4) (-2.6, 0.1) (-3,0.02)( -4,0.03) (-4.5, 0.18) (-5,0.04) (-6,0.04)  (-8,0)} ;
     
     \end{axis}

     \node at (-1.7,4.1) {\small{\textsf{(e)}}};

 \node at (4.9,1.4) {\normalsize $\mathsf{,}$};
\node at (10.6,1.4) {\normalsize $\mathsf{, \cdots}$};

\end{tikzpicture}

\vspace{-2.05cm} 

\begin{tikzpicture}[scale=0.6]

 \hspace{4.5cm}
       \begin{axis}[xmin=-7,
        xmax=7,
        ymin=0,
        ymax=4.5,
        ytick={0,4},
        yticklabels={,},
        xtick={-5,0,5},
        xticklabels={,0,},
        width = 0.35\textwidth,
    	height = 0.25\textwidth]
     
     \draw[-stealth,very thick,darkred] (0,0) -- (0,2.8) ;

     \draw[-stealth,thick,darkred] (3,0) -- (3,0.6) ;
\draw[-stealth,thick,darkred] (-3,0) -- (-3,0.6) ;
  \draw[-stealth, thick,darkred] (6,0) -- (6,0.4) ;
\draw[-stealth, thick,darkred] (-6,0) -- (-6,0.4) ;

    \draw[darkred, yscale=0.8] plot[smooth] coordinates {(0.06,3) (0.1,0.8) (0.2,0.35) (0.4,0.2) (0.6, 0.1) (0.9, 0.05) (1.2, 0.03) (1.6, 0.06) (2.3,0.15) (3,0.4) (3.5, 0.1) ( 4,0.03) (6, 0.18) (7,0.04)   (8,0)} ;

    \draw[darkred, yscale=0.8] plot[smooth] coordinates {(-0.06,3) (-0.1,0.8) (-0.2,0.35) (-0.4,0.2) (-0.6, 0.1) (-0.9, 0.05) (-1.2, 0.03) (-1.6, 0.06) (-2.3,0.15) (-3,0.4) (-3.5, 0.1) ( -4,0.03) (-6, 0.18) (-7,0.04)   (-8,0)} ;

     \end{axis}

\end{tikzpicture}

\vspace{-2.05cm}

\begin{tikzpicture}[scale=0.6]
 \hspace{8.7cm}
       \begin{axis}[xmin=-7,
        xmax=7,
        ymin=0,
        ymax=4.5,
        ytick={0,4},
        yticklabels={,},
        xtick={0},
        xlabel={\Large\textsf{Frequency} $\omega$},
        width = 0.35\textwidth,
    	height = 0.25\textwidth]
     
     \draw[-stealth,very thick,darkred] (0,0) -- (0,4) ;

 \draw[darkred, yscale=2] plot[smooth] coordinates {(0.06,1.5) (0.2,0.5) (0.8, 0.17)  (1.4, 0.11)  (3, 0.08)   (8,0.02)} ;

 \draw[darkred, yscale=2] plot[smooth] coordinates {(-0.06,1.5) (-0.2,0.5) (-0.8, 0.17)  (-1.4, 0.11)  (-3, 0.08)   (-8,0.02)} ;

     \end{axis}

\end{tikzpicture}

    \caption{Characteristics of the Fourier spectrum of entanglement dynamics for a sample spin in a disordered Heisenberg chain. Each panel schematically illustrates the amplitude of the Fourier transform of a sample spin's entanglement evolution as a function of frequency ($\omega$). (a) In a finite chain, the entanglement dynamics of each spin is governed by a finite number of discrete frequency modes, which correspond to linear combinations of the system Hamiltonian’s eigenvalues. If the entanglement evolution is observed over an infinitely long time \(T\), these modes appear as sharp delta functions in the frequency spectrum (indicated by orange arrows) (b) When the observation time \(T\) is finite, the Fourier spectrum becomes inherently smooth, as delta functions are broadened \cite{exp6}. Moreover, a finite \(T\) effectively limits the correlation length in the chain, meaning that each spin primarily interacts with its nearby neighbors, leaving no fundamentally distinguishing difference between finite and infinite chains. (c)–(d) Possible scenarios in the limit \(T \to \infty\) and in infinite chains: (c) As the correlation length diverges, sharp delta functions can emerge in the Fourier spectrum, which signals many-body localization (MBL). The presence of such delta functions indicates persistent quasi-periodic oscillations in entanglement dynamics, preventing entanglement saturation and thereby inhibiting quantum thermalization. (d) Alternatively, if no sharp delta functions appear even as \(T \to \infty\), entanglement can saturate, indicating quantum thermalization. However, this saturation may occur after an exponentially slow buildup of entanglement. Thus, (c) and (d) distinguish between truly localized dynamics and slow thermalization—an ongoing debate in the field. In this work, we analytically prove that MBL, characterized by emergent singularities as in (c), can occur in sufficiently strongly disordered chains. (e) If one averages over multiple disordered realizations, each exhibiting localized dynamics as in (c), the resulting averaged spectrum becomes smooth, resembling thermal behavior as in (d). This effect arises due to the non-ergodicity of the MBL phase, highlighting that ensemble averaging fails to capture true MBL properties. As a consequence, misleading features, such as logarithmic entanglement growth, may emerge in the averaged spectrum.  {Additional numerical simulations in Figure \ref{fig:figgy4} illustrate the delta functions in the frequency spectrum of the MBL dynamics.} }

    \label{fig222}
\end{figure*}

\par Our approach to detecting {many-body localization (MBL)} is based on analyzing the {entanglement evolution} of individual spins in the {frequency domain}, specifically, the Laplace domain. As highlighted in the Introduction, the presence of singularities (at non-zero frequencies) in the frequency spectrum of individual spins serves as a hallmark of MBL dynamics. This is because, unlike thermalizing systems, MBL systems exhibit non-saturating entanglement growth at long time scales and in the thermodynamic limit. The existence of singularities in the frequency spectrum of entanglement evolution prevents this saturation, thereby leading to MBL.

\par To quantify subsystems' entanglement dynamics, we use the  {newly introduced Quantum Correlation Transfer Function (QCTF) formulation \cite{QCTF, azodi2024measuringentanglementexploitingantisymmetric}. This formulation is summarized in Appendix \ref{intro-qctf}. Using QCTF, the {Laplace transform} of the determinant of the {reduced density matrix} ($\rho_{\mathcal{M}}$) corresponding to individual spins (called subsystem \(\mathcal{M}\)) can be obtained, while avoiding the evaluation of the system's many-body state's evolution. This quantity is defined as}

\begin{equation}\label{lapltr}
    \tilde{\mathcal{Q}}_{\mathcal{M}}(s)=\int_0^T e^{-st} \det{\rho_{\mathcal{M}}(t)} \, dt,
\end{equation}
where $s$ is the Laplace variable and \(T\) is the {simulation time}, the time duration after a quench (from the  {anti-}ferromagnetic order) during which entanglement dynamics is observed.  {Notably, this quantity is an entanglement monotone, and is equivalent to the second-order R\'enyi entropy, fully capturing the subsystem's entanglement evolution (see Appendix \ref{intro-qctf}).} By {singularities in the Laplace domain}, we refer to {Laplace poles}, which can be best introduced through a simple and well-known example: the cosine waveform, $ \cos(ft)=\frac{e^{ift}+e^{-ift}}{2} $. This function transforms into the Laplace domain as $\frac{s}{s^2 + f^2} = \frac{s}{(s + if)(s - if)}$, revealing two Laplace poles at frequencies $\pm if$. The purely imaginary nature of these poles highlights the periodic behavior of the waveform in the time domain. In the {Fourier domain}, these poles correspond to delta functions located at frequencies \( \pm f \). Due to this correspondence, we may use the phrases ``poles" or ``delta functions" interchangeably, depending on the context, to best convey the intended message.
\par The behavior of the frequency spectrum of (\ref{lapltr}) depends critically on the simulation time \(T\), with two main pathways influencing the spectrum: 
 First, the resolution of the transformation scales with $1/T $; Hence, when $T$ is finite, the frequency spectrum is inherently smooth \cite{exp6}.
    Second, the correlation length in the chain grows over time.  {Here, correlation length is a measure of the spatial extent over which a spin becomes entangled with others during time $T$, and grows as correlations spread throughout the system—a process constrained by the finite-speed bounds implied by the Lieb-Robinson theorem \cite{lieb1972finite}}. For finite $T$, each subsystem is effectively interacting within a growing correlation length. Only in the limit $T\rightarrow \infty$ the full many-body dynamics can be studied. 
    Therefore, it is important to distinguish the following two regimes (when $J \ll W$, or in strong disorder regime):

\begin{itemize}
    \item when \(T \sim \mathcal{O}(\hbar W /J^2)\), the time-scale when disordered Hamiltonian dominates \cite{exp7}, the entanglement frequency spectrum \(\tilde{\mathcal{Q}}_{\mathcal{M}}(s)\) primarily incorporates interactions with neighboring spins, featuring a finite number of semi-sharp spikes (they are not fully sharp due to the inherent smoothness of the spectrum for finite $T$) (Fig \ref{fig222}.b). These features can be captured via low-order perturbations of the Hamiltonian (when the interaction Hamiltonian in (\ref{Hamil}) is considered as perturbation to the local Hamiltonian). As $T$ increases, the frequency resolution improves, causing the peaks to sharpen and converge toward delta functions.
    \item When \(T \to \infty\), the correlation length grows (i.e., diverges), and the many-body nature of the system is revealed. Additionally, as the frequency resolution approaches zero, sharp singularities (delta functions) might emerge in the frequency spectrum; \textit{The MBL dynamics can be reliably identified through the presence of delta functions (singularities) in the frequency spectrum (Fig \ref{fig222}.c), signifying quasi-periodic features reminiscent of few-body dynamics that persist despite the influence of many-body interactions.} In thermalizing systems, the frequency spectrum must be smooth, except at the origin (Fig \ref{fig222}.d). To fully capture this region, contributions from all orders of Hamiltonian perturbations must be included.
\end{itemize}
Given the explanation above, surviving singularities in the frequency spectrum of \(\tilde{\mathcal{Q}}_{\mathcal{M}}(s)\) in the $T\rightarrow \infty$ limit signify MBL dynamics. In the following Sections, our goal is to analyze and prove the stability of these singularities in the thermodynamic limit. Figure \ref{fig222} summarizes the above criterion for the stability of MBL.
\par Given the randomness in the chain, the entanglement measure frequency spectrum \(\tilde{\mathcal{Q}}_{\mathcal{M}}(s)\) is characterized through probability distribution functions.

\section{Spins' entanglement dynamics}\label{sec3}
\par In the previous Section, we explained how MBL is identified through the singularities in the frequency spectrum of subsystems' entanglement dynamics. Here, we will examine the behavior of the entanglement spectrum in the two regimes: finite $T$ and \( T \to \infty \), respectively corresponding to the low (second and fourth)-order and higher-order perturbations of the Hamiltonian (\ref{Hamil}). 
\subsection{Low-order perturbation of entanglement dynamics $\tilde{\mathcal{Q}}_{\mathcal{M}}(s)$} \label{lowoo}
\par For spins in an $\epsilon (<1)$- ideally disordered region, the low-order approximation of $\tilde{\mathcal{Q}}_{\mathcal{M}}(s)$, corresponding to entanglement dynamics within a short simulation times \(T \sim \mathcal{O}(\hbar W/J^2)\), has the structure (see the derivation in Appendix \ref{lowo})
\begin{equation}\label{Q22}
\begin{split}
     \tilde{\mathcal{Q}}_{\mathcal{M}}(s)= 
     \frac{2(a_1+a_2)}{s}
     &-a_1 (s-\frac{i}{h}f_1)^{-1}-a_1 (s+\frac{i}{h}f_1)^{-1}\\&-a_2 (s-\frac{i}{h}f_2)^{-1}-a_2 (s+\frac{i}{h}f_2)^{-1}
     \\&+ \text{H.O.Ts},
\end{split}   
\end{equation}
where, $\pm if_{1,2}/\hbar$ are the random, non-zero \textit{Laplace poles}, and $a_{1,2} \in \mathbb{R}^+$ are their corresponding random \textit{amplitudes}. 
This approximation only includes the \textit{dominant poles}, arising from the second-order perturbation of the system's Hamiltonian  {in the QCTF-based derivation of the entanglement dynamics}, which are shown to be exponentially (in disorder) stronger in amplitude compared to the other poles, when the simulation time ($T$) is relatively short. 

\par  {The PDFs describing the random dominant frequencies ($f_{1,2}$) and their amplitudes ($a_{1,2}$), respectively denoted by $P_{F}^{(2)}(f)$ and $P_A^{(2)}(a)$, are derived in Appendices (\ref{A200}, \ref{app45}), illustrated in Figure \ref{fig:PDFs}, and numerically verified in Figure \ref{fig:55}. Details regarding the numerical simulation is included in Figures' captions and in Appendix \ref{appcc}.}

\usetikzlibrary{shapes}
\begin{figure}
    \begin{tikzpicture}[spy using outlines={rectangle, magnification=3.5,connect spies,thin}]
    
    \node at (-1,4.5) {{(a)}};
        \begin{axis}[xmin=5,
        xmax=45,
        ymin=0,
        ymax=0.16,
        ytick={0,0.1},
        xlabel={Frequency $f(\frac{J}{\hbar})$},
        ylabel={$\bar{P}_F^{(2)}(f)$},
        width = 0.45\textwidth,
    	height = 0.35\textwidth]
    
        \node at (10,0.15) {\scriptsize\textcolor{bluegrey}{W=10}};
        \node at (10,0.14) {\scriptsize\textcolor{darkred}{W=15}};
       \node at (10,0.13)
       {\scriptsize\textcolor{darkblue}{W=20}};

       \draw [thin,black] (14.5,0.145)--(16.5,0.145) node[right]{\hspace{1pt} \scriptsize Bulk -Theory};    
       \draw [thin,black,dashed] (14.5,0.135)--(16.5,0.135) node[right]{\hspace{1pt} \scriptsize Edge -Theory};
       
     \node [right] at (29,0.135) {\ding{108} \scriptsize Edge -Numerical};
       \node [right] at (29,0.145) {$\triangle$ \scriptsize Bulk -Numerical};
     \addplot [no marks,very thin, dashed,darkblue]    table [x=xdatt, y=ydatt]{spinDataedgew20.txt};
        
        \addplot [bluegrey,only marks,mark size=0.8pt,opacity=0.6]    table [x=xdat, y=ydat]{spinDataedgew10.txt};
        \addplot [bluegrey,no marks, very thin,dashed]    table [x=xdatt, y=ydatt]{spinDataedgew10.txt};
        \addplot [only marks, darkred,mark size=0.8pt,opacity=0.6]    table [x=xdat, y=ydat]{spinDataedgew15.txt};
        \addplot [no marks, very thin, dashed,darkred]    table [x=xdatt, y=ydatt]{spinDataedgew15.txt};
        \addplot [only marks,darkblue,mark size=0.8pt,opacity=0.6]    table
        [x=xdat, y=ydat]{spinDataedgew20.txt};
        
                \addplot [bluegrey,only marks,mark=triangle,mark size=0.8pt,opacity=0.6]    table [x=xdat, y=ydat]{spinDatabulkw10.txt};
        \addplot [bluegrey,very thin,no marks]    table [x=xdatt, y=ydatt]{spinDatabulkw10.txt};
        \addplot [only marks,darkred,mark=triangle,mark size=0.8pt,opacity=0.6]    table [x=xdat, y=ydat]{spinDatabulkw15.txt};
        \addplot [very thin,no marks, darkred]    table [x=xdatt, y=ydatt]{spinDatabulkw15.txt};
        \addplot [only marks,darkblue,mark=triangle,mark size=0.8pt,opacity=0.6]    table
        [x=xdat, y=ydat]{spinDatabulkw20.txt};
        \addplot [very thin,no marks,darkblue]    table [x=xdatt, y=ydatt]{spinDatabulkw20.txt};

        
        \coordinate (spypoint) at (20.2,0.005);
        \coordinate (spyviewer) at (12,0.105);
        \spy[width=2cm,height=1cm] on (spypoint) in node [fill=white,thin] at (spyviewer);

        \coordinate (spypointt) at (30.2,0.005);
        \coordinate (spyviewerr) at (25,0.105);
        \spy[width=2cm,height=1cm] on (spypointt) in node [fill=white,thin] at (spyviewerr);

        \coordinate (spypointtt) at (40.2,0.005);
        \coordinate (spyviewerrr) at (38,0.105);
        \spy[width=2cm,height=1cm] on (spypointtt) in node [fill=white,thin] at (spyviewerrr);
        
        \end{axis}
    \end{tikzpicture}
    
    \qquad\qquad
        \begin{tikzpicture}
        
        \node at (-1,4.5) {(b)};
        \begin{axis}[xmin=0,
        xmax=3,
        ymin=0,
        ymax=0.13,
        ytick={0,0.1},xtick={0,1/8,1,2,3,4,5},
        xticklabels={0,$\frac{1}{8}$,1,2,3,4,5},
        xlabel={Amplitude $a \big((\frac{J}{W})^2\big)$ },
        ylabel={$\bar{P}_A^{(2)}(a) $},
        width = 0.45\textwidth,
    	height = 0.35\textwidth,legend style={nodes={scale=0.75, transform shape}}]
       
        \addplot [bluegrey,only marks,mark size=1pt]    table [x=xdat, y=ydat]{spinpdfw10-m.txt};
        \addlegendentry{W=10-Numerical};
        \addplot [bluegrey,no marks, dashed]    table [x=xdat, y=ydatt]{spinpdfw10-m.txt};
        \addlegendentry{W=10-Theory};
        \addplot [only marks,darkred,mark size=1pt]    table [x=xdat, y=ydat]{spinpdfw15-m.txt};
        \addlegendentry{W=15-Numerical};
        \addplot [no marks, dashed,darkred]    table [x=xdat, y=ydatt]{spinpdfw15-m.txt};
        \addlegendentry{W=15-Theory}
        \addplot [only marks,darkblue,mark size=1pt]    table
        [x=xdat, y=ydat]{spinpdfw20-m.txt};
        \addlegendentry{W=20-Numerical};
        \addplot [no marks, dashed,darkblue]    table [x=xdat, y=ydatt]{spinpdfw20-m.txt};
        \addlegendentry{W=20-Theory}
        \end{axis}
    \end{tikzpicture}
    
    \caption{\textbf{Numerically constructed PDFs ${P}_F^{(2)}(f)$ and ${{P}}_A^{(2)}(a)$ compared with QCTF theoretical predictions.} This figure shows the discretized PDFs for second-order frequencies and amplitudes ( {in Eq. \ref{SH}}). Note that due to the numerical discretization, vertical axes, which are denoted by $\bar{P}_F^{(2)}(f)$ and $\bar{P}_A^{(2)}(a)$, are scaled (by a function of the (adaptive) length of the bins). In this simulation, the coupling and disorder strengths are $J=1$, $W=10,15 \text{ and } 20$. Sub-figure (a) shows the re-constructed $\bar{P}_F^{(2)}(f)$, for spins at the edges and in the bulk of the chain. The insets show the details of these PDFs at their high-frequency corners, where the edge spins are expected to behave differently than the bulk spins due to their differing first-order energy correction (see Appendix \ref{A200} for more detail). As expected, the different statistics between these two classes of spin subsystems is more noticeable at the lowest disorder strength ($W=10$). With the help of a large number of random samples ($10^6$), this difference is captured. Sub-figure (b) shows the re-constructed $\bar{P}_A^{(2)}(a)$. The precision of the theoretical prediction generally increases with the disorder strength; In particular, the features match accurately near the sharp peaks. The deviation between the numerical PDFs and their respective theoretical predictions at low disorder strengths are due to the fourth-order contributions, which smoothly fade away at higher disorder strengths. }
    \label{fig:55}
\end{figure}

\par Another set of poles on the entanglement frequency spectrum are the fourth-order poles are at located in the vicinity of $\pm 2 f_1$ and $\pm 2 f_2$ in the frequency spectrum, with the general structure

\begin{equation}\label{SH}
\begin{split}
        &a_3 (s-\frac{2i}{h}f_1)^{-1}+a_3 (s+\frac{2i}{h}f_1)^{-1}
        \\ +&a_4 (s-\frac{2i}{h}f_2)^{-1}+a_4 (s+\frac{2i}{h}f_2)^{-1}.
\end{split}
\end{equation}
This class of Laplace poles, referred to as \textit{second harmonics}, will be the focus of our analysis due to their high tractability \cite{exp2}. By {tractability}, we specifically mean the ability to characterize the full set of higher-order poles in the vicinity of the second-harmonic poles. This characterization enables a detailed investigation of the {smoothness}—that is, the presence or absence of singularities— in the frequency spectrum as $T\rightarrow \infty$, within the immediate vicinity of the second-harmonic poles.

\subsection{Higher-order Laplace poles of $\tilde{\mathcal{Q}}
_{\mathcal{M}}(s)$}\label{higho}
\par As the simulation time ($T$) approaches infinity, the correlation length within the chain increases, resulting in the emergence of higher-order poles in the Laplace spectrum of $\tilde{\mathcal{Q}}_{\mathcal{M}}(s)$. These poles correspond to interactions involving spins located at increasingly distant positions. In particular, higher-order perturbations of the Hamiltonian (\ref{Hamil}) give rise to \textit{side-poles} situated near the second-harmonic poles (\ref{SH}). These side-poles contribute to the smoothing of the frequency spectrum of $\tilde{\mathcal{Q}}_{\mathcal{M}}(s)$. In this section, we discuss how these side-poles are characterized in our  {QCTF} analysis, as detailed in Appendices (\ref{high-order}-\ref{unfold}).

\par Each {Laplace pole} of the entanglement measure \( \tilde{\mathcal{Q}}_{\mathcal{M}}(s) \) corresponds to a {linear combination} of four eigenvalues of the system's Hamiltonian (\ref{Hamil}). The {side-poles} associated with the {second harmonic} (\ref{SH}) are identified by finding linear combinations of eigenvalues that are {close} to the second-harmonic poles. These can be systematically determined through a {perturbative scheme} that incorporates all possible perturbation orders, corresponding to interactions between spins spanning the entire chain. As the chain length increases, the number of side-poles grows exponentially, driven by the combinatorial increase in the number of perturbative interactions along the chain, leading to a gradual smoothing of the entanglement frequency spectrum near the second-harmonic poles.

For spins within an {\(\epsilon\)-ideally disordered region}, the {amplitude} and {frequency} of the side-poles can be {upper-bounded} by analyzing the effects of perturbations originating from a given spin and extending outward to further spins, thereby encoding the many-body interactions in the chain. The {\(\epsilon\)-ideally disordered criterion} (\ref{500c}) ensures that the energy of the resulting side-poles asymptotically approaches that of the second-harmonic poles as the perturbations extend further from the initial spin. Hence, through the described systematic procedure, the (infinitely many) side poles of the second-harmonics can be characterized.

\par The analysis can be extended to include side-poles induced by rare Griffiths regions. In these regions, which are assumed to be clean (i.e., disorder free), all spins experience the same local Hamiltonian. Under these conditions, the Hamiltonian can be locally diagonalized using the quasi-particle picture \cite{alcaraz1987surface}. The resulting integrable modes produce side-poles near the second-harmonic poles of the entanglement dynamics. Using perturbation theory, we show that the amplitude and frequency of these side-poles can also be upper-bounded. Importantly, the statistical characteristics of side-poles arising from rare Griffiths regions differ significantly from those generated in $\epsilon$-ideally disordered regions. For a full characterization of side-poles from rare Griffiths regions, refer to the Appendix \ref{griff}.

\section{Smoothness of entanglement frequency spectrum versus MBL} \label{smoothness}
Building on the discussion of the Laplace poles' characterization in the frequency spectrum of the entanglement measure $\tilde{\mathcal{Q}}_{\mathcal{M}}(s)$ in subsection \ref{spectrum}, we now examine how the singularities (non-smooth features) in the frequency spectrum relate to the stability of many-body localized (MBL) dynamics in the thermodynamic limit. The criteria presented in this paper regarding the connection between the spectrum's smoothness, as influenced by side-poles, and MBL dynamics is as follows:  \par  A precondition for thermalization is the existence of a saturation value for the entanglement of spins. Due to the final value theorem, the existence of this saturation value depends on the smoothness of $s\tilde{\mathcal{Q}}_{\mathcal{M}}(s)$. If the side-poles near the second harmonics fail to constitute a smooth behavior, the entanglement measure cannot saturate, and the system remains many-body localized.

It is more illustrative to examine this property in the Fourier domain, i.e., by studying $\eval{\tilde{\mathcal{Q}}_{\mathcal{M}}(s)}_{s=i\omega }$. However, there are analytical complications in this transition. In the Fourier domain, simple imaginary poles of the entanglement measure become Dirac delta functions:
\begin{equation}\label{lap2f}
    a(s-\frac{i}{\hbar}f)^{-1}\Leftrightarrow a \delta(\omega -\frac{f}{\hbar}),
\end{equation}
and this is an immediate signature of quasi-periodic dynamics since all Laplace poles are imaginary, causing the final value theorem to fail. This complication can only be potentially resolved in the thermodynamic limit, where an infinite number of higher-order side-poles exist in any neighborhood of the individual poles, particularly the second harmonic poles. In this limit, it becomes possible to describe the dense accumulation of delta functions via a density function $D(\omega)$ and rewrite the entanglement measure as a continuous function of frequency in the Fourier domain:

\begin{equation}
    \sum_{i=1}^\infty a_i \delta(\omega-\omega_i)\rightarrow \int D(\omega_0) \delta(\omega -\omega_0) d\omega_0=D(\omega).
\end{equation}

This mechanism—where dominant delta functions are smoothed by the contributions of side-poles—is critical for thermal equilibration. In the thermal phase, the Fourier spectrum is expected to become completely smooth, with no singularities other than at $\omega = 0$. In essence, as the disorder strength decreases, all delta functions should ``melt" into a continuous spectrum, signaling the onset of thermalization. If this fails to happen, and residual singularities (delta functions at non-zero frequencies in the Fourier spectrum) persist, then the system must behave many-body localized (these two scenarios are illustrated and contrasted in Figure \ref{fig222} c,d).

To prove the stability of the MBL phase in the next Section, it is sufficient to locate \textit{at least one} persisting singularity (a delta function in the Fourier domain) in the infinite-time ($T\rightarrow \infty$ entanglement spectrum of \textit{at least one} spin in the chain, in the thermodynamic limit $N\rightarrow \infty$. Using this criterion, it is sufficient to prove the persistence of the second-harmonic poles in the thermodynamic limit to demonstrate MBL.

\section{Stability of the MBL dynamics}\label{stabil}
\par Given the above criteria for the stability of the many-body localized (MBL) phase, this section analyzes the smoothness of the frequency spectrum of a spin's entanglement dynamics in the vicinity of the second harmonic poles. 

The critical question to address is: Can the spins' entanglement spectrum with singularities (i.e., Dirac delta functions) persist in the thermodynamic limit? Alternatively, one might hypothesize that, irrespective of disorder strength, the entanglement spectrum becomes smooth in the thermodynamic limit. These two possibilities differentiate between the genuinely quasi-periodic dynamics, and slow growth of entanglement. 

To test the possibility above, we examine the total energy carried by the side-poles and compare it to the energy of the dominant second-harmonic poles. If, by increasing the disorder strength, the total energy carried by the side poles can be made arbitrarily smaller that the energy carried by the second-harmonic poles, then the second harmonic poles can survive in the thermodynamic limit when disorder is sufficiently large.

To consider these factor, two different analyses are performed regarding the effects of $\epsilon$- ideally disordered regions and rare Griffiths regions. 

Notably, since demonstrating the {quasi-periodic} nature of entanglement dynamics for even a {single spin} in the chain suffices to establish {MBL} (as further justified in {Section} \ref{nonerg}), we focus on the entanglement dynamics of a {chosen spin} within an \(\epsilon\)-ideally disordered region. To account for the rare {Griffiths regions}, we will incorporate their effects only on the entanglement dynamics of this selected spin (discussed in {Section} \ref{gr}).

\par  {The central result of this section—and the main goal of this paper—is to demonstrate that in sufficiently strongly disordered chains, the entanglement dynamics of individual subsystems remain quasi-periodic, owing to the stability of isolated peaks in their Fourier spectra. }

 \subsection{Stability of MBL in fully $\epsilon$-ideally disordered regions}\label{stabil-eps}

In Appendices \ref{high-order}-\ref{unfold}, we analyze the entanglement frequency spectrum of spins in fully $\epsilon$-ideally disordered regions. Each side-pole is uniquely identified using two labels representing sets of perturbation Hamiltonians. We show that the asymptotic total energy carried by the side-poles a second-harmonic pole is bounded above the following scale
\begin{equation}\label{50122}
\epsilon^{4i_{min}+6},
\end{equation}
where $i_{min} \geq 1$. 

For $\epsilon \rightarrow 0$, the value of (\ref{50122}), approaches $0$, at least as fast as $\sim \epsilon ^{10}$.
\par Additionally, we find that the energy carried by individual second-harmonic poles scales as $\sim\epsilon^8$ as $\epsilon \rightarrow 0$ (see Appendix \ref{lowo}).
\par In the limits $\epsilon \rightarrow 0$, the energy carried by a second-harmonic pole can be made arbitrarily larger relative to the cumulative energy carried by their corresponding side-poles. Conversely, for $\epsilon \rightarrow 1$, the side-pole energy can become arbitrarily large. By the intermediate value theorem, there must exist a critical value $\epsilon^*$ that marks the transition between the regions where side-poles and the second-harmonics are more energetic than each other. 

Consequently, there must exist a spin in a fully $\epsilon$- ideally disordered region, such that for sufficiently small $\epsilon$ (i.e., large disorder), the continuity of $\mathcal{Q}(i\omega)$ must break down near the second-harmonics.

 {Figure \ref{fig:figgy4} presents a numerical simulation (exact diagonalization) of a disordered Heisenberg chain of $15$ spins at disorder strength $W=10$, up to the simulation time $T=40\frac{\hbar}{J}$. The stability of the dominant poles, in particular the second harmonics, is evident in this numerical simulation. }

\subsection{Effects of rare Griffiths regions on the Stability of the MBL dynamics}\label{gr}
As disorder decreases, the probability of encountering regions in the chain with comparable local fields $h_i$ (on the scale of the coupling strength $J$) increases. These regions, characterized by low field variation, become dominated by the interaction Hamiltonian in (\ref{Hamil}) and can exhibit thermal behavior. Accounting for these rare regions is crucial, as they may trigger avalanche thermal effects \cite{morningstar2022avalanches}. In this subsection, we outline how their impact on the stability of the MBL phase is analyzed in Appendix \ref{griff}. The statistics of encountering $\epsilon$- ideally disordered regions and rare Griffiths regions is analyzed in Appendix \ref{stattt}.

To examine the destabilizing effects of rare regions, we consider their most disruptive scenario—assuming they are completely free of disorder, or ``clean," following the approach of \cite{leonard2023probing, rubio2019many}. Under this assumption, eigenstates within these regions become fully delocalized, meaning they can no longer be described as dressed eigenstates of local Hamiltonians, unlike in $\epsilon$-ideally disordered regions.

By adopting this extreme assumption, the Hamiltonian within these rare regions can be locally diagonalized in the quasi-particle picture, allowing us to perturbatively analyze their effect on the entanglement frequency spectrum of spins. In Appendix \ref{griff}, we characterize the side poles induced by clean rare regions, determine their energy statistics, and establish upper bounds  {using the QCTF formulation}. As in the previous section, we demonstrate that the energy of the side-poles near the second-harmonics induced by the rare regions can be made arbitrarily smaller than that of the second-harmonic poles for sufficiently large disorder strength $W$ and in the $\epsilon \rightarrow 0$ limit. Consequently, the destabilizing effects of rare regions diminish in the strong disorder regime, consistent with experimental observations in \cite{leonard2023probing, rubio2019many}, thereby reaffirming the stability of the MBL phase, even in the presence of rare regions.

 \usetikzlibrary{shapes}
\begin{figure}
    \flushleft
\pgfplotsset{scaled x ticks=false}

    \begin{tikzpicture}
             \node [left] at (-3,4) {(a)};
    
    \begin {scope} [yshift=75, xshift=-40]
    \begin{axis}[ultra thin, xmin=0,
    axis x line=middle,
    axis y line*=left,
        xmax=16.75,
        ymin=-11,
        ymax=11,
        ytick={-10,0,10},
        xtick=\empty,
        xlabel={$ i$ },
        ylabel={$h_i$},
        width = 0.322\textwidth,
    	height = 0.15\textwidth,legend style={nodes={scale=0.75, transform shape}}]
       
        \addplot [black,only marks,mark size=2pt]    table [x=xdata, y=ydata]{hdata.txt};
        \addplot [darkred,only marks,mark size=2pt]   (1, 6.29447372786358);
        \addplot [bluegrey,only marks,mark size=2pt]   (4, 8.26751712278039);
        \addplot [blue,only marks,mark size=2pt]   (8, 0.937630384099677);
        \end{axis}
        \end{scope}
   \begin{scope} [yshift=-55, xshift=-76]
        \foreach \u in {2,3,4,4.5,5}{
            \draw[->, thick]  (\u-0.25,4.5) -- (\u-0.25,4) ;
            \draw[->, thick]  (\u,4) -- (\u,4.5) ;
            }; 
            \draw[->,darkred,thick]  (1.5,4) -- (1.5,4.5) ;
        \draw[->,bluegrey,thick]  (2.25,4.5) -- (2.25,4) ;
        \draw[->,thick]  (2.5,4) -- (2.5,4.5) ;
        \draw[->,blue,thick]  (3.25,4.5) -- (3.25,4) ;
        \draw[->,thick]  (3.5,4) -- (3.5,4.5) ;  
   \end{scope}

    \end{tikzpicture}

     \qquad\qquad  
         
         \pgfplotsset{every axis/.append style={width=6cm,extra x ticks={-20,-10,0,10,20},extra x tick style={major tick length=-3pt}}}
    \begin{tikzpicture}[spy using outlines={rectangle, magnification=2.5,connect spies,thin}]
       
         \node [right] at (-1,2) {(b)};

    \node [above,rotate=90] at (-0.5,-0.8) {${|\tilde{\mathcal{Q}}_\mathcal{M}(i2\pi f)|}$};
    \begin{groupplot}[
    group style={
        group name=my fancy plots2,
        group size=1 by 3,
        xticklabels at=edge bottom,
        vertical sep=0pt
    },
    xmin=-24, xmax=24,
    width = 0.44\textwidth,
    height = 0.18\textwidth
    ]

    \nextgroupplot[ymin=0,ymax=0.08,
    ytick={0,0.04,0.08},
    scaled y ticks=true,
    xtick=\empty,
    extra x tick labels=\empty,
    axis x line*=middle,
    axis y line*=left
    ]

    \addplot [darkred]    table [x=x, y=y]{f2.txt};
    \draw[black,opacity=0.8]  (-2.68,0) -- (-2.68,0.0314) node[below] {};
       \draw[black,opacity=0.8]  (2.68,0) -- (2.68,0.0314) node[below] {};
              \draw[black,opacity=0.8]  (5.36,0) -- (5.36,9.85e-4) node[below] {};

       \draw[black,opacity=0.8]  (0,0) -- (0,0.0628) node[below] {};
    
    \coordinate (spypoint) at (5.1,0.001);
        \coordinate (spyviewer) at (16,0.0305);
        \spy[width=1cm,height=0.8cm, magnification=5] on (spypoint) in node [fill=white,thin] at (spyviewer)  ;
    \node [above,yshift=10pt] at (spyviewer){}; 

    \nextgroupplot[ ymin=0,ymax=0.02,
    ytick={0,0.01},
    yticklabels={0, 1},
        scaled y ticks=false,
    xtick=\empty,
        extra x tick labels=\empty,
    axis x line*=middle,
    axis y line*=left
    ]
           \addplot [bluegrey]    table [x=x, y=y]{f22.txt};
       \draw[black,opacity=0.8]  (-16.72,0) -- (-16.72,0.0009) node[below] {};
       \draw[black,opacity=0.8]  (16.72,0) -- (16.72,0.0009) node[below] {};
       \draw[black,opacity=0.8]  (-6.62,0) -- (-6.62,0.0057) node[below] {};
       \draw[black,opacity=0.8]  (6.62,0) -- (6.62,0.0057) node[below] {};

              \draw[black,opacity=0.8]  (0,0) -- (0,0.0132) node[below] {};
         \coordinate (spypointt) at (16.72,0.001);
        \coordinate (spyviewerr) at (16,0.0105);
        \spy[width=1cm,height=0.8cm] on (spypointt) in node [fill=white,thin] at (spyviewerr);
    \nextgroupplot[ymin=0,ymax=0.04,
    ytick={0,2e-2},
        yticklabels={0, 2},
        scaled y ticks=false,
    axis x line*=middle,
        xtick=\empty,
    axis y line*=left,
    xlabel={Frequency $f(\frac{J}{\hbar})$ }
    ]

    \addplot [blue!80]    table [x=x, y=y]{f33.txt};
    \draw[black,opacity=0.8]  (-6.36,0) -- (-6.36,0.0062) node[below] {};
       \draw[black,opacity=0.8]  (6.36,0) -- (6.36,0.0062) node[below] {};
       \draw[black,opacity=0.8]  (-7.21,0) -- (-7.21,0.0048) node[below] {};
       \draw[black,opacity=0.8]  (7.21,0) -- (7.21,0.0048) node[below] {};
              \draw[black]  (0,0) -- (0,0.0220) node[below] {};
      \coordinate (spypointtt) at (6.58,0.003);
        \coordinate (spyviewerrr) at (16,0.02);
        \spy[width=2cm,height=0.8cm] on (spypointtt) in node [fill=white,thin] at (spyviewerrr);
        
    \end{groupplot}
    
    \end{tikzpicture}
    
    \caption{This figure shows the quenched entanglement measure in the frequency domain $|\tilde{\mathcal{Q}}_\mathcal{M}(s=i2\pi f)|$ for three particles in a disordered ($W=10$) Heisenberg chain of fifteen spins, initialized at an anti-ferromagnetic state. The random fields $h_i$ and the location of the three particular spins are shown in sub-figure (a). The Fourier transforms of the corresponding entanglement measures for each particle from a $40\frac{\hbar}{J}$-long exact numerical simulation are shown in sub-figure (b) in the same color-codes. The black lines show the predicted dominant frequency components (including the second-harmonics) in the entanglement measure analysis (Appendix \ref{lowo}). Note that all of the non-zero frequencies appear in positive and negative pairs. Smaller peaks have been magnified in the insets. As shown in the top plot in sub-figure (b), the entanglement measure for the edge spin has only one dominant (non-zero) frequency pair, accompanied by their corresponding second-harmonics (shown in the inset). In the bottom plot, which corresponds to the spin in the middle of the chain (blue), fourth order components are present, although their contribution is small. The main reason for this behavior is that $|h_{10}-h_9|/J \ll 1$ which strongly reinforces higher order interferences for particles 9 and 8. The frequency spectra are dominated by the few second-order frequency components, which are predicted by the QCTF formulation. This property ultimately leads to quasi-periodicity of the entanglement measures, as illustrated in Figure \ref{fig222}.}
    \label{fig:figgy4}
\end{figure}

\section{Quasi-periodicity versus logarithmic growth:\\ non-ergodicity and failure of ensemble averaging} \label{nonerg}

In the previous section, we demonstrated that at sufficiently strong disorder, there must exist spins within $\epsilon$-ideally disordered regions that exhibit quasi-periodic entanglement dynamics, a hallmark of the many-body localized (MBL) phase.  {Importantly, the observation of quasi-periodicity in the entanglement evolution of individual spins is sufficient to establish the MBL. This is because the entanglement of any larger-than-one-spin subsystem arises from correlations between its internal constituents and the spins outside the subsystem. When each constituent displays quasi-periodic behavior, the subsystem’s entanglement must necessarily inherit this quasi-periodic dynamics.}
\par However, detecting MBL through quasi-periodicity is obscured in numerical simulations due to a fundamental limitation, which we now discuss. In general, we will demonstrate that various form of averaging, e.g., ensemble averaging, fundamentally obscure quasi-periodicity due to an inherent obstruction tied to {non-ergodicity} of the MBL phase.

In what follows, we will focus on the commonly used ensemble averaging. However, the non-ergodicity limitation is not limited to this particular form of averaging. Ensemble averaging is when a dynamical variable (such as imbalance \cite{sierant2022challenges} or entanglement entropy \cite{bardarson2012unbounded, serbyn2013universal}) is obtained (numerically, experimentation, or analytically) for a specific spin or subsystem and then averaged over different disorder realizations of the chain (e.g., by sampling different values of the random fields $h_i$). Similarly, sometimes the dynamical variable is averaged over various subsystems of the same disordered realization.

\subsection{How Ensemble Averaging Masks Quasi-Periodicity}\label{nonegr1}

Consider a scenario where each individual spin in an MBL phase exhibits quasi-periodic entanglement dynamics. This implies the presence of sharp singularities (delta functions) in the frequency spectrum of entanglement evolution. However, because of disorder, these singularities occur at {random frequencies} for different spins.

 {If one performs ensemble averaging—whether over spins, disorder realizations, or different subsystems—the resultant spectrum is significantly altered. The Fourier transform of the averaged entanglement corresponds to the averaged Fourier spectra of individual realizations. Since each realization contains delta functions at randomly distributed frequencies, their sum produces an increasingly \textit{smooth} spectrum as the number of ensembles grows.}  { The only delta peak that survives the averaging process is the one shared across all spectra—located at zero frequency (the DC component).}  { This effectively erases the quasi-periodic signature present in any single realization (see Figures \ref{fig222}. e and \ref{fig:aver}). } {The only remaining sharp peak in the averaged spectrum is the delta peak at zero frequency (DC component), which is common to all realizations. As a result, the averaged spectrum, dominated by a single stable peak at zero frequency, reflects thermalizing behavior—namely, the saturation of subsystem entanglement.}

To illustrate this phenomenon, consider a simple biased coin model. Suppose one \textit{randomly} (with $50-50$chance) selects between two biased coins: the first always lands on heads, and the second always lands on tails. If one flips the chosen coin multiple times (analogous to time averaging within a single realization), the outcome is completely deterministic—always heads for the first coin, always tails for the second. However, if one instead considers the probability distribution over both coins (analogous to ensemble averaging), the overall outcome is a 50-50 probability of heads or tails.

In this analogy, each individual coin’s behavior is deterministic, yet the probability distribution over all possible choices does not reflect this determinism. Similarly, in the MBL phase, each spin undergoes well-defined quasi-periodic evolution, but ensemble averaging over disorder realizations obscures this structure. Fundamentally, this is a direct consequence of {non-ergodicity}: the time average of a single realization (always heads for the first coin) does not match the ensemble average (50-50 probability).

\subsection{Challenging Conventional Views of MBL} \label{nonerg2}

The demonstrated quasi-periodicity challenges conventional descriptions of MBL, which typically emphasize the {logarithmic growth} of subsystem entanglement, as reported in \cite{serbyn2013universal, bardarson2012unbounded, sierant2022challenges, PhysRevB.77.064426}, that must be followed by saturation to a ``thermal" value (which can  {tend} to infinity in infinite chains) in the $T\rightarrow \infty$ limit. However, probing this behavior via ensemble averaging, used in the majority of numerical studies in the literature, relies on two key assumptions:
\begin{enumerate}
    \item A saturation value exists (even if it tends to infinity).
    \item If it exists, the saturation value (obtained via time averaging) is equal to the ensemble-averaged entanglement at a fixed, yet large time, over different disorder realizations.
\end{enumerate}
Both of these assumptions {break down} due to quasi-periodicity (which prevents saturation) and the resultant non-ergodicity of MBL dynamics.

Ultimately, the non-ergodicity of MBL is encoded in the entanglement dynamics of individual subsystems in each random realization of the system. However, ensemble averaging—whether over disorder realizations or different spins—smooths out these singularities, removing quasi-periodic signatures and yielding a qualitatively different spectrum, that lead to alterative observations such as the logarithmic growth.

A crucial remark should be made regarding averaging over {large subsystems}.  {When studying the entanglement of a larger subsystem (e.g., half-chain entanglement), the quasi-periodicity must persist, given the quasi-periodicity of entanglement of it's subsystems; however, detecting this can be more challenging. Importantly, the total entanglement of a large subsystem aggregates all entangling contributions between spins inside and outside of it. Consequently, multiple quasi-periodic features in the entanglement evolution may superimpose, potentially making the frequency spectrum appear smoother, compared to the case of studying smaller subsystems. Furthermore, averaging the entanglement of such large subsystems over various disorder realizations further {erases quasi-periodicity}, leading to observations such as the unbounded logarithmic growth of entanglement.}

\par
 {Figure~\ref{fig:aver} demonstrates how thermal-like features, such as the logarithmic growth of entanglement, can misleadingly emerge due to ensemble averaging. While this effect can already be seen in numerical simulation of single-spin entanglement dynamics, here we focus on {half-chain entanglement entropy} to better align with the broader literature, where this quantity is more commonly analyzed~\cite{Sierant_2025}.}

 {Since the subsystems in this case are not two-level systems, the entanglement measure introduced earlier in (\ref{lapltr}) is no longer applicable. Instead, we use an extended form of this measure (denoted by \(\tilde{\mathcal{Q}}_{hf} (i 2\pi f) = \mathcal{F}[\mathcal{Q}_{hc}(t)]\), where \(\mathcal{F}\) denotes the Fourier transform) based on the {Laplace transform of the sum of principal minors} of the reduced density matrix, as introduced in \cite{QCTF}. This generalized measure reduces to (\ref{lapltr}) in the single-spin case, and is monotonically related to the second-order R\'enyi entropy (for a full definition, refer to Appendix \ref{half-chain}).}

 {In Figure \ref{fig:aver}, we show the ensemble-averaged evolution of half-chain entanglement over $1000$ realizations of the disordered Heisenberg chain (\ref{Hamil}) with disorder strength $W = 10$. Despite the fact that each individual realization (see insets for an example) undergoes quasi-periodic evolution, the averaged dynamics exhibit a clear logarithmic growth in entanglement. In the Fourier domain, in sub-figure \ref{fig:aver}(b), an individual realization (inset) shows multiple sharp peaks at non-zero frequencies, while the ensemble-averaged spectrum exhibits a single sharp peak at zero frequency, mimicking spectral features of a thermalizing system. This mechanism was also schematically illustrated in Figure \ref{fig222}.}

\par  {We note that prior work \cite{evers2023internal} introduced the notion of an “internal clock,” emphasizing that each disorder realization exhibits its own distinct dynamical time scale. This study highlights the significance of treating each realizations individually, an idea that resonates with the emphasis of our work. However, our conclusions fundamentally diverge in one key respect: while \cite{evers2023internal} shows that ensemble-averaged entanglement can act as a universal clock governing the behavior of ensemble-averaged observables such as imbalance, our findings indicate that ensemble averaging obscures essential features, particularly the quasi-periodic structure—that are only visible in single realizations. Despite this divergence, a deeper comparison between the two approaches may yield fruitful insights in future work.}

\usetikzlibrary{shapes}
\begin{figure}

            \begin{tikzpicture}
        
        \node at (0,5.5) {(a)};
        \begin{axis}[xmin=0,
        xmax=7,
        ymin=0,
        ymax=0.07,
        ytick={0,0.05},
        xtick={0,1,2,4,6,8},
        xlabel={$\log \big(t (\frac{\hbar}{J})\big )$ },
        ylabel={\text{Half-chain entanglement} $\mathcal{Q}_{hf} (t)$},
        width = 0.5\textwidth,
    	height = 0.35\textwidth]

       \addplot [darkblue, very thick]    table [x=T, y=V10]{dataave.txt};

        \end{axis}
\begin{scope}[yshift=1.2cm, xshift=2.9cm]
 \begin{axis}[xmin=0,
        xmax=2000,
        ymin=0,
        ymax=0.15,
        ytick={0,0.1},
        xtick={0,1000,2000},
        xticklabels={0 ,1, 2},
        xlabel={$t (10^3\frac{\hbar}{J})$ },
        ylabel={$\mathcal{Q}_{hf} (t)$},
        width = 0.3\textwidth,
    	height = 0.15\textwidth]
          \addplot [darkred]    table [x=T1, y=T3]{dataindv.txt};  
        \end{axis}

\end{scope}
        
    \end{tikzpicture}

        \qquad\qquad
    \begin{tikzpicture}[]
    
    \node at (0,5.5) {{(b)}};
        \begin{axis}[xmin=-1000,
        xmax=1000,
        ymin=0,
        ymax=50.6,
        ytick={0,45},
        yticklabels={,$5$},
        xtick={-1000,-500,0,500,1000},
        xticklabels={$-2\pi$,,0,, $2\pi$},
        xlabel={Frequency $f(\frac{J}{\hbar})$ },
        ylabel={\text{Fourier transformed} $|\tilde{\mathcal{Q}}_{hf} (i2\pi f)|$},
        width = 0.5\textwidth,
    	height = 0.35\textwidth]
    
        \addplot [darkblue, very thick]    table [x=F, y=Fd]{fourierdataave.txt};
        \draw[->, darkblue, very thick] (0,0)--(0,45);
        
        \end{axis}
        \node at (0.45,5) {$.10^{-2}$};
        \begin{scope}[yshift=1.2cm, xshift=4.55cm]
 \begin{axis}[xmin=-1000,
        xmax=1000,
        ymin=0,
        ymax=120,
        ytick={0,100},
        yticklabels={,5},
        xtick={-1000,-500, 0,500, 1000},
        xticklabels={$-2\pi$,,0,, $2\pi$},
        xlabel={$f (\frac{J}{\hbar})$ },
        ylabel={$|\tilde{\mathcal{Q}}_{hf} (i2\pi f)|$},
        ylabel style={ anchor=west, xshift=-30pt, yshift=0pt}, 
        width = 0.23\textwidth,
    	height = 0.25\textwidth]
          \addplot [darkred, very thick]    table [x=T2, y=T4]{dataindv.txt};  
        \end{axis}
        \node at (0.4,3.1) {$.10^{-2}$};

\end{scope}

    \end{tikzpicture}


    \caption{ {Emergence of evident logarithmic entanglement growth due to ensemble averaging. The plot shows the evolution of half-chain entanglement for a $10$ spin disordered Heisenberg chain (\ref{Hamil}) with disorder strength $W=10$. The ensemble average over $1000$ quasi-periodic realizations is shown in the main panels of (a) and (b), in the time and frequency domains respectively, while insets display the evolution of a sample single realization. Although each realization exhibits purely quasi-periodic dynamics, the ensemble-averaged entanglement misleadingly shows logarithmic growth. In the frequency domain (b), the single realization displays multiple sharp, stable peaks at non-zero frequencies, while the ensemble-averaged spectrum essentially collapses to a single peak at zero frequency, mimicking thermal behavior. This demonstrates a central claim of this work: the true entanglement dynamics in the MBL phase is quasi-periodic, and the apparent logarithmic growth reported in the literature arises from unwittingly using averaging procedures, which is also demonstrated schematically in Figure \ref{fig222}.}}
    \label{fig:aver}
\end{figure}

\section{Conclusion}\label{conc}
In this study, we introduced a novel analytical approach to establish the stability of Many-Body Localization (MBL) using the Quantum Correlation Transfer Function (QCTF) \cite{QCTF}. This framework revealed the quasi-periodicity of the MBL phase, persisting even in the infinite-length regime and at asymptotically long time scales. By identifying and characterizing singularities in the frequency domain of subsystems' entanglement dynamics, we demonstrated that MBL dynamics remain stable under sufficiently strong disorder, resolving a longstanding controversy in the field. 

Our findings refutes the conventional view that logarithmic entanglement growth is a defining feature of the MBL phase. We showed that such growth arises from ensemble averaging, which is incapable of capturing the intrinsic non-ergodic, quasi-periodic nature of MBL. Instead, we established that MBL is better characterized by persistent delta functions in the Fourier spectrum of subsystems' dynamical features of individual realizations of the disordered system, marking its quasi-periodicity. This wrong conclusion is an example of logical \textit{fallacy of division} \cite{kelley2013art}. Additionally, we highlighted multiple studies that had observed this quasi-periodicity in numerical and experimental analyses but had not recognized it as a defining feature of MBL. This observation, in addition to our proof in subsection \ref{gr}, rules out the claim that Griffiths regions destabilize the MBL phase through avalanches.

\begin{acknowledgments}
P.A acknowledges support from the U.S Department Of Energy (DOE) grant (DE-FG02-02ER15344) and the Princeton Program in Plasma Science and Technology (PPST). H.R acknowledges support from the Army Research Office (ARO) grant (W911NF-19-0382).
\end{acknowledgments}

\onecolumngrid

\section*{Appendix} \label{appendixlabel}
\appendix

The Appendices contain detailed proofs and derivations of several items stated in the main text.
\section{ {Laplace Transform Techniques}}\label{lap-intro}

In this appendix, we briefly summarize the Laplace-transform-based methods used throughout the main text to analyze entanglement dynamics and diagnose thermalization versus localization. Our aim is to provide a self-contained reference for the analytical tools we employ. For a more in-depth introduction refer to \cite{widder2015laplace}.

\paragraph{Laplace Transform Definition.} For a time-dependent function $f(t)$ defined on the interval $t \in [0, T]$, where $T$ is the total simulation time, we define the (one-sided) Laplace transform as 
\begin{equation}
    \mathcal{L}[f](s) = \int_0^T e^{-st} f(t) \, dt,
\end{equation}
where $s \in \mathbb{C}$ is the Laplace variable. The transformed function $\mathcal{L}[f](s)$ captures the spectral structure of $f(t)$ and plays a central role in diagnosing the presence or absence of thermalization.

\paragraph{Poles and Quasi-Periodic Dynamics.} The structure of $\mathcal{L}[f](s)$ encodes the long-time behavior of $f(t)$. A \emph{pole} of a complex function is a point where the function diverges due to a zero in the denominator of its analytic expression. In particular:
\begin{itemize}
    \item Poles on the negative real axis correspond to exponentially decaying modes.
    \item Poles on the positive real axis indicate exponentially growing behavior, typically unphysical in bounded quantum dynamics.
    \item Poles on the imaginary axis correspond to non-decaying oscillatory behavior, such as persistent quasi-periodicity.
    \item Complex poles with negative real parts correspond to damped oscillations.
\end{itemize}
To provide a concrete example, the Laplace transform of a cosine function, $f(t) = \cos(\omega t)$, is given by:
\begin{equation}
    \mathcal{L}[\cos(\omega t)](s) = \frac{s}{s^2 + \omega^2},
\end{equation}
which has poles at $s = \pm i\omega$. These purely imaginary poles signify undamped, persistent oscillations.

\paragraph{Final Value Theorem (FVT).} A central analytic tool we use is the Final Value Theorem:
\begin{equation}
    \lim_{t \to \infty} f(t) = \lim_{s \to 0} s \mathcal{L}[f](s),
\end{equation}
provided that the following necessary conditions are satisfied:
\begin{itemize}
    \item All poles of $s \mathcal{L}[f](s)$ must lie strictly in the left half of the complex plane (i.e., \text{Re}$(s) < 0$).
    \item There must be no poles or branch cuts on the imaginary axis or in the right half-plane.
    \item $f(t)$ must converge to a finite value as $t \to \infty$.
\end{itemize}
Violation of any of these conditions invalidates the theorem. In particular, the presence of poles on the imaginary axis obstructs convergence and is indicative of persistent oscillations, which are incompatible with thermal saturation.

\paragraph{Application to Entanglement Dynamics.} In the Quantum Correlation Transfer Function (QCTF) formulation, introduced in the next appendix, we study the Laplace transformation of the subsystems' entanglement dynamics. Establishing sharp poles in such Laplace transformations proves quasi-periodicity of subsystem's entanglement. If the system is thermal, then each subsystems' entanglement's Laplace transformation should only have sharp poles at the origin of the complex domain. Otherwise, stable quasi-periodic features must persist in the subsystems' entanglement evolution.


\section{Summary of the QCTF formulation}\label{intro-qctf}
The evolution of a quantum system in the Quantum Correlation Transfer Function (QCTF) formulation is described via a complex function, i.e. the QCTF, through what we refer to as chronological and structural frequency components. The chronological frequency component describes the system's time dependence, while the structural frequencies encode the variation of the system's state, in an arbitrary basis for the underlying Hilbert space. In this formulation, the evolution of correlations between the constituents of the many-body quantum system can be obtained by finding the residues of the QCTF. In this subsection, we briefly introduce the mathematical formulation of QCTF and limit the discussion to the case where the subsystem of interest consists of one spin. and how the entanglement's evolution of a subsystem can be obtained. For more details, readers are referred to \cite{QCTF, Azodi_2024, azodi2024measuringentanglementexploitingantisymmetric}.

\par Consider a closed quantum system with discrete energy levels, consisting of a two-level particle (referred to as subsystem $\mathcal{M}$) that interacts with an accompanying $d$-dimensional quantum subsystem $\mathcal{R}$. The bipartite quantum system evolves according to the Hamiltonian $\mathbf{H}$ from the initial state $\ket{\psi_0}=\ket{\psi(t=0)}$. If we denote the reduced density matrix of the subsystem $\mathcal{M}$ by $\rho_{\mathcal{M}}(t)=\mathbf{\Tr}_{\mathcal{R}}\{\dyad{{\psi(t)}}\}$, then $\mathcal{Q}_{\mathcal{M}}(t)= \det (\rho_{\mathcal{M}}(t))$ is a time-dependent entanglement measure of subsystem $\mathcal{M}$, which is also monotonically related to the second-order R\'enyi entanglement entropy through $ \mathcal{S}_2({\mathcal{M}})=-\ln (1-2\mathcal{Q_{M}})$. Equivalently, we will use the Laplace transformation of this variable, $\tilde{\mathcal{Q}}_{\mathcal{M}}(s)=\mathcal{L}\{\mathcal{Q}_{\mathcal{M}}(t)\}$, as the \textit{dynamical entanglement measure} in the analysis. 
\par The entanglement measure $\tilde{\mathcal{Q}}_{\mathcal{M}}(s)$ can be obtained from the QCTF, which is defined on an off-diagonal block of the system's density matrix, $\tilde{\rho}(s)$. Given any basis for the quantum system which is constructed from the \textit{arbitrary} basis vectors $\{\ket{+}, \ket{-}\}$ for $\mathcal{M}$ and $\{\ket{l},l=0,...,d-1\}$ ($d$ can be countably infinite) for $\mathcal{R}$, we define the QCTF on the off-diagonal block $\mel{+}{\tilde{\rho}(s)}{-}$, as follows,

\begin{subequations}
\begin{equation}\label{QCTFE}
       \mathcal{K}(z_d,z_a,s)=\expval{\mathcal{H}}{\psi_0},
\end{equation}
\begin{equation}\label{generator}
 \mathcal{H}=\textbf{G}^\dagger(s^*)\bigg( \sum_{ l,k}{z_a}^{l+k}{z_d}^{l-k}\ket{-\otimes k}{\bra{ +\otimes l}\bigg)\star \textbf{G}(s)},
\end{equation}
\end{subequations}    
where $\mathbf{G}(s)=(s+\frac{i}{\hbar}\mathbf{H})^{-1}$ is the resolvent of the system and $z_d$, $z_a$ and $s$ are complex variables.

\par Given this QCTF, the dynamical entanglement measure ($\tilde{\mathcal{Q}}_{\mathcal{M}}(s)$) can be obtained as follows,

\begin{equation}\label{main}
\begin{split}
    \Tilde{\mathcal{Q}}_M(s)=&\underset{{\substack{z_d=0\\z_a=0}}}{\mathbf{Res}}\big((z_d z_a)^{-1}{\mathcal{K}(z_d,z_a,s)\star \mathcal{K}^*(1/z^*_d,1/z^*_a,s^*)} \big)-{\mathcal{K}_d(s)\star\mathcal{K}_d^*(s^*)},
    \end{split}
\end{equation}
with $\mathcal{K}_d(s)=\eval{\underset{{\substack{z_d=0}}}{\mathbf{Res}}\big(z_d^{-1}{\mathcal{K}(z_d,z_a,s)}\big )}_{z_a=1}$, where $\underset{{{z=a}}}{\mathbf{Res}}(f(z))$ is the residue of the function $f$ at $a$. Also, the operation $\star$ is defined by the ordinary product operation in the $z_d$ and $z_a$ domains and the following convolution operation in the $s$ domain. If $\mathbf{F}_1(s)$ and $\mathbf{F}_2(s)$ are functions in the Laplace domain, then:
\begin{equation}\label{conv}
    \mathbf{F}_1(s)\star \mathbf{F}_2(s)\doteq \frac{1}{2\pi i} \int_{-\infty}^{\infty} {\mathbf{F}_1(\sigma + i \omega) \mathbf{F}_2(s-\sigma -i \omega) d\omega},
\end{equation}
for some real $\sigma$ in the region of convergence (ROC) of $\mathbf{F}_1 (s)$. 
As an important special case of equation (\ref{conv}), we have $(s+i\omega_1)^{-1}\star (s+i\omega_2)^{-1}=(s+i(\omega_1+\omega_2))^{-1}$. The first term in (\ref{main}) corresponds to the Frobenius norm of the off-diagonal sub-matrix $\mel{+}{\tilde{\rho}(s)}{-}$, while the second term corresponds to the summation of the cross-correlation of its diagonal. These two quantities are identical when the subsystems $\mathcal{M}$ and $\mathcal{R}$ are not entangled. 

\subsection*{ {Measure of larger subsystems' (e.g., half-chain) entanglement $\mathcal{Q}(t)$}}\label{half-chain}
In Section \ref{nonerg}, we presented numerical simulations of the half-chain entanglement entropy in disordered Heisenberg chains. Note that the measure introduced in (\ref{lapltr}), which is based on the determinant of the reduced density matrix, is no longer suitable for characterizing entanglement in larger subsystems. Instead, for such subsystems, the QCTF formulation, introduced above, gives the sum of all $2 \times 2$ principal minors of the reduced density matrix $\rho_{\mathcal{M}}$ (in this case, corresponding to the half-chain). This quantity is equal to the second elementary symmetric polynomial of the eigenvalues $\{ \lambda_i(t) \}$ of $\rho_{\mathcal{M}}$ \cite{QCTF}:
\begin{equation}\label{esp}
    \mathcal{Q}(t) = \sum_{1 \le j_1 < j_2 \le n} \lambda_{j_1}(t)\lambda_{j_2}(t).
\end{equation}
This measure is monotonically related to the second-order R\'enyi entropy $\mathbf{S}^{(2)}$ via:
\begin{equation}
    \mathcal{Q}(t) = \frac{1}{2} \left(1 - \mathrm{Tr}[\rho_{\mathcal{M}}^2] \right) = \frac{1}{2} \left(1 - 2^{-\mathbf{S}^{(2)}} \right).
\end{equation}

\section{$\epsilon$- ideally disordered regions} \label{epsil}
For a sub-chain of \( n+1 \) spins labeled as \( \{m_i, m_{i+1}, \dots, m_{i+n}\} \), with associated random fields \( \{h_i, h_{i+1}, \dots, h_{i+n}\} \), define the following requirement

\begin{equation}\label{500c}
\left|\sum_{j=i}^{i+n} (-1)^{j} h_{j}\right| > \frac{nJ}{2\epsilon}.
\end{equation}

This condition ensures that the alternating sum of the random fields within the sub-chain is large enough—greater than a threshold proportional to the length of the sub-chain size \( \frac{nJ}{2\epsilon} \). The presence of the alternating sign \( (-1)^j \) emphasizes the role of local fluctuations in the random fields, and is embedded to reflect the initial state of the chain, which is the anti-ferromagnetic order.

A sub-chain is called $\epsilon$-ideally disordered if all of its smaller sub-chains (including those with fewer spins) satisfy the above criterion. 

In particular, for sub-chains consisting of just two neighboring spins, the condition simplifies to

\begin{equation}
|h_i - h_{i+1}| > \frac{J}{2\epsilon}.
\end{equation}

Note that the classification of the chain into parts is not unique and regions in the chain with variable size can be labeled as an \(\epsilon\)-{ideally disordered region} if a suitable \(\epsilon\) value is chosen. However, regions with \(\epsilon < 1\) are particularly important, as they lead to a convergent entanglement perturbation expansion, as will be demonstrated later.

\subsection{Statistics of $\epsilon$-ideally disordered regions}\label{stattt}

\par Using the $\epsilon$-ideally disordered criterion, the chain can be divided into contiguous segments that alternate between $\epsilon$-ideally disordered regions and Griffiths regions. Naturally, as the overall disorder decreases, the density of Griffiths regions increases. For instance, if the probability of observing an $\epsilon=1$-ideally disordered region of size $n_b$ is approximated by requiring only neighboring spins to satisfy the criterion, we find it scales as $(1-\frac{J}{W})^{n_b-1}$. Consequently, the mean length of an $\epsilon$-ideally disordered region is approximately $ \left(\frac{W}{J}\right)^2 - 1$. Conversely, the probability of finding an isolated rare Griffiths region of size $n_g$ decreases as $\sim \left(\frac{J}{W}\right)^{n_g-1}$.

Thus, in strongly disordered chains, long $\epsilon$-ideally disordered regions are interspersed with smaller rare Griffiths regions. It is important to note that this investigation into the arrangement of these regions is dependent on the chosen value of $\epsilon$. $\epsilon<1$ leads to a convergent entanglement perturbation series, as will be discussed in the next Sections.

\section{Low-order perturbation of $\tilde{\mathcal{Q}}_{\mathcal{M}}(s)$}\label{lowo}
 
\par We consider individual spins (subsystem $\mathcal{M})$ inside an $\epsilon (< 1)$- ideally disordered region (defined in subsection \ref{epsiloni}) and take the product of local eigenstates of $\sigma_z$, denoted by $\ket{\uparrow}$ and $ \ket{\downarrow}$, as both the unperturbed eigenstates and the basis for the QCTF transformation (an introduction to the QCTF formulation can be found in Appendix \ref{intro-qctf}). We will label the product states and denote them by $\ket{l}$ or in the form $\ket{\uparrow \otimes l}$ ($\ket{\downarrow \otimes l}$) to emphasize that spin of subsystem $\mathcal{M}$ is $\uparrow$ ($\downarrow$). The quantum system is initially in the Ne\'el state $\ket{\uparrow \downarrow \uparrow \downarrow \cdots}$, denoted by $\ket{0}$. Without loss of generality, we assume that the considering spin (subsystem $\mathcal{M}$) is initially in state $\ket{\uparrow}$. Similarly, we label some of the product states as shown in Figure \ref{fig:spingraph}. For simplicity, we use the notations $\mathbf{J}_k=J\mathbf{S}_k\mathbf{S}_{k+1}$ and $\mathbf{J}=\sum \mathbf{J}_k$.
\begin{figure*}
       \centering
\begin{tikzpicture}[scale=0.6,ultrthick/.style={very thick}]
\coordinate (Origin)   at (0,0);
\coordinate (XAxisMin) at (-3,0);
\coordinate (XAxisMax) at (5,0);
\coordinate (YAxisMin) at (0,-2);
\coordinate (YAxisMax) at (0,5);
\coordinate (num-1) at (-5,-2);
\coordinate (num-11) at (-8,-3.8);
\coordinate (num-12) at (-2.5,-3.8);
\coordinate (num1) at (5,-2);
\coordinate (num11) at (8,-3.8);
\coordinate (num12) at (2.5,-3.8);
\coordinate (num-2) at (-5,-5);
\coordinate (num2) at (5,-5);
\coordinate (num-3) at (-5,-8);
\coordinate (num3) at (5,-8);

    \draw[ultrthick,-latex,black] (-1,-0.5) -- (-1,0.5);
    \draw[ultrthick,-latex,black] (-0.5,0.5) -- (-0.5,-0.5) ;
    \draw[ultrthick,-latex,black] (0,-0.7) -- (0,0.7);
    \draw[ultrthick,-latex,black] (0.5,0.5) -- (0.5,-0.5) ;
    \draw[ultrthick,-latex,black] (1,-0.5) -- (1,0.5);
    \draw[ultrthick,-latex,black] (2.2,0.5) -- (2.2,-0.5) ;
    \draw[ultrthick,-latex,black] (2.7,-0.5) -- (2.7,0.5);
        \draw[ultrthick,-latex,black] (-2.2,0.5) -- (-2.2,-0.5) ;
    \draw[ultrthick,-latex,black] (-2.7,-0.5) -- (-2.7,0.5);
    \node at (1.7,0) {\ldots};
    \node at (-1.7,0) {\ldots};
    \node at (0,-1) {$\ket{ 0}, E_{ 0}$};
    
    \draw [decorate,decoration={brace,amplitude=2pt},xshift=0pt,yshift=0pt]
(-0.3,0.7) -- (0.3,0.7) node [black,above,xshift=-0.2cm] 
{\footnotesize $\mathcal{M}$};
    \draw[thick,dashed,-latex,gray] (1.5,-0.4) --node [above, shift={(0.1,-0.1)},black]{$\mathbf{J}_0$}  (5,-1.2) ;
    \draw[thick,dashed,-latex,gray] (-1.5,-0.4) --node [above, shift={(-0.2,-0.12)},black]{$\mathbf{J}_{-1}$}  (-5,-1.2) ;
    \draw[thick,dashed,-latex,gray] (-5,-2.8) --node [left,black]{$\mathbf{J}_{1}$}  (-5,-4.2) ;
    \draw[thick,dashed,-latex,gray] (5,-2.8) --node [right,black]{$\mathbf{J}_{-2}$}  (5,-4.2) ;
    
     \draw[ thick,dashed,-latex,gray] (-6.2,-2.5) --node [above,black]{\small {$\mathbf{J}_{0}$}}  (-8,-3.3) ;
    \draw[thick,dashed,-latex,gray] (-4,-2.5) --node [above,black]{$\mathbf{J}_{-2}$}  (-2.4,-3.3) ;
   \draw[ thick,dashed,-latex,gray] (6.2,-2.5) --node [above,black]{\small {$\mathbf{J}_{-1}$}}  (8,-3.3) ;
    \draw[thick,dashed,-latex,gray] (4,-2.5) --node [above,black]{$\mathbf{J}_{1}$}  (2.4,-3.3) ;

    \begin{scope}[shift=(num1)]
    \draw[ultrthick,-latex,black] (-1,-0.5) -- (-1,0.5);
    \draw[ultrthick,-latex,black] (-0.5,0.5) -- (-0.5,-0.5) ;
    \draw[ultrthick,-latex,bluegrey] (0,0.7) -- (0,-0.7);
    \draw[ultrthick,-latex,bluegrey]   (0.5,-0.5) --(0.5,0.5);
    \draw[ultrthick,-latex,black] (1,-0.5) -- (1,0.5);
    \node  at (1.7,0) {\ldots};
    \node at (-1.7,0) {\ldots};
    \end{scope}     \node at (9.5,-2) {$\ket{1},E_{ 1}$};

   \begin{scope} [shift=(num-1)]
    \draw[ultrthick,-latex,black] (1,-0.5) -- (1,0.5);
    \draw[ultrthick,-latex,black] (0.5,0.5) -- (0.5,-0.5) ;
    \draw[ultrthick,-latex,bluegrey] (0,0.7) -- (0,-0.7);
    \draw[ultrthick,-latex,bluegrey]   (-0.5,-0.5) --(-0.5,0.5);
    \draw[ultrthick,-latex,black] (-1,-0.5) -- (-1,0.5);
    \node  at (1.7,0) {\ldots};
    \node at (-1.7,0) {\ldots};
    \end{scope}
    \node at (-9.5,-2) {$\ket{ -1}, E_{ -1}$};
               
    \begin{scope}[scale=0.8] 
    \begin{scope}[shift=(num11)]
    
    \draw[ultrthick,-latex,black] (-1,-0.5) -- (-1,0.5);
    \draw[ultrthick,-latex,darkred] (-0.5,0.5) -- (-0.5,-0.5) ;
    \draw[ultrthick,-latex,darkred] (0,0.7) -- (0,-0.7);
    \draw[ultrthick,-latex,black]   (0.5,-0.5) --(0.5,0.5);
    \draw[ultrthick,-latex,black] (1,-0.5) -- (1,0.5);
    \node  at (1.8,0) {\ldots};
    \node at (-1.8,0) {\ldots};
    \end{scope}

    \begin{scope}[ shift=(num12)]
    \draw[ultrthick,-latex,black] (-1,-0.5) -- (-1,0.5);
    \draw[ultrthick,-latex,black] (-0.5,0.5) -- (-0.5,-0.5) ;
    \draw[ultrthick,-latex,black] (0,0.7) -- (0,-0.7);
    \draw[ultrthick,-latex,darkred]   (0.5,-0.5) --(0.5,0.5);
    \draw[ultrthick,-latex,darkred] (1,-0.5) -- (1,0.5);
     \node  at (1.8,0) {\ldots};
    \node at (-1.8,0) {\ldots};
    \end{scope}
       
       \begin{scope}[shift=(num-11)]
         \draw[ultra thick,-latex,black] (1,-0.5) -- (1,0.5);
    \draw[ultrthick,-latex,darkred] (0.5,0.5) -- (0.5,-0.5) ;
    \draw[ultrthick,-latex,darkred] (0,0.7) -- (0,-0.7);
    \draw[ultrthick,-latex,black]   (-0.5,-0.5) --(-0.5,0.5);
    \draw[ultrthick,-latex,black] (-1,-0.5) -- (-1,0.5);
      \node  at (1.8,0) {\ldots};
    \node at (-1.8,0) {\ldots};
  
       \end{scope}

    \begin{scope}[shift=(num-12)]
       \draw[ultra thick,-latex,black] (1,-0.5) -- (1,0.5);
    \draw[ultrthick,-latex,black] (0.5,0.5) -- (0.5,-0.5) ;
    \draw[ultrthick,-latex,black] (0,0.7) -- (0,-0.7);
    \draw[ultrthick,-latex,darkred]   (-0.5,-0.5) --(-0.5,0.5);
    \draw[ultrthick,-latex,darkred] (-1,-0.5) -- (-1,0.5);
                \node  at (1.8,0) {\ldots};
    \node at (-1.8,0) {\ldots};
 
    \end{scope}
              \end{scope}    
    
    \begin{scope}[shift=(num2)]
    \draw[ultrthick,-latex,bluegrey] (-1,0.5)-- (-1,-0.5) ;
    \draw[ultrthick,-latex,bluegrey] (-0.5,-0.5)-- (-0.5,0.5)  ;
    \draw[ultrthick,-latex,black] (0,0.7) -- (0,-0.7);
    \draw[ultrthick,-latex,black]   (0.5,-0.5) --(0.5,0.5);
    \draw[ultrthick,-latex,black] (1,-0.5) -- (1,0.5);
                    \node  at (1.7,0) {\ldots};
    \node at (-1.7,0) {\ldots};
    \end{scope}
        \node at (9.5,-5) {$\ket{ 2}, E_{ 2}$};

     \begin{scope}[ shift=(num-2)]
     \draw[ultrthick,-latex,bluegrey] (1,0.5)-- (1,-0.5) ;
    \draw[ultrthick,-latex,bluegrey, shift=(num-2)] (0.5,-0.5)-- (0.5,0.5)  ;
    \draw[ultrthick,-latex,black, shift=(num-2)] (0,0.7) -- (0,-0.7);
    \draw[ultrthick,-latex,black, shift=(num-2)]   (-0.5,-0.5) --(-0.5,0.5);
    \draw[ultrthick,-latex,black, shift=(num-2)] (-1,-0.5) -- (-1,0.5);
                       \node  at (1.7,0) {\ldots};
    \node at (-1.7,0) {\ldots};
     \end{scope}
                     \node at (-9.5,-5) {$\ket{ -2}, E_{ -2}$};

\end{tikzpicture}
\caption{{Network illustration of interference between unperturbed (product) eigenstates due to local exchange interactions, for a spin in the chain's bulk.} This figure shows how the unperturbed states ($\ket{\uparrow}_{\mathcal{M}}\otimes \ket{l}$), interfere locally to produce the exact eigenstates of the spin chain during the perturbation process. The spin $\mathcal{M}$, for which we study its entanglement dynamics, is drawn slightly taller. The ket notation and zeroth order energy of each chain are denoted adjacently. The local interaction Hamiltonians $\mathbf{J}_k=J\mathbf{S}_k\mathbf{S}_{k+1}$, written on the vertices, flip anti-parallel spins ($k$ and $k+1$), and leave the parallel spins unchanged. Upon applying the perturbations at each step, the flipped spins are colored in blue, and the unchanged ones are colored in orange. }
\label{fig:spingraph}
\end{figure*}
The unperturbed resolvent, i.e., $\mathbf{G}(s)=(s\mathbf{I}-\frac{i}{\hbar}\mathbf{H})^{-1}$, corresponding to the unperturbed Hamiltonian ${\sum_{k=1}^{N}{h_k{S}_{k}^{z}}}$ in (\ref{Hamil}), can now be written as
\begin{equation}\label{greenef}\begin{split}
    \mathbf{G}^0&= (s-\frac{i}{\hbar}E_{ 0})^{-1} \op{ 0}+ (s-\frac{i}{\hbar}E_{ 1})^{-1} \op{ 1}+\cdots
\end{split}\end{equation}
where $E_{l}$, each a linear combination of $h_k$'s, are unperturbed energies corresponding to the unperturbed product states $\ket{l}$. This summation has $2^N$ terms corresponding to the unperturbed basis introduced above.  We denote the $o$th order corrected eigenstate corresponding to the product state $\ket{j}$ by $\ket{j}^{(o)}$, which can be expanded as follows
\begin{equation}\label{co}
    \ket{j}^{(o)}=\sum_{i} {c_{ji}^{(o)}\ket{i}},
\end{equation}
with $c_{ii}^{(o)}=1$ for $o=1,2$. Using the introduced notations, we can perform the QCTF transformation given by (\ref{QCTFE}-\ref{generator}):
\begin{subequations}
\begin{equation}\label{QCTFE2}
       \mathcal{K}(z_d,z_a,s)=\expval{\mathcal{H}}{\psi_0},
\end{equation}
\begin{equation}\label{generator2}
 \mathcal{H}=\textbf{G}^\dagger(s^*)\bigg( \sum_{ l,k}{z_a}^{l+k}{z_d}^{l-k}\ket{\downarrow\otimes k}{\bra{ \uparrow\otimes l}\bigg)\star \textbf{G}(s)},
\end{equation}
\end{subequations}    
where $\mathbf{G}(s)$ is now the $o$th order corrected resolvent, obtained by substituting the corrected eigenstates (\ref{co}) in (\ref{greenef}). 
Accordingly, using equations (\ref{QCTFE2}-\ref{generator2}), the QCTF is obtained as:
\begin{equation}\label{190}
\begin{split}
       &\mathcal{K} (z_d,z_a,s)= \sum_{l, k}{\vphantom{\sum}}'{z_a^{l+k}z_d^{l-k}} \sum_{i,j}{c_{i0}^*c_{il}c_{j0}c_{jk}^* \big(s+\frac{i}{\hbar}(E_{i}^{(1)}-E_j^{(1)})\big )^{-1}}, 
\end{split}
\end{equation}
where $\sum '$ denotes the summation over $l$'s and $k$'s corresponding to products states with subsystem $\mathcal{M}$ in $\uparrow$ and $\downarrow$ respectively. $E_l^{(1)}$ is the first-order corrected eigenvalue corresponding to the eigenvector $\ket{l}$. We should note that with this choice of basis kets, the diagonal part of QCTF, $\mathcal{K}_d$, vanishes \cite{exp1}. Based on this perturbative form of the QCTF, the entanglement measure is obtained using relation (\ref{main}), which is (equal to the coefficient of $(z_d z_a)^0$ in the series expansion of $\mathcal{K}\star \mathcal{K^*}$): 
\begin{equation}\label{200}
\begin{split}
    \tilde{\mathcal{Q}}_{\mathcal{M}}(s)=\sum _{l,k}{\vphantom{\sum}}'\sum_{i,j,i',j'}\frac{c_{i0}^*c_{il}c_{j0}c_{jk}^*c_{i'0}c_{i'l}^*c_{j'0}^*c_{j'k}}{
    s+\frac{i}{\hbar}  (E_{i}^{(1)}-E_j^{(1)}-E_{i'}^{(1)}+E_{j'}^{(1)})}.
\end{split}
\end{equation}
The most significant (lowest perturbation order) frequency contributions correspond to $l=0$ and $k=\pm 1$ (see Figure
\ref{fig:spingraph}), which are the second order contributions to $\tilde{\mathcal{Q}}_\mathcal{M}(s)$ at frequencies $f_{1,2}=\pm |E_0^{(1)}-E_{\pm 1}^{(1)}|/\hbar$ (for example, setting $i=i'=j'=0, j=\pm1$ gives the middle term):
\begin{equation}\label{qform2}\begin{split}
    \tilde{\mathcal{Q}}_\mathcal{M}^{(2)}(s)&=\frac{|c_{0\pm 1}|^2+|c_{\pm 1 0}|^2}{s}
    + c_{\pm 10}c_{0\pm 1}\big(s+\frac{i}{\hbar}(E_{0}^{(1)}-E_{\pm 1}^{(1)})\big )^{-1}+c_{\pm 10}^*c_{0\pm 1}^*\big(s+\frac{i}{\hbar}(E_{\pm 1}^{(1)}-E_0^{(1)})\big )^{-1},
\end{split}\end{equation}

where $\tilde{\mathcal{Q}}_\mathcal{M}^{(2n)}(s)$ denotes the contributions of order $2n$ to the entanglement measure $\tilde{\mathcal{Q}}_\mathcal{M}(s)$ (also we have $c_{0\pm 1}=-c_{\pm 10}^*$). Therefore, (\ref{Q22}) is obtained. 
\par The fourth-order contributions appear for a collective variety of indices in the summations, in (\ref{200}). Non-zero frequencies which appear as fourth-order contributions are $\pm|E_{-1}^{(1)}-E_{1}^{(1)}|/\hbar$, $\pm|2E_{0}^{(1)}-E_{\pm 1}^{(1)}-E_{\mp 1}^{(1)}|/\hbar$, $|E_{\pm 2}^{(1)}-E_{0}^{(1)}|/\hbar$, $\pm2|E_{\pm 1}^{(1)}-E_{0}^{(1)}|/\hbar$ and $\pm|E_{\pm 1}^{(1)}-E_{0}^{(1)}|/\hbar$. For instance $l=0 (i=0,i'=\pm 1)$ with $k=1  (j=\pm1,j'=0)$ produce poles at frequencies $2f_{1,2}=2(E_{0}^{(1)}-E_{\pm1}^{(1)})/\hbar$, and $l=0 (i=\pm 1,i'=0)$ with $k=1  (j=0,j'=\pm1)$ produce poles at frequencies $-2(E_{0}^{(1)}-E_{\pm1}^{(1)})/\hbar$ as follows
\begin{equation}\label{SHH}
     c_{\pm 10}^2c_{0\pm 1} c^*_{\pm 1 0}\big(s+\frac{2i}{\hbar}(E_{0}^{(1)}-E_{\pm 1}^{(1)})\big )^{-1}+ {c^*_{\pm 10}}^2 c_{0\pm 1}^* c_{\pm 1 0}\big(s-\frac{2i}{\hbar}(E_{0}^{(1)}-E_{\pm 1}^{(1)})\big )^{-1}
\end{equation}
 which give the second-harmonic poles in (\ref{SH}).

\par According to equation (\ref{qform2}), the second-order frequencies ($f_{1,2}$) and their corresponding amplitudes ($a\doteq 2|c_{\pm 10}|^2$) can be described via a set of probability density functions, denoted by $P_{F}^{(2)}(f)$ and $P_A^{(2)}(a)$ respectively, as proved and numerically illustrated in Appendices \ref{A200} and \ref{app45}.

\section{Derivation of $P_F^{(2)}(f)$}\label{A200}
If the disorder coefficients ($h_i$) are drawn randomly from the PDF $P_h(x)$, denoted by $h_i \sim P_h(x)$, then we have:
\begin{equation}\label{220}
    E_{\uparrow 0}-E_{\downarrow\pm 1}=E^{(1)}_0-E^{(1)}_{\pm 1}+\expval{\mathbf{J}}_{0}-\expval{\mathbf{J}}_{\pm 1}\sim P_h(x)*P_h(-x),
\end{equation}
where $*$ is the convolution operator. The states $\ket{0}$ and $\ket{ \pm 1}$ describe the same orientation of spins in the chain, except for the two of the spins, which are flipped (see Figure \ref{fig:spingraph}). Consequently, the difference on the L.H.S of this equation is equal to the subtraction $h_i-h_{i\pm 1}$. Finally, the PDF of the term ($h_i-h_{i\pm 1}$) is obtained from the convoluted form on the R.H.S. Note that the expectation values in the middle term correspond to the first order energy corrections, which are equal to $\expval{\mathbf{J}}_{0}-\expval{\mathbf{J}}_{\pm 1}=-J$. 
\\ For the uniform distribution form of $P_h(x)$ on the interval $[-W,W]$, we have $P_h(x)*P_h(-x)=\frac{1}{2W}-\frac{|x|}{4W^2};|x|\le 2W$, which is shown in Figure \ref{fig:appfig}(a). Therefore, the PDF for the first order energy differences $E^{(1)}_0-E^{(1)}_{\pm 1}\sim P_{E^{(1)}_0-E^{(1)}_{\pm 1}}(x)$ is obtained by shifting $P_h(x)*P_h(-x)$ by $J$, which is shown in Figure \ref{fig:appfig}(b). Finally, the PDF of the second-order frequencies $f=\pm |E_0^{(1)}-E_{\pm 1}^{(1)}|/\hbar \sim P_{F}^{(2)}(f)$, is obtained by adding the value of $P_{E^{(1)}_0-E^{(1)}_{\pm 1}}(x)$ for positive energies to the reflected (about the vertical axis) values for negative energies (due to the absolute value), which is shown by a dashed blue line in Figure \ref{fig:appfig}(b), followed by scaling the horizontal and vertical axes by $1/\hbar$ and $\hbar$ respectively, which is shown in Figure \ref{fig:appfig}(c). This PDF is given by
\begin{eqnarray} 
\label{2333}
    P_{F}^{(2)}(f) &&= 
     \begin{cases}
       \hbar\big(\frac{1}{W}-\frac{J}{2W^2}\big) &\quad 0\le f\le \frac{J}{\hbar}, \\
       \hbar\big(\frac{1}{W}-\frac{f}{2W^2}\big ) & \frac{J}{\hbar}<f\le \frac{2W-J}{\hbar}, \\
       \hbar\big(\frac{J+2W-f}{4W^2}\big ) &\frac{2W-J}{\hbar}<f\le \frac{2W+J}{\hbar},\\
       0 &\quad\text{otherwise.} \\ 
     \end{cases}
      \hspace{-7px} \overset{J\ll W}{\approx}\frac{\hbar}{W}-\frac{\hbar f}{2W^2}; \hspace{0.7cm} 0\le f\le \frac{2W}{\hbar}
\end{eqnarray}    
\par The presented proof holds for a spin in the bulk of the chain. In the case of edge spins, we have $\expval{\mathbf{J}}_{0}-\expval{\mathbf{J}}_{\pm 1}=-J/2$, therefore the resulting PDF, $P_{F}^{(2)}(f)$, must be slightly different, which can be obtained by substituting $J$ by $\frac{J}{2}$ in the derivations and plots.

\begin{figure}
    \centering

\begin{tikzpicture}[domain=0:10,scale=0.4,thick]
\usetikzlibrary{calc}   

\def\dint{5}          

\def\gr{\x,{\dint-abs(\x-5)}}

    \coordinate (dint) at (0,{\dint});
    \coordinate (br1) at  (0,0);
    \coordinate (br2) at  (10,0);
    \coordinate (br31) at  (5,5);
    \coordinate (br3) at  (4,5);

    \draw[thick,color=bluegrey] plot (\gr);

    \draw[->] (-2,0) -- (12,0) node[right] {$x$};
    \draw[->] (5,0) -- (5,6) node[above] {$ P_h(x)*P_h(-x)$};
        
    \draw[dashed] (br1) -- (br1) node[below] {$-2W$};
    \draw[dashed] (br2) -- (br2) node[below] {$2W$};
    \draw[dashed,thin] (br31) -- (br3) node[left] {$\frac{1}{2W}$};
    \draw[dashed] (0,5) -- (0,5) node[below] {(a)};
     
\end{tikzpicture}
\qquad\qquad
\begin{tikzpicture}[domain=0:10,scale=0.4,thick]
\usetikzlibrary{calc}   
 
\def\dint{5}          

\def\gr{\x+1,{\dint-abs(\x-5)}}
\def\grd{\x/2.5+5,{\dint-(\x/2.5+1)}}

    \coordinate (dint) at (0,{\dint});
    \coordinate (br1) at  (1,0);
    \coordinate (br2) at  (11,0);
    \coordinate (br31) at  (6,5);
    \coordinate (br3) at  (4,5);

    \draw[thick,color=bluegrey] plot (\gr);
    \draw[dashed,thick,color=bluegrey] plot (\grd);

    \draw[->] (-2,0) -- (12,0) node[right] {$x$};
    \draw[->] (5,0) -- (5,6) node[above] {$ P_{E^{(1)}_0-E^{(1)}_{\pm 1}}(x)$};
        
    \draw[dashed] (br1) -- (br1) node[below] {$-2W+J$};
    \draw[dashed] (br2) -- (br2) node[below] {$2W+J$};
    \draw[dashed,thin] (br31) -- (br3) node[left] {$\frac{1}{2W}$};
    \draw[dashed,thin] (6,5) -- (6,0) node[below] {$J$};
    \draw[dashed] (0,5) -- (0,5) node[below] {(b)};
\end{tikzpicture}
\qquad\qquad
\begin{tikzpicture}[domain=0:10,scale=0.5,thick]
\usetikzlibrary{calc}   
 
\def\dint{4.37}          
\def\dintt{5} 
\def\dinttt{2.8125} 
\def\dslp{-0.5}         
\def\dslpp{-0.6250}
\def\dslppp{-0.6250/2}
\def\sint{1.2}          
\def\sslp{0.8}          
 
\def\pfc{2.5}           
 
\def\demand{\x/10,{\dint}}
\def\demandd{\x*0.6+1,{\dslpp*(\x*0.6+1)+\dintt}}
\def\demanddd{\x/5+7,{\dslppp*(\x/5+7)+\dinttt}}
\def\supply{\x,{\sslp*\x+\sint}}
 
    \coordinate (ints) at ({(\sint-\dint)/(\dslp-\sslp)},{(\sint-\dint)/(\dslp-\sslp)*\sslp+\sint});
    \coordinate (ep) at  (0,{(\sint-\dint)/(\dslp-\sslp)*\sslp+\sint});
    \coordinate (eq) at  ({(\sint-\dint)/(\dslp-\sslp)},0);
    \coordinate (dint) at (0,{\dint});
    \coordinate (sint) at (0,{\sint});
    \coordinate (pfq) at  ({(\pfc-\dint)/(\dslp)},0);
    \coordinate (pfp) at  ({(\pfc-\dint)/(\dslp)},{\pfc});
    \coordinate (sfq) at  ({(\pfc-\sint)/(\sslp)},0);
    \coordinate (sfp) at  ({(\pfc-\sint)/(\sslp)},{\pfc});
    \coordinate (2w+f) at  ({9},{0});
    \coordinate (br1) at  (1,\dint);
    \coordinate (br10) at  (1,0);
    \coordinate (br2) at  (7,-1*\dslpp);
    \coordinate (br20) at  (7,0);
    \coordinate (br3) at  (9,0);

    \draw[thick,color=bluegrey] plot (\demand);
    
    \draw[thick,color=bluegrey] plot (\demandd) ;
    \draw[thick,color=bluegrey] plot (\demanddd) ;
    \draw[dashed] (-4,4) -- (-4,4) node[below] {(c)};
    
 
    \draw[->] (0,0) -- (11,0) node[right] {$f$};
    \draw[->] (0,0) -- (0,5.5) node[left] {$ P_{F}^{(2)}(f)$};
        
    \draw[dashed,thin] (br1) -- (br10) node[below] {$\frac{J}{\hbar}$};
    \draw[dashed,thin] (br2) -- (br20) node[below] {\hspace{-0.2cm}$\frac{2W-J}{\hbar}$};
    \draw[dashed,thin] (br3) -- (br3) node[below] {\hspace{0.2cm}$\frac{2W+J}{\hbar}$};
    \draw[dashed,thin] (dint) -- (dint) node[left] {$\frac{\hbar}{W}-\frac{\hbar J}{2W^2}$};
     
 
\end{tikzpicture}

    \caption{{Illustration of the probability density functions in the proof in Appendix \ref{A200}}. Each sub-figure corresponds to a step in the proof of equation \ref{2333}. Sub-figure (a) shows the PDF for $E_{\uparrow 0}-E_{\downarrow\pm 1} \sim P_h(x)*P_h(-x)$. The PDF for the corrected energy difference $E^{(1)}_0-E^{(1)}_{\pm 1}$ is obtained by shifting $P_h(x)*P_h(-x)$ by $J$ due to equation \ref{220}, which is shown in the sub-figure (b). The dashed line shows the reflected (about the vertical axis) probability densities for negative energies. Sub-figure (c) shows the PDF for the second-order frequencies in the entanglement measure $f=\pm |E_0^{(1)}-E_{\pm 1}^{(1)}|/\hbar \sim P_{F}^{(2)}(f)$, which is obtained by adding probability densities on the positive-energy side of the sub-figure (b), as a result of the absolute value, followed by scaling the horizontal and vertical axes by a factor of $\frac{1}{\hbar}$ and $\hbar$, respectively.}
    \label{fig:appfig}
\end{figure}

\section{Derivation of $P_A^{(2)}(a)$}\label{app45}
Here, we obtain the PDF $P_A^{(2)}(a)$ of the amplitudes $2|c_{\pm1 0}|^2$ of second-order poles, discussed in Appendix \ref{lowo}. We start by finding the Cumulative Density Function (CDF) $R_A(a)=\int_{-\infty}^a {P_A^{(2)}(a')da'}$:
\begin{equation}\label{cumu}
\begin{split}
    R_A(a)=&\mathbf{P}\Big\{2|c_{\pm 1 0}|^2\le a\Big\}=\mathbf{P}\Big\{2|\frac{\mel{\pm 1}{\mathbf{J}}{0}}{E_{\pm1}-E_0}|^2 \le a\Big\}=\mathbf{P}\Big\{|E_{\pm1}-E_0|\ge \frac{J}{\sqrt{2a}}\Big\},    
\end{split}
\end{equation}
where $\mathbf{P\{.\}}$ is the probability measure and we have used the fact that $|\mel{\pm 1}{\mathbf{J}}{0}|=\frac{J}{2}$. This probability can be obtained from $P_{E_{\pm1}-E_0}(x)$ calculated in the previous subsection. Figure \ref{fig:ordd} shows the area corresponding to this probability (in blue color), which is equal to:
\begin{equation}
     R_A(a)=(1-\frac{J}{2W\sqrt{2a}})^2.
\end{equation}
Now we can obtain the PDF $P_A^{(2)}(a)$ by differentiating $R_A(a)$:
\begin{equation}
    P_A^{(2)}(a)=\frac{d}{da}R_A(a)=\frac{\frac{W}{{J}}\sqrt{8a}-1}{8(\frac{W}{{J}}{a})^2};\hspace{0.7cm} a\ge\frac{1}{8}(\frac{{J}}{W})^2.
\end{equation}
\begin{figure}
    \centering
\begin{tikzpicture}[domain=0:10,scale=0.4,thick]
\usetikzlibrary{calc}   

\def\dint{5}          

\def\gr{\x,{\dint-abs(\x-5)}}

    \coordinate (dint) at (0,{\dint});
    \coordinate (br1) at  (0,0);
    \coordinate (br2) at  (10,0);
    \coordinate (br31) at  (5,5);
    \coordinate (br3) at  (4,5);

    \draw[thick,color=bluegrey] plot (\gr);

    \draw[->] (-2,0) -- (12,0) node[right] {$x$};
    \draw[->] (5,0) -- (5,6) node[above] {$ P_{E_{\pm1}-E_0}(x)$};
        
    \draw[dashed] (br1) -- (br1) node[below] {$-2W$};
    \draw[dashed] (br2) -- (br2) node[below] {$2W$};
    \draw[dashed,thin] (br31) -- (br3) node[left] {$\frac{1}{2W}$};
    \draw[bluegrey, fill=bluegrey,fill opacity=0.4, draw opacity=0.4] (7,0) -- (7,3) -- (10,0)  -- cycle;
     \draw[bluegrey, fill=bluegrey,fill opacity=0.4, draw opacity=0.4] (3,0) -- (3,3) -- (0,0)  -- cycle;
     \draw[dashed] (7,0) -- (7,0) node[below] {$\frac{J}{\sqrt{2a}}$};
     \draw[dashed] (3,0) -- (3,0) node[below] {$-\frac{J}{\sqrt{2a}}$};

\end{tikzpicture}
    \caption{{Schematic illustration of the CDF (\ref{cumu}).} This figure demonstrates the CDF $R_A(a)$, as the area below the PDF curve for the difference between the unperturbed energy levels, $P_{E_{\pm1}-E_0}(x)$, which was previously derived in Appendix \ref{app45}.  }
    \label{fig:ordd}
\end{figure}

Figure \ref{fig:PDFs} illustrated the PDFs $P_{F}^{(2)}(f)$ and $P_A^{(2)}(a)$, derived in the Appendices \ref{A200} and \ref{app45}. In the Appendix \ref{appcc}, these PDFs are numerically verified.
\begin{figure}
    \begin{tikzpicture}[domain=0:10,scale=0.58,thick, xscale=1.1]
\usetikzlibrary{calc}   
 
\def\dint{4.37}          
\def\dintt{5} 
\def\dinttt{2.8125} 
\def\dslp{-0.5}         
\def\dslpp{-0.6250}
\def\dslppp{-0.6250/2}
\def\sint{1.2}          
\def\sslp{0.8}          
 
\def\pfc{2.5}           
 
\def\demand{\x/10,{\dint}}
\def\demandd{\x*0.6+1,{\dslpp*(\x*0.6+1)+\dintt}}
\def\demanddd{\x/5+7,{\dslppp*(\x/5+7)+\dinttt}}
\def\supply{\x,{\sslp*\x+\sint}}
 
    \coordinate (ints) at ({(\sint-\dint)/(\dslp-\sslp)},{(\sint-\dint)/(\dslp-\sslp)*\sslp+\sint});
    \coordinate (ep) at  (0,{(\sint-\dint)/(\dslp-\sslp)*\sslp+\sint});
    \coordinate (eq) at  ({(\sint-\dint)/(\dslp-\sslp)},0);
    \coordinate (dint) at (0,{\dint});
    \coordinate (sint) at (0,{\sint});
    \coordinate (pfq) at  ({(\pfc-\dint)/(\dslp)},0);
    \coordinate (pfp) at  ({(\pfc-\dint)/(\dslp)},{\pfc});
    \coordinate (sfq) at  ({(\pfc-\sint)/(\sslp)},0);
    \coordinate (sfp) at  ({(\pfc-\sint)/(\sslp)},{\pfc});
    \coordinate (2w+f) at  ({9},{0});
    \coordinate (br1) at  (1,\dint);
    \coordinate (br10) at  (1,0);
    \coordinate (br2) at  (7,-1*\dslpp);
    \coordinate (br20) at  (7,0);
    \coordinate (br3) at  (9,0);

    \draw[thick,color=bluegrey,dashed] plot (\demand);
    
    \draw[thick,color=bluegrey,path fading=west,fading angle=0] plot (\demandd) ;
     \draw[thick,color=bluegrey,opacity=0.55] plot (\demandd) ;
    \draw[dashed,color=bluegrey,path fading=east] plot (\demandd) ;
    \draw[thick,color=bluegrey] plot (\demanddd) ;

    \draw[->] (0,0) -- (10,0) node[right] {$f$};
    \draw[->] (0,0) -- (0,5.8) node[left] {$ P_{F}^{(2)}(f)$};

    \draw[dashed,thin] (br1) -- (br10) node[below] {$\frac{J}{\hbar}$};
    \draw[dashed,thin] (br2) -- (br20) node[below] {$\frac{2W-J}{\hbar}$};
    \draw[dashed] (br3) -- (br3) node[below] {$\frac{2W+J}{\hbar}$};
    \draw[dashed] (0,\dint) -- (0,\dint) node[left] {$\hbar(\frac{1}{W}-\frac{J}{2W^2}) $};
    \draw[dashed] (-2.5,6.4) -- (-2.5,6.4) node[below] {(a)};
\end{tikzpicture}
\begin{tikzpicture}[domain=0:2.5,scale=2.2,thick,samples=150]
\usetikzlibrary{calc}   
 
\def\dint{4.37}          
\def\dintt{5} 
\def\dinttt{2.8125} 
\def\dslp{-0.5}         
\def\dslpp{-0.6250}
\def\dslppp{-0.6250/2}
\def\sint{1.2}          
\def\sslp{0.8}          
 
\def\pfc{2.5}           
 
\def\demandd{\x+0.1111,{1*(2*1.5*(\x+0.1111)^0.5-1)/(2*1.5*(\x+0.1111))^2}}
\def\demanddd{\x/1.18+0.5111,{0.5*(1.5)^(-1)*(\x/1.18+0.5111)^(-1.5)}}
\def\supply{\x,{\sslp*\x+\sint}}
 
    \coordinate (ints) at ({(\sint-\dint)/(\dslp-\sslp)},{(\sint-\dint)/(\dslp-\sslp)*\sslp+\sint});
    \coordinate (ep) at  (0,{(\sint-\dint)/(\dslp-\sslp)*\sslp+\sint});
    \coordinate (eq) at  ({(\sint-\dint)/(\dslp-\sslp)},0);
    \coordinate (dint) at (0,{\dint});
    \coordinate (sint) at (0,{\sint});
    \coordinate (pfq) at  ({(\pfc-\dint)/(\dslp)},0);
    \coordinate (pfp) at  ({(\pfc-\dint)/(\dslp)},{\pfc});
    \coordinate (sfq) at  ({(\pfc-\sint)/(\sslp)},0);
    \coordinate (sfp) at  ({(\pfc-\sint)/(\sslp)},{\pfc});
    \coordinate (2w+f) at  ({9},{0});
    \coordinate (br1) at  (0.9975,0.9492);
    \coordinate (br10) at  (0,0.9492);
    \coordinate (br2) at  (0.1975,0.9492);
    \coordinate (br20) at  (0.1975,0);
    \coordinate (br3) at  (0.1111,0);

    \draw[thick,color=darkblue] plot (\demandd) ;
   \draw[dashed,color=darkred] plot (\demanddd) ;

    \draw[->] (0,0) -- (2.9,0) node[right] {$a$};
    \draw[->] (0,0) -- (0,1.5) node[left] {$ P_{A}^{(2)}(a)$};

     \draw[dashed,thin] (br2) -- (br20) node[below] {$\text{\hspace{0.7cm}}    a_{max}$};
     \draw[dashed,thin] (br10) -- (br2) node[left] {$\text{\hspace{-2.6cm}}(\frac{27}{32})(\frac{W}{{J}})^2$};
    \draw[dashed] (br3) -- (br3) node[below] {$a_0$};
    \draw[dashed] (br1) -- (br1) node[below] {$\frac{{\sqrt{2}J}}{4W} a^{-\frac{3}{2}}$};
        \draw[dashed] (-0.7,1.7) -- (-0.7,1.7) node[below] {(b)};
        \draw (1.8,1.3) node [draw,thin] {$a_0=\frac{1}{8}(\frac{{J}}{W})^2$, $a_{max}=(\frac{\sqrt{2}{J}}{3W})^2$};
        
    \end{tikzpicture}
    \caption{
    {Illustration of the PDFs for (a) the frequency components and (b) their amplitudes in the lowest (second) order perturbation of the entanglement measure $\tilde{\mathcal{Q}}_\mathcal{M}^{(2)}(s)$ for a spin in the bulk of the chain.} Sub-figure (a) shows the PDF for the second order frequencies $P_{F}^{(2)}(f)$ in (\ref{2333}). These frequencies have the most significant contribution to the entanglement dynamics of the spin $\mathcal{M}$. For frequencies close to and smaller than $\frac{J}{\hbar}$ (plotted with a dashed line), the local energy levels are vulnerable to strong mixing with the other energy levels (corresponding to the neighboring sites), which interferes with the perturbative treatment; but, the probability of this subset of frequencies contributing is approximately $J/W$, which is negligibly small in the MBL phase ($J \ll W$). Sub-figure (b) shows the PDF for second order amplitudes, $ P_{A}^{(2)}(a)$. In the MBL phase, small amplitudes (i.e., of order $(\frac{J}{W})^2$) are highly probable, with a lower cut-off at $a_0$. As shown in this sub-figure, the most probable amplitude ($a_{max}$) is located very close to the lower cut-off frequency, when taking into account the slow convergence of this PDF at high amplitudes (i.e., scaling as $a^{-\frac{3}{2}}$, plotted as a red dashed line). The lower cut-off implies the existence of a {minimum amplitude} in the entanglement measure in the MBL phase.}
    \label{fig:PDFs}
\end{figure}

\section{Numerical verification of $P_{F}^{(2)}(f)$ and $P_A^{(2)}(a)$ in Figure \ref{fig:PDFs}} \label{appcc}

Here, we numerically iterate the unitary evolution of $10^6$ random realizations of the disordered Heisenberg spin chain (\ref{Hamil}) with $J=1$ and $W=10,15$ and $20$ from an anti-ferromagnetic state. Followed by evaluating the system's state, we obtain the density matrix for spins at the edge and in the bulk (with at least two sites away from the edges) of chains with twelve spins. In this numerical simulation, our focus is on the finite-length and finite-time effects, therefore we study chains of twelve spins for the duration of $40 \frac{J}{\hbar}$. Eventually, we reconstruct the PDFs $P_{F}^{(2)}(f)$ and $P_A^{(2)}(a)$ through extracting the peaks in the Fourier transformation (at the locations predicted by our analysis) of the entanglement measure, as shown in Figure \ref{fig:55}. Due to the difference in first-order energy corrections for an edge spin compared with a spin in the bulk of the chain, the PDFs of the second-order frequencies are slightly different, particularly at the high-frequency corner of the PDFs (refer to Appendix \ref{app45} for more detail). This phenomenon is captured in the numerical simulation, as shown in the insets in Figure \ref{fig:55}(a), which affirms the effectiveness of the QCTF perturbative analysis. The deviation between the numerical and theoretical PDFs, which results from higher order contributions, fades away as the disorder strength increases, as shown in Figure \ref{fig:55}(b).

\section{Higher (than 4th)-order QCTF pole analysis- Introduction}\label{high-order}
\par In this Section, we present the notation used in the QCTF analysis of higher-order poles (discussed in \ref{higho}), and we derive the QCTF entanglement form that includes higher (than 4th)-order poles, to be used in the analysis in the following subsections.
\par To simplify the notation, a different form of labeling of product states is used, compared to the analysis in \ref{lowo}. We define $\hat{\mathbf{J}}_{k}= {S}_k^{+}{S}_{k+1}^{-}+{S}_k^{-}{S}_{k+1}^{+} $, which is the normalized (by $\frac{2}{J}$) off-diagonal part of ${\mathbf{J}}_{k}$. Using this operator is advantageous for two main reasons: Firstly, due to its zero diagonal, this operator uniquely projects each product state to exactly one other product state or returns zero. Secondly, the mentioned feature is compatible with the perturbation analysis presented later, where only projections to other product states are relevant to the leading terms in high-order corrections. The product state produced from applying an ordered product of operators ${\hat{\mathbf{J}}}_{i_1}{\hat{\mathbf{J}}}_{i_2}\cdots {\hat{\mathbf{J}}}_{i_k}$ (also denoted by $\hat{{\mathbf{J}}}_{i_1}: \hat{{\mathbf{J}}}_{i_k}$ for brevity) to the Ne\'el state is labeled by $\ket{l_{i_1:i_k}}$. 
\par From basic linear algebra, we know that for any two sets of basis vectors for the Hilbert space of the chain, one can always propose a bijective mapping between them. In particular, we use this feature in the case where one of the sets is the product states of eigenstates of $\mathbf{S}_z$ and the other set is the eigenstates of the system. In an entirely $\epsilon$-ideally-disordered chain, there is a \textit{trivial} candidate for this mapping because the eigenstates are assumed to be dressed product states. Therefore the trivial mapping is through perturbing the product states. The eigenstate corresponding to the product state $\ket{l_{i_1:i_k}}$ is meant to be through this trivial mapping and is denoted by $\ket{\overline{l_{i_1:i_k}}}$, with energy $E_{l_{i_1:i_k}}$. The case where the map between the sets is not trivial will also be considered when studying rare Griffiths regions in Section \ref{griff}.
\par Our goal is to obtain the perturbative QCTF \textit{centered around the second harmonic poles} in the entanglement measure. Without loss of generality, we will consider one of the second harmonic poles in our analysis, corresponding to $2 f_1$ (introduced in Section \ref{lowo}). Let us call this pole by $S$. The frequency spectrum around $S$ is formed by finding the convolution between the QCTF transformation ($\mathcal{K}$), centered around the dominant poles at $f_1$ and $-f_1$ (according to (\ref{main}), and due to the complex conjugation involved). This vicinity of poles of the QCTF transformation corresponds to the eigenmodes related to the product states of the forms $\hat{\mathbf{J}}_{m+i_1}: \hat{\mathbf{J}}_{m+i_k} \ket{0}$ and $ \hat{\mathbf{J}}_{m+i_1}: \hat{\mathbf{J}}_{m+i_k} \hat{\mathbf{J}}_m\ket{0}$ (denoted by $\ket{\overline{l_{m+i_1:m+i_k}}}$ and $\ket{\overline{l_{m+i_1:m+i_k,m}}}$ respectively) with $|i_j|\ge2$. Since each pair in these two families have equal energy difference up to the first order in perturbation ($E_{0}^{(1)}-E_{ 1}^{(1)}=h_{m}-h_{m+1}-J$), the emergent modes are adjacent to the dominant pole ($f_1$). Before writing down the QCTF, let us emphasize that the elements of the resolvent functions in (\ref{generator}) corresponding to these pairs of product states are:

\begin{equation}\label{43}
\begin{split}
    &(s+\frac{i}{\hbar}E_{l_{m+i_1:m+i_k}})^{-1} \ket{\overline{l_{m+i_1:m+i_k}}} \bra{\overline{l_{m+i_1:m+i_k}}},
    \\
        &(s+\frac{i}{\hbar}E_{l_{m+i_1:m+i_k,m}})^{-1} \ket{\overline{l_{m+i_1:m+i_k,m}}} \bra{\overline{l_{m+i_1:m+i_k,m}}}.
\end{split}
\end{equation}
Given this form of resolvent functions, the expectation value in (\ref{QCTFE}) at the initial Ne\'el state will give the inner products $\braket{\overline {l_{m+i_1:m+i_k}}}{0}$ and $\braket{0}{\overline {l_{m+i_1:m+i_k,m}}}$. Also, the star product in the QCTF (\ref{generator}) gives:
\begin{equation}\label{44}
\begin{split}
    (s-\frac{i}{\hbar}E_{l_{m+i_1:m+i_k,m}})^{-1} \star (s+\frac{i}{\hbar}E_{l_{m+i_1:m+i_k}})^{-1}
    =(s+\frac{i}{\hbar}\big(E_{l_{m+i_1:m+i_k}}-E_{l_{m+i_1:m+i_k,m}})\big)^{-1}
     \doteq (s+\frac{i}{\hbar}\Delta E_{\substack{m+i_1:m+i_k\\m+i_1:m+i_k,m}})^{-1}.
\end{split}
\end{equation}
Using (\ref{43}-\ref{44}) and (\ref{QCTFE},\ref{generator}), the QCTF centered around the aforementioned dominant pole ($f_1$) is

\begin{eqnarray}\label{q123}
\mathcal{K} \Big | _{f_1}=\sum_{k', l'} \sum_{k=0}\sum_{\{i_1:i_k\}}\frac{\braket{\overline{ {l_{m+i_1:m+i_k}}}}{0}\braket{0}{\overline {l_{m+i_1:m+i_k,m}}} \braket{\uparrow k'}{\overline{ {l_{m+i_1:m+i_k}}}}\braket{\overline {l_{m+i_1:m+i_k,m}}}{\downarrow l'}z_d^{l'-k'}z_a^{l'+k'}}{s+\frac{i}{\hbar}\Delta E_{\substack{m+i_1:m+i_k\\m+i_1:m+i_k,m}}},
\end{eqnarray}
where the first summation $(k',l')$ is understood to be over any arbitrary basis for the rest of the system (product states in subsystem $\mathcal{R}$, including all spins except $m$). There is no emphasis on the structure of this basis; Shortly we will see that after finding the residues of QCTF, it gives rise to a basis-independent quantity. Note that, similar to the previous section, the structure of the Hamiltonian dictates that the diagonal part of the QCTF, $\mathcal{K}_d (s)$ is zero \cite{exp1}. The QCTf transformation at $-f_1$ can be analogously found by interchanging the eigenstates in the left and right resolvent functions, therefore leading to the Laplace pole $- \Delta E_{\substack{m+i_1:m+i_k\\m+i_1:m+i_k,m}}=\Delta E_{\substack{m+i_1:m+i_k,m\\m+i_1:m+i_k}}$ in the denominator. Thus, using QCTF (\ref{q123}) and (\ref{main}), the entanglement frequency spectrum around the second harmonic poles $S$ is given by:
\begin{equation}\label{161}
\begin{split}
   \mathcal{Q}(s)= \sum_{k', l'}\sum_{\substack{k_1=0\\k_2=0}}\sum_{\substack{\{i_1:i_{k_1}\}\\\{i'_1:i'_{k_2}\}}}&\frac{\braket{\overline{ {l_{m+i_1:m+i_{k_1}}}}}{0}\braket{0}{\overline {l_{m+i_1:m+i_{k_1},m}}} \braket{0}{\overline{ {l_{m+i'_1:m+i'_{k_2},m}}}}\braket{\overline {l_{m+i'_1:m+i'_{k_2}}}}{0}}{(s+\frac{i}{\hbar}(\Delta E_{\substack{m+i_1:m+i_{k_1}\\m+i_1:m+i_{k_1},m}}-\Delta E_{\substack{m+i'_1:m+i'_{k_2},m\\m+i'_1:m+i'_{k_2}}}))}
\\ &\times \braket{\uparrow k'}{\overline{ {l_{m+i_1:m+i_{k_1}}}}}\braket{\overline {l_{m+i_1:m+i_{k_1},m}}}{\downarrow l'} \braket{\overline{ {l_{m+i'_1:m+i'_{k_2},m}}}}{\uparrow k'}\braket{\downarrow l'}{\overline {l_{m+i'_1:m+i'_{k_2}}}}.   
\end{split}
\end{equation}
\par We should note that, eigenstates of the system used in this formula, e.g., $\ket{\overline{ {l_{m+i_1:m+i_{k_1}}}}}$, are not limited to the case where the whole chain is $\epsilon-$ ideally disordered, but involve Griffiths regions as well, as will be discussed. Namely, using the two types of trivial and non-trivial mapping between the sets of basis vectors, discussed earlier, these eigenstates correspond to weakly or highly disordered regions, respectively. We will analyze effects of both types of regions separately. 
\par In what follows, our goal is to find and analyze an alternative form of this entanglement measure. In this alternative form, each side pole is substituted by its upper bound amplitude (or its strongest possible contribution), both for its frequency and amplitude. As a result, showing localized behavior from this alternative entanglement measure will sufficiently prove localization from the generic QCTF (\ref{161}).
\par Formula (\ref{161}) indicates three different mechanisms that contribute to the amplitude or frequency of a pole. The most trivial of these three is the numerator of the fraction, which is simply the inner products of four eigenkets with the initial Ne\'el state. The other two mechanisms, namely the term on the second line and the denominator of the fraction correspond to the localized behavior of the eigenmodes of the Hamiltonian and require more cautious analysis. To help determine these quantities, it is convenient to define the set of common indices $\tilde{\bigcap}\doteq \{m+i_1:m+i_{k_1}\}\bigcap \{m+i'_1:m+i'_{k_2}\}$ and the symmetric difference between the set of indices $\tilde{\Delta}\doteq\{m+i_1:m+i_{k_1}\}\Delta \{m+i'_1:m+i'_{k_2}\}$. Based on these definitions, in Appendix \ref{unfold}, the three aforementioned mechanisms are translated to the language of $\epsilon$-ideally disordered assumption through a perturbation expansion scheme, discussed in Section \ref{pertepsi}. Hence, it is shown that the upper-bound version of the entanglement measure (\ref{161}) can be rewritten in terms of $\tilde{\Delta}$ and $\tilde{\bigcap}$ sets as

\begin{equation}\label{264}
   \bar{\mathcal{Q}}(s)= \sum_{\tilde{\bigcap}=\{m+i^{\tilde{\bigcap}}_j\}}\sum_{k=1}\sum_{\tilde{\Delta}=\{m+i_{1}:m+i_k\}} q_{\tilde{\Delta},\tilde{\bigcap}},
\end{equation}
where 
\begin{equation}\label{364}
   {q}_{\tilde{\Delta},\tilde{\bigcap}}= \frac{\epsilon^{2|\tilde{\Delta}|+4|\tilde{\bigcap}|+2+\max\{2i_j-1|i_j>1\}+\max\{2|i_j|-1|i_j<-1\}}}{s\pm\frac{iJ}{2\hbar}(2f_1+\epsilon^{2(\min\{|i_j|\}-1)-2}+\epsilon^{2(\min\{|i^{\tilde{\bigcap}}_j|\}-1)-2})}.
\end{equation}
In this equation, we have $\min \{|i_j|\}< \min \{|i^{\tilde{\bigcap}}_j|\}$. In rewriting the QCTF (\ref{264}), we should note that the poles appear in positive and negative pairs because there is an anti-symmetry in exchanging the sets of indices in the first summation in (\ref{161}), which is disregarded in defining the sets $\tilde{\bigcap}$ and $\tilde{\Delta}$. Equation (\ref{364}) has important physical implications about the mechanism of broadening of each dominant pole, and consequently the localization mechanism, which is described in the next paragraph.
\par Summation (\ref{264}) implies that each pair of poles in the vicinity of the second-harmonic can be characterized by a choice of sets $\tilde{\bigcap}$ and $\tilde{\Delta}$, the frequency and amplitude of which are described by the denominator and numerator of (\ref{364}), respectively. To understand how the set of indices affects the features of the poles, or equivalently how broadening takes place, we start with the simple case of $\tilde{\bigcap}=\emptyset$ and $\tilde{\Delta}=\{m+i_1\}$ (Note that $|i_1|$ characterizes the distance between the local perturbation and spin $m$). In this case, the (maximum) amplitude of the pair of poles decays according to $\epsilon^{3+2|i_1|}$, as one considers farther perturbations. The (maximum) frequencies of the poles are $\pm \frac{J}{\hbar} \epsilon^{2|i_1|-4}$, which is, at its strongest, in the order of $\frac{J}{\hbar}$, and decay with power $\epsilon^2$ per site, which is the same power that it's amplitude decays with when considering farther perturbations. Adding more indices to set $\tilde{\Delta}$, that are bigger in their absolute value (corresponding to farther perturbations) than the initial index, will not change the frequency of the new pair of poles, up to the leading order of perturbation, due to the ``$\min$" in the denominator; However, the new resulting pair of poles is exponentially weaker in their amplitude, because of the ``$\max $" and $|\tilde{\Delta}|$ in the exponent of the numerator. Consequently, by keeping the closest (smallest index) perturbation in the set $\tilde{\Delta}$ fixed, one creates ``side-poles" by adding more indices to this set. Adding indices to the set $\tilde{\bigcap}$ produces weaker side-poles, since their amplitude decay by $\epsilon^4$ per perturbation, while the frequencies of the poles are primarily (up to the leading order) determined by the set $\tilde{\Delta}$. 

\par To summarize this subsection, the entanglement frequency spectrum of individual spins was obtained (\ref{161}). This formula provides the amplitudes (numerator) and frequency (the zeros of the denominator) of each frequency mode. These modes are associated with the sets $\tilde{\bigcap}$ and $\tilde{\Delta}$, which identify the set of perturbation operators incorporated in the QCTF formulation. To simplify the formula, an ``extreme" version of the entanglement measure was obtained in (\ref{364}) by replacing the amplitudes and frequencies with their upper bounds. Finally, the relation between the indices in $\tilde{\bigcap}$ and $\tilde{\Delta}$ and pole structures of (\ref{364}) was explained. Through a similar set of derivations, the effect of rare Griffiths regions on the entanglement measure of spins is studied in Appendix \ref{griff}. The analysis in this section will be used in Section \ref{infinitec}, to prove the quasi-periodicity and stability of the MBL phase. Meanwhile, in Section (\ref{unfold}) the three mechanisms in (\ref{161}), discussed above, are unfolded and Equation (\ref{364}) is derived.

\section{Unfolding the three mechanisms in (\ref{161})} \label{unfold}

\par We begin by analyzing the first mechanism, which involves establishing an upper bound on the magnitude of the inner products in the numerator of Eq.(\ref{161}). To achieve this, we employ the perturbation scheme outlined in Section\ref{pertepsi} along with the $\epsilon$-ideally disordered assumption from Eq.(\ref{epsiloni}). Based on the definition of the eigenstates $\ket{\overline{l_{i_1:i_k}}}$ provided in Section\ref{higho}, the minimal set of perturbation operators required to connect each of these states to $\ket{0}$ corresponds to a permutation of the set ${\mathbf{J}_{i_1:i_k}}$.

In Section~\ref{pertepsi}, we establish the following bounds:
\begin{equation}\label{h11}
\begin{split}
|\braket{\overline{ {l_{m+i_1:m+i_k}}}}{0}| \le \epsilon^k, 
|\braket{0}{\overline {l_{m+i_1:m+i_k,m}}}| \le \epsilon^{k+1}.
\end{split}
\end{equation}
Consequently, the magnitude of the numerator in Eq.~(\ref{161}) is bounded above by
\begin{equation}\label{175}
\epsilon^{2(k_1 + k_2 + 1)} = \epsilon^{2|\tilde{\Delta}| + 4|\tilde{\bigcap}| + 2},
\end{equation}
where we have utilized the fact that $\tilde{\Delta}$ and $\tilde{\bigcap}$ are disjoint, yielding $k_1 + k_2 = |\tilde{\Delta}| + 2|\tilde{\bigcap}|$.

\par Next, we address the second mechanism in Eq.~(\ref{161}), which pertains to the inner products in the second line and the summations over $k'$ and $l'$. Decomposing the eigenstates as
\begin{equation}
\ket{\overline{l_{m+i_1:m+i_k}}} = \ket{ \overline{l_{\uparrow m+i_1:m+i_k}}}\ket{\uparrow}^{m} + \ket{ \overline{l_{\downarrow m+i_1:m+i_k}}}\ket{\downarrow}^{m},
\end{equation}
we find that each summation over $k'$ and $l'$ corresponds to a basis-invariant quantity known as the \textit{local overlap} \cite{QCTF}. For instance, the summation over $k'$ results in the following expression for the local overlap:
\begin{equation}
\begin{split}
\braket{\overline{ {l_{\uparrow m+i_1:m+i_{k_1}}}}}{\overline{l_{\uparrow m+i'1:m+i'{k_2}}}} = -\braket{\overline{l_{\downarrow m+i'1:m+i'{k_2}}}}{\overline{ {l_{\downarrow m+i_1:m+i_{k_1}}}}}^*.
\end{split}
\end{equation}

To evaluate these local overlaps, consider the case where $\tilde{\Delta} = {i_1}$. In this scenario, one side of the local overlap corresponds to the eigenstate $\ket{\overline{ {l_{\tilde{\bigcap}}}}}$, while the other side corresponds to $\ket{\overline{ {l_{\tilde{\Delta} \cup \tilde{\bigcap}}}}}$. Although these eigenstates are orthogonal, their local overlap is not necessarily zero. Hence, starting from their respective product states $\ket{ {l_{\tilde{\bigcap}}}}$ and $\ket{ {l_{\tilde{\Delta} \cup \tilde{\bigcap}}}}$, all perturbation orders must generate orthogonal linear combinations of product states.

Since the initial difference between these product states is localized at $\mathbf{J}_{\tilde{\Delta}} = \mathbf{J}_{m+i_1}$, the leading-order perturbation that yields a non-vanishing local overlap at spin $m$ involves the shortest sequence of perturbation operators acting on spin $m$, non-zero for only one of the corresponding product states. When $i_1 > 1$, exactly one of these product states exhibits a non-zero perturbation magnitude for certain permutations of the Hamiltonians ${\mathbf{J}_m, \mathbf{J}_{m+1}, \ldots, \mathbf{J}_{m+i_1-1}}$. Applying the same permutation followed by $\mathbf{J}_{m+i_1}$ to the other product state results in the leading-order perturbation. Hence, counting all the required perturbations, this procedure requires $2i_1 - 1 = 2(i_1 - 1) + 1$ perturbation orders to achieve a non-zero local overlap.

This approach generalizes to cases where $\tilde{\Delta} = {i_1, \ldots, i_k}$. For each side, identify the maximal difference (local perturbation of spins) between the product states, given by $\max\{i_j | i_j > 1\}$ and $\max\{-i_j | i_j < -1\}$. Apply the previous procedure to each of these values (if the sets are not empty). Given the fact that the largest perturbation set (including the furthest perturbation) also includes operators connecting closer distortions to spin $m$, the local overlaps require at least 
\begin{equation}\label{201}
\max\{2i_j - 1 | i_j > 1\} + \max\{-1 - 2i_j | i_j < -1\}
\end{equation}
perturbation steps (the number of perturbation Hamiltonians used in the process).

The third mechanism concerns the denominator of the fraction in Eq.(\ref{161}), which determines the frequency of each pole in the second-harmonic QCTF. Utilizing the sets $\tilde{\Delta}$ and $\tilde{\bigcap}$ proves advantageous. Basic reasoning indicates that the upper bound on the magnitude of pole frequencies should not depend on the set of common indices, $\tilde{\bigcap}$, at least to the leading order of perturbation. This is because the denominator of Eq.(\ref{161}) is a linear combination (with alternating signs) of four eigenvalues, all sharing the common indices set.

Starting with $\tilde{\Delta} = {i_1}$, we evaluate the energy difference
\begin{equation}
E_{\tilde{\bigcap}} - E_{\tilde{\bigcap}, m} - E_{\tilde{\bigcap}, m+i_1} + E_{\tilde{\bigcap}, m+i_1, m} \approx E_{0} - E_{m} - E_{m+i_1, m} + E_{m+i_1}.
\end{equation}
Up to the zeroth order of perturbation (if $i_1$ did not exist), this linear combination of energies give the second-harmonic frequency $2f_1$ (as discussed in Appendix \ref{lowo}). The higher-than-second, leading-order energy difference arises from the minimal set of connected perturbation operators, due to perturbation $i_1$, that induce non-zero energy shifts for some product states (in the linear combination) while leaving others unaffected.

For $|i_1| > 2$, the minimal connecting set contains at least $i_1 - 1$ operators. Considering the fact that energy corrections stem from operator permutations returning to the initial product state (see Section \ref{pertepsi}), each perturbation Hamiltonian must be repeated twice in the minimal set of perturbations. Hence, using the perturbation scheme introduced in Section \ref{pertepsi}, the leading-order energy correction is bounded by
\begin{equation}
\frac{J}{2} \epsilon^{2(|i_1| - 1) - 2}.
\end{equation}
In the general case $\tilde{\Delta} = \{i_1, \ldots, i_k\}$, the leading energy difference is dictated by the closest distortion, bounded by
\begin{equation}\label{213}
J \epsilon^{2(\min{|i_j|} - 1) - 2}.
\end{equation}

\par To integrate all three mechanisms, we rewrite the summations in Eq.~(\ref{161}) using $\tilde{\Delta}$ and $\tilde{\bigcap}$:
\begin{equation}\label{261}
\mathcal{Q}(s) = \sum_{\tilde{\bigcap}} \sum_{k=1} \sum_{\tilde{\Delta} = {m+i_1, \ldots, m+i_k}} q_{\tilde{\Delta}, \tilde{\bigcap}}(s),
\end{equation}
where $q_{\tilde{\Delta}, \tilde{\bigcap}}(s)$ recasts the original summation labels in terms of these sets. Notably, poles appear in positive-negative pairs.

Using this notation, we adopt an alternative summation strategy: select members of $\tilde{\Delta}$ (with cardinality $|\tilde{\Delta}| = k_1 > 0$), then add common members to $\tilde{\bigcap}$. Importantly, $\tilde{\bigcap}$ does not affect local overlaps (second mechanism) or energy differences (third mechanism) at leading order, while each member reduces the numerator (first mechanism) by $\epsilon^4$. Thus, $\tilde{\bigcap}$ members induce exponentially weak broadening of peaks tied to $\tilde{\Delta}$.

In the thermodynamic limit, each $i_j^{\tilde{\bigcap}} \in \tilde{\bigcap}$, farther than the nearest $\tilde{\Delta}$ member, broadens near-zero peaks, shifting them by at most $\frac{J}{2} \epsilon^{2(|i_j^{\tilde{\bigcap}}| - 1) - 2}$ with reduced magnitude (by up to $\epsilon^4$). For large $i_j$, these effects become negligible.

\par Analyzing $\tilde{\Delta}$, the strongest near-zero pole corresponds to $k=1$, $i_1=-2$, producing a pole at $\pm J$ with magnitude bounded by $\epsilon^3$. Moving distortions further reduces both frequency and magnitude by at least $\epsilon^2$ per step. Adding more indices to $\tilde{\Delta}$ introduces weak side peaks, with each new member contributing at least a factor of $\epsilon^{2|i_j|+1}$ weaker than the original.

Summarizing, we rewrite the entanglement measure's upper bound:
\begin{equation}\label{362}
\overline{q}_{\tilde{\Delta}, \tilde{\bigcap}}(s) = \frac{\epsilon^{2|\tilde{\Delta}| + 4|\tilde{\bigcap}| + 2 + \max\{2i_j - 1 | i_j > 1\} + \max\{2|i_j| - 1 | i_j < -1\}}}{s \pm \frac{iJ}{2\hbar}(2f_1+\epsilon^{2(\min{|i_j|} - 1) - 2} + \epsilon^{2(\min{|i_j^{\tilde{\bigcap}}|} - 1) - 2})},
\end{equation}
which proves (\ref{364})

\begin{figure}
    \centering
    \begin{tikzpicture}

\node[draw, circle] at (-4,0) {$\ket{0}$};
\node[draw, circle] at (-2,2) {$\ket{l_{i_1}}$};
\node[draw, circle] at (1,1) {$\ket{l_{i_1:i_2}}$};
\node[draw, circle] at (2.5,-1.5) {$\ket{l_{i_1:i_k}}$};

\node[rotate=-60] at (1.7,-0.3) {$\cdots$};
\draw[->, >=latex] (-3.75,0.27) node[above, xshift=0.8cm,yshift=0.8cm, rotate=46] {$\hat{\mathbf{J}}_{i_1}$} node[below, xshift=0.7cm,yshift=0.8cm, rotate=46] {$E^{(0)}_0-E^{(0)}_{l_{i_1}}$} -- (-2.35,1.75); 

\draw[->, >=latex]  (-1.6,1.8)node[above, xshift=0.8cm,yshift=-0.3cm, rotate=-25] {$\hat{\mathbf{J}}_{i_2}$}  -- (0.43,1.1); 

\draw[->, >=latex] (-3.6,0)  node[below, xshift=1.85cm,yshift=0.4cm, rotate=10] {$E^{(0)}_0-E^{(0)}_{l_{i_1:i_2}}$} -- (0.43,0.8); 

\draw[->, >=latex] (-3.6,-0.1)  node[below, xshift=3cm,yshift=-0.7cm, rotate=-12] {$E^{(0)}_0-E^{(0)}_{l_{i_1:i_k}}$} -- (1.9,-1.4); 

\end{tikzpicture}
    \caption{This figure illustrates how the leading order perturbations can be obtained from a series of perturbation operators $\hat{\mathbf{J}}_{i_1},\cdots, \hat{\mathbf{J}}_{i_k}$. Arrows labeled by operators $\hat{\mathbf{J}}_{i_j}$ show how perturbation operators connect the product states and arrows connecting $\ket{0}$ to the other product states is accompanied by the energy differences that appear in the denominator of the perturbation coefficient. To obtain the $k$th order correction to the eigenvalue corresponding to $\ket{0}$, we require that this series of perturbations return back to $\ket{0}$, or equivalently create a loop.  }
    \label{pertdiag}
\end{figure}

\section{MBL phase transition in infinite chains}\label{infinitec}

In infinite chains, side-poles inevitably approach the original dominant poles (e.g., second harmonics) exponentially closely; however, no mechanism exists to align them exactly with the dominant pole's frequencies. Consequently, within any punctured neighborhood around the second-harmonic pole $s=2f_1$, there exists an infinite number of poles. In the Fourier domain, each pole is represented as a delta function at the corresponding frequency (see \ref{lap2f}). The magnitude of these delta functions diminishes exponentially as they converge towards $s=2f_1$, according to (\ref{364}). This distinctive behavior allows for the frequency spectrum of the entanglement measure to be recast as a continuous function $f(\omega)$ in Fourier space, rather than as an infinite series of densely packed delta functions. This approximation holds true only within an infinitesimal region around each dominant pole.

A key principle in the proof of the stability of MBL is that for entanglement to saturate and reach a finite limit as $t \rightarrow \infty$, the final value theorem requires the existence of the limit $\lim_{\omega \rightarrow 0} \omega f(\omega)$. This implies that the infinite series of delta functions must collectively yield a well-behaved continuous function $f(\omega)$. In an $\epsilon$- ideally disordered region, as detailed in Appendix \ref{lowo} (Eq. (\ref{SHH})), the fourth-order second-harmonic component consists of two non-zero-frequency delta functions, each with a magnitude that must scale as $\sim \epsilon^4$ to leading order in perturbation. An alternative approach to derive this scaling is as follows: (In subsection \ref{app45}, we demonstrated that the magnitude of the dominant second-order poles scales no slower than $(\frac{J}{W} \sim \epsilon^2)$. Consequently, the fourth-order poles, which include the second-harmonics, must scale no slower than $\sim \epsilon^4$). This scaling dictates a minimum energy that scales as $\sim\epsilon^4$ (quadratic in the amplitude of delta functions) carried by the second-harmonic modes in the entanglement dynamics. In what follows, we will employ this energy scaling to prove the stability of the MBL phase. We achieve this goal by comparing the energy carried by individual second-harmonic poles and the cumulative energy carried by its side poles. 

(Note that in this analysis, we normalize the (infinite) energy carried by a delta function by dividing it by the infinite simulation time ($T$ in \ref{lapltr}) after the quench. This ensures that, according to Parseval's theorem, a unit delta function in the frequency spectrum carries exactly one unit of energy.)
 
Before proceeding to calculate the cumulative energy carried by the side-poles, we utilize a key property of delta functions to simplify the derivations. This property pertains to the effect of delta functions generated by labels in the set $\tilde{\Delta}$. In Appendix \ref{161}, we demonstrated that each member of $\tilde{\Delta}$ produces a side-pole that is exponentially close to the original side-poles induced by the set $\tilde{\bigcap}$. Crucially, this process leads to a reduction in the amplitude of the original side-poles (associated with $\tilde{\bigcap}$) by a factor denoted by $\delta$, while simultaneously generating a new side-pole with an amplitude not exceeding $\delta$, as a consequence of the perturbation process. Since the energy carried by a delta function is proportional to the square of its amplitude, the energy reduction from the original side-pole exceeds the energy contribution of the newly emerged side-pole resulting from the indices in $\tilde{\Delta}$. To illustrate, if the original side-pole amplitude is normalized to 1 and $\delta < 1$, we find that $1 - (1 - \delta)^2 > \delta^2$. Therefore, because our objective is to determine the scaling behavior of the maximum energy carried by the side-poles—and the inclusion of the set $\tilde{\Delta}$ only diminishes the total energy—we can neglect the effects of this set in our analysis.

To obtain the energy scaling of the side-poles of the second-harmonic pole, Using (\ref{364}) and Parseval's theorem, the energy within a neighborhood of the second harmonic $s=2f_1$. This neighborhood is defined through an adjustable parameter $i_{min}$ such that all the perturbation labels $\{i_j\}$ in (\ref{364}) satisfy $\min {|i_j|} > i_{min}$ (requiring that they are closer to the second-harmonic pole than a pole corresponding to a perturbation at $i_{min}$). Therefore, using (\ref{362}), the cumulative energy of side-poles can be obtain by the summation over the squared amplitudes of the side poles, given by

\begin{equation}\label{490}
\sum_{\substack{\tilde{\Delta}=\{i_j\}\\ \min {|i_j|}>i_{min}}} \epsilon^{2(2|\tilde{\Delta}|+2+\max\{2i_j-1|i_j>1\}+\max\{1-2i_j|i_j<-1\})},
\end{equation}

This sum spans indices on both sides of spin $\mathcal{M}$ (indicated by positive and negative $i_j$), with the closest index beyond $i_{min}$ and the farthest approaching infinity ($i_{max}\doteq \max\{|i_j|\}\rightarrow \infty$). Hence, the number of indices per side ranges from zero to ${i_{max}} -  {i_{min}} + 1$. Here, let us consider perturbations on one side of $\mathcal{M}$, i.e., when $i_j >0$. This simplification is possible because by having perturbations on both sides of $\mathcal{M}$, the energies are exponentially smaller (note $\max\{2i_j-1|i_j>1\}+\max\{1-2i_j|i_j<-1\}$). In this case, for each $i_{max}$, and each $|\tilde\Delta|=1,2,\cdots, i_{max}-i_{min} +1$, there exist $\binom{i_{max}-i_{min} +1}{|\tilde{\Delta} |}$ choices of indices in the set $\Delta$. Hence, (\ref{490}) can be rewritten as
\begin{equation}
\epsilon^{2} \sum_{i_{max}=i_{min}}^\infty \epsilon ^{4 i_{max}}\sum_{|\tilde{\Delta}|=1}^{i_{max}-i_{min}+1} {\binom{i_{max}-i_{min} +1}{|\tilde{\Delta }|} \epsilon^{4 |\tilde{\Delta}|}}.
\end{equation}
After rewriting the summations using binomial theorem and Taylor expansion, i.e., $\sum_{n=n_0} a^n= \frac{a^{n_0}}{1-a} (|a|<1)$, we obtain

\begin{equation}
    \epsilon^{4 i_{min}+2}\Big (\frac{1+\epsilon^4}{1-\epsilon^4 (1+\epsilon^4)}-\frac{1}{1-\epsilon^4}  \Big )\approx \epsilon^{4 i_{min}+6,
    }
\end{equation}
which approaches faster than $\epsilon^{10}$ towards $0$ for small $\epsilon$.
In other words, for any $i_{min}$, this decay rate is faster than $\sim \epsilon^8$, which is the rate at which the energy of the second-harmonic pole decay. Conversely, as $\epsilon \rightarrow 1$, second-harmonic frequency modes can carry arbitrarily large energy. By the intermediate value theorem, there must exist an $\epsilon^*$ separating regions where second-harmonic poles dominate energetically over the cumulative energy of its side poles.

Thus, by decreasing $\epsilon$ (increasing disorder strength), the influence of the second-harmonic frequency spectrum can be made arbitrarily weaker compared to their corresponding second-harmonic mode. This proves that the side-poles can not diminish the second-harmonic pole and lead to a continuous frequency spectrum of the entanglement measure. As a result, the frequency spectrum of the spin under consideration must remain discontinuous in a sufficiently strong disordered chain. We also provided a logical proof that discontinuity of one spin suffices for the discontinuity of any finite subsystem that includes that spin (in Section \ref{nonerg}). Therefore, due to the final value theorem, the entanglement evolution of the corresponding spin can not saturate and remains quasi-periodic. This proof, in the context presented in the paper, proves that that many-body localization (MBL) is a stale phase, and the entanglement dynamics of individual spins exhibit quasi-periodicity. 



\section{Localization in presence of rare Griffiths regions}\label{griff}

We examine the delocalizing effects arising from a poorly disordered region adjacent to an $\epsilon$-ideally disordered region. Specifically, we aim to calculate the energy transferred by side-poles generated through interactions with rare Griffiths regions of size $n_g$, separated from the target spin (subsystem $\mathcal{M}$) by $d_g$ $\epsilon$-ideally disordered spins. Due to the inherently non-local structure of eigenstates within these rare regions, the perturbative approach employed in our earlier analysis is no longer applicable.

To facilitate the calculations, we adopt the following assumptions, which reflect fully non-local behavior:
\begin{itemize}
    \item The initial N\'eel-ordered state of the spins within the rare region is uniformly projected across all $2^{n_g}$ local degrees of freedom within the rare region.
    \item The degrees of freedom (local eigen-modes) within the rare region (corresponding to the eigenstates of the isolated subsystem) interact uniformly with those outside the region, consistent with volume-law entanglement.
\end{itemize}

These assumptions impose strongly non-local characteristics on the rare regions, representing the strongest thermalizing effects that can emerge from such regions.

In the presence of Griffiths regions, the mapping between unperturbed product states and the system's eigenstates is non-trivial. Nevertheless, one can define a mapping from labels $l_g = 1 : 2^{n_g}$ (serving as basis labels for the sub-Hilbert space of the region $\mathcal{R}_g$, e.g., product states within this region) to the set of actual eigenstates $\ket{\bar{l_g}}$ corresponding to the isolated rare region. 

As a result, the direct product of this mapping with the trivial mapping (acting on the adjacent $\epsilon$-ideally perturbed region) yields a new mapping $\mathcal{I}(l_g, l_{i_1:i_k})$ from a set of unperturbed eigenstates 
\begin{equation}\label{mapping}
    \ket{l_{g, i_{1:k}}} \doteq \ket{l_g} \otimes \ket{l_{i_1:i_k}}
\end{equation}

to the eigenstates of the compound system (comprising the rare region and its adjacent $\epsilon$-ideally disordered region) 
\[
\ket{\overline{l_{g, i_{1:k}}}}.
\]
Similarly, using this mapping, the unperturbed energy of the compound region can be expressed as 
\[
E^{(0)}_{l_g, l_{i_1:i_k}} = E_{l_g} + E^{(0)}_{l_{i_1:i_k}}.
\]
With the introduced labeling and notation, we can modify equation (\ref{161}) to derive the QCTF entanglement measure function in the presence of a Griffiths region as follows:

\begin{equation}\label{380}
\begin{split}
     \eval{\mathcal{Q}(s)}_{2f_1}= \sum_{\substack{l_g=1:2^{n_g}\\l'_g=1:2^{n_g}}}
     \sum_{\substack{\{i_1:i_{k_1}\}\\\{i'_1:i'_{k_2}\}}} B \frac{\braket{\overline{ {l_{\uparrow l_g, m+i_1:m+i_{k_1}}}}}{\overline{l_{\uparrow l'_g, m+i'_1:m+i'_{k_2},m}}}\braket{\overline{l_{\downarrow l'_g, m+i'_1:m+i'_{k_2}}}}{\overline{ {l_{\downarrow l_g, m+i_1:m+i_{k_1},m}}}}}{s+\frac{i}{\hbar}(\Delta E_{\substack{l_g,m+i_1:m+i_{k_1},m \\ l_g,m+i_1:m+i_{k_1}}}-\Delta E_{\substack{l'_g,m+i'_1:m+i'_{k_2}\\{l'_g,m+i'_1:m+i'_{k_2},m}}})},
     \\ B=  \braket{\overline{ {l_{l_g,m+i_1:m+i_{k_1}}}}}{0} \braket{0} {\overline{ {l_{l'_g,m+i'_1:m+i'_{k_2}}}}} \braket{0}{\overline{ {l_{l_g,m+i_1:m+i_{k_1},m}}}} \braket {\overline{ {l_{l'_g,m+i'_1:m+i'_{k_2},m}}}}{0}.
\end{split}
\end{equation}
Similar to our analysis before, we have to consider the three mechanisms, embedded in (\ref{380}), that contribute to the side poles.

\par 
To characterize the eigenmodes of this compound system—specifically, the eigenvalues in the denominator of (\ref{380}) and those in its numerator—we consider an $\epsilon$-ideally disordered region adjacent to a rare Griffiths region. The disordered region consists of at least \( n_d + 1 \) spins, including the spins indexed by \( \{i_d \in \mathcal{R}_d = m - n_d : m\} \). The rare Griffiths region has size \( n_g \) and includes the spins indexed by \( \{i_g \in \mathcal{R}_g = m - n_d - n_g : m - n_d - 1\} \), where the lattice sites exhibit minimal disorder (\( h_i \approx h_j \) in units of \( J \)). In the extreme case, we assume a disorder-free condition within \( \mathcal{R}_g \), setting \( h_i = h_g \) for all \( i \in \mathcal{R}_g \).

Under this assumption, and in the absence of interactions with the rest of the system, the eigenstates of the sub-lattice \( \mathcal{R}_g \) are characterized by a set of momenta \( \{q_{1:r}\} \) corresponding to magnon quasi-particles. We focus on the highest density of magnons, \( r = \lfloor n_g / 2 \rfloor + 1 \), which aligns with the Néel order. This density ensures that the spin adjacent to \( \mathcal{R}_g \) in \( \mathcal{R}_d \) (i.e., spin \( m - n_d \)) initially resides in the state \( \ket{\downarrow} \).

Given the open boundary conditions in \( \mathcal{R}_g \), the unperturbed eigenstates in \( \mathcal{R}_g \) are described by the following standing-wave Bethe ansatz \cite{alcaraz1987surface}:
\begin{equation}\label{bethe}
\begin{split}
    &\ket{q_{1:r}} = \mathcal{N}_r^{-1} \sum_{n_1 < n_2 < \cdots < n_r,\, n_i \in \mathcal{R}_g}  a^{q_{1:r}}(n_{1:r})\ket{n_{1:r}}, \\
    &\ket{n_{1:r}} = \left(\prod \mathbf{S}^{+}_{n_i}\right) \ket{\downarrow, \dots, \downarrow}, \\
    &a^{q_{1:r}}(n_{1:r}) = \sum_{\mathcal{P}^{q_{1:r}}} \epsilon_{\mathcal{P}} A(\mathcal{P}^{q_{1:r}}_1, \dots, \mathcal{P}^{q_{1:r}}_r) e^{i\sum  \mathcal{P}^{q_{1:r}}_i n_i },
\end{split}
\end{equation}
where \( \mathcal{P}^{q_{1:r}} \) represents all possible permutations and negations of the momenta \( q_{1:r} \), and \( \epsilon_{\mathcal{P}} \) denotes the sign of a particular permutation and negation (\(-1\) if the total number of such transformations is odd, and \( +1 \) otherwise). The function \( A(.) \) captures the interactions between the standing-wave magnon modes.

The momenta \( q_{1:r} \) must satisfy a nonlinear coupled Bethe equation, making the problem highly intricate. However, rather than solving these equations explicitly, we will establish upper bounds for the relevant inner products in the QCTF analysis.

The unperturbed eigenvalues (corresponding to the isolated Hamiltonian for \( \mathcal{R}_g \)) are given by \cite{alcaraz1987surface}:
\begin{equation}\label{245}
    E_{q_{1:r}} = \left( 2r - n_g \right) \frac{h_g}{2} + \frac{J}{2} (n_g - 1) + 2J \sum (\cos(q_i) - 1).
\end{equation}

\par Using the fact that the eigenmodes, when localized inside the rare Griffiths region, can be characterized through the labels \( q_{1:r} \), we can use this label instead of \( l_g \) to address the eigenmodes. Hence, for simplicity, we rewrite the mapping (\ref{mapping}) as:
\begin{equation}
    \ket {l_{q_{1:r},i_{1:k}}} \doteq \ket{q_{1:r}} \otimes \ket{l_{i_1:i_k}}.
\end{equation}
Eigenstates of the compound system, \( \ket{\overline {l_{q_{1:r},i_{1:k}}}} \), can be obtained by applying perturbation Hamiltonians in the ideally disordered region. 

Further steps in obtaining the entanglement measure for spin \( m \) involve calculating the relevant local overlaps and energy differences that appear in the QCTF function in the presence of Griffiths regions. Using the introduced labeling and notation, one can modify (\ref{161}) to derive the QCTF entanglement measure function in the presence of the Griffiths region:

\par Using the fact that the eigenmodes, when localized inside the rare Griffiths region, can be characterized through the labels \( q_{1:r} \), we can use this label instead of \( l_g \) to address the eigenmodes. Hence, for simplicity, we rewrite the mapping (\ref{mapping}) as:
\begin{equation}
    \ket {l_{q_{1:r},i_{1:k}}} \doteq \ket{q_{1:r}} \otimes \ket{l_{i_1:i_k}}.
\end{equation}
Exact eigenstates of the compound system, \( \ket{\overline {l_{q_{1:r},i_{1:k}}}} \), can be obtained by applying perturbation Hamiltonians in the $\epsilon$-ideally disordered region and where the two regions connect.

Further steps in obtaining the entanglement measure for spin \( m \) involve calculating the relevant local overlaps and energy differences that appear in the QCTF function (\ref{380}).

In order to study the strongest set of effects from the Griffiths region, we consider the local distortions only from this region and not from the ideally perturbed region. That is, we begin with the case \( k_1 = k_2 = 0 \). Similar to the previous case, three different variables need to be calculated: the inner products in \( B \), the local overlaps in the numerator, and the energy differences in the denominator. 

We start by calculating \( B \). Note that, using the notation in (\ref{bethe}), the initial Néel state \( \ket{0} \) is given by:
\begin{equation}
    \ket{0} = \ket{n_i^{\text{Ne\'el}} = m - n_d - n_g + 2(i-1)} \otimes \ket{\downarrow \uparrow, \cdots},
\end{equation}
which leads to the following bounds for the first (similarly for the second) and third (similarly for the fourth) inner products in \( B \):
\begin{equation}
\begin{split}
    |\braket{\overline{ {l_{q_{1:r}}}}}{0}| &= \mathcal{N}^{-1} |a^{q_{1:r}}(n_i^{\text{Ne\'el}})| \leq 1, \\
    |\braket{0}{\overline{ {l_{q_{1:r},m}}}}| &= \epsilon \mathcal{N}^{-1} |a^{q_{1:r}}(n_i^{\text{Ne\'el}})| \leq \epsilon,
\end{split}
\end{equation}
which gives the upper bound:
\begin{equation}\label{bbb}
    |B| \leq \epsilon^2,
\end{equation}
for every set of labels $q_{1:r}$ and $q'_{1:r}$. This upper bound is very generous, but avoids the highly nonlinear behavior of \( B \) as a function of the set of momenta. Moreover, as will be demonstrated, other mechanisms have stronger delocalizing effects in the presence of a Griffiths region.

Similar to the calculations for the case of fully $\epsilon$- ideally disordered region, to obtain the local overlaps and energy differences in (\ref{380}), all ordered sets of perturbation operators that connect region \( \mathcal{R}_g \) to spin \( m \) must be taken into account, as they are the only source of any non-vanishing effect in both of these variables. Most importantly, the effect from the perturbation operator \( \mathbf{J}_{m-n_d-1} \), which connects the two regions \( \mathcal{R}_g \) and \( \mathcal{R}_d \) should be explicitly calculated. To this end, we need to obtain the transition amplitude:
\begin{equation}\label{2823}
    \mel{q'_{1:r'}\otimes \uparrow}{\mathbf{J}_{m-n_d-1}}{q_{1:r}\otimes \downarrow}
\end{equation}
for different sets of momenta on each side. The arrow represents the state of the spin adjacent to \( \mathcal{R}_g \) in \( \mathcal{R}_d \). This transition amplitude appears in the perturbation series (with perturbation Hamiltonian \( \mathbf{J}_{m-n_d-1} \)) of the unperturbed ket \( \ket{q_{1:r}\otimes \downarrow} \).

It is important to note that the perturbation due to the diagonal part of \( \mathbf{J}_{m-n_d-1} \) is not relevant to our analysis, as it requires \( q'_{1:r} = q_{1:r} \) (due to the orthogonality of eigenstates in (\ref{bethe})), leading to exactly zero frequencies in QCTF. Instead, we are interested in near-zero frequencies. 

The off-diagonal part of the perturbation Hamiltonian in the inner product (\ref{2823}) vanishes when applied to the ket \( \ket{q_{1:r}\otimes \downarrow} \), except when \( n_r = m - n_d - 1 \) in the Bethe wave function (\ref{bethe}). In this case, it projects \( \ket{q_{1:r}\otimes \downarrow} \) onto the sector with \( r' = r - 1 \) excitations, where \( n_{r'} < m - n_d - 1 \). Therefore, one can rewrite (\ref{2823}) as:
\begin{equation}
\begin{split}
\frac{J}{2\mathcal{N}_r\mathcal{N}_{r-1}} \sum_{n_1<n_2<\cdots<n_{r-1}}  \Bigg (a^{q_{1:r}}(n_{1:r}) {a^{q'_{1:r-1}}}^*(n_{1:r-1}) \sum_{\mathcal{P}^{q_{1:r}}} \sum_{\mathcal{P}^{q'_{1:r-1}}} e^{i\sum  (\mathcal{P}^{q_{1:r-1}}_i-\mathcal{P}^{q'_{1:r-1}}_i) n_i}  e^{i \mathcal{P}^{q_{1:r}}_r n_r} \\
\times \epsilon_{\mathcal{P}}\epsilon_{\mathcal{P'}} A(\mathcal{P}^{q_{1:r}}_1,\cdots, \mathcal{P}^{q_{1:r}}_r) A^*(\mathcal{P}^{q'_{1:r-1}}_1,\cdots, \mathcal{P}^{q'_{1:r-1}}_{r-1})\Bigg ).
\end{split}
\end{equation}

The outer summation reflects the orthogonality of eigenstates in $r'=r-1$ sector, which requires that the sets of momenta $q_{1:r-1}$ and $q'_{1:r-1}$ be identical. To simplify, let us use the upper-bound $|A|\le 1$. Also we can use the identity:
\begin{equation}\label{perm}
    \mathcal{P}^{q_{1:r}}=\bigcup_i \{\mathcal{P}^{q_{1:i-1},q_{i+1:r}},\mathcal{P}^{q_{1:r}}_r= q_i\},
\end{equation}
for the summation over $\mathcal{P}^{q_{1:r}}$. Intuitively, we can pick any of the members of the set $\{q_{1:r}\}$ and assign it as the last member of the permutation $\mathcal{P}^{q_{1:r}}$, followed by all possible permutations of the remaining $r-1$ members. Imposing the orthogonality condition, requires that for $\{q_{1:i-1},q_{i+1:r}\}=\{q'_{1:r-1}\}$, which then gives $\mathcal{N}_{r-1}e^{i \mathcal{P}^{q_{1:r}}_r n_r}$ for the summations. We can repeat this procedure for each individual member of $\{q_{1:r}\}$. We can start by excluding $q_r$ (first set of permutations) and require $\{q_{1:r-1}\}=\{q'_{1:r-1}\}$. Building up on this, proceeding with excluding $q_i$ can be thought of as exchanging $q_i$ and $q_r$ in all of the first set of permutation, which requires $q_r=q_i$ due to the orthogonality condition. Therefore, by proceeding further, the number of times we get a non-zero inner product is equal to the number of repeated momenta in $\{q_{1:r-1}\}$ plus one (for the first set of permutations). Hence, after taking the absolute value of the inner product,
\begin{equation} \label{314}
\begin{split}
      |\mel{q'_{1:r'}\otimes \uparrow}{\mathbf{J}_{m-n_d-1}}{q_{1:r}\otimes \downarrow}|&\le \frac{J}{2\mathcal{N}_r} \sum_{i} \delta_{q_i,q_r}
      \\ &\le \frac{J}{2\mathcal{N}_r}  (1+\max \{\#_{q_i=q_j}\})
      \\ &\le \frac{J r}{2\mathcal{N}_r}  
\end{split}
\end{equation}
where $  \max \{\#_{q_i=q_j}\}$ the highest number of repeated momenta in the set $q_{1:r-1}$. We should remind that the set of momenta should satisfy the non-linear Bethe equation with a set of Bethe quantum numbers which restricts repeated momenta in a an admissible set; Therefore, the upper-bound is very generous. Also, note that using (\ref{245}) the zeroth order energy difference between the kets in (\ref{2823}) is:
\begin{equation}\label{315}
    h_g-h_{m-n_d}+2J(-1+\cos(q_i)),
\end{equation}
which we take to be larger than $\epsilon^{-1} J$ in magnitude; otherwise the adjacent spin would be considered to be inside the rare region. This energy difference appears in the perturbation expansion.
\par Having calculated an upper-bound for (\ref{2823}), local overlaps and energy differences can be obtained. Similar to our calculations for $\epsilon$- ideally disordered regions, the energy differences in the QCTF, up to the leading order, are equal to the energy corrections corresponding to the perturbation Hamiltonians that connect spin $m$ to region $\mathcal{R}_g$. Such connected perturbation include Hamiltonians $\mathbf{J}_{m-1}, \mathbf{J}_{m-2},\cdots, \mathbf{J}_{m-n_b-1}$ and each should appear at least twice (to form a close loop and return to the initial un-perturbed state, as explained in the perturbation scheme in Appendix \ref{pertepsi}). Using the fact that there are least $n_b-2$ perturbation steps that connect spin $m$ to adjacent spin to $\mathcal{R}_g$, similar to (\ref{213}), there is a contribution $\epsilon^{2n_b-4}$. This should be followed by the last step of perturbation that connects the series to $\mathcal{R}_g$. For this step, $\mathbf{J}_{m-n_b-1}$ should be used twice as the perturbation operator (to form a close loop). Therefore using (\ref{314}-\ref{315}) we have:
\begin{equation}\label{474}
    |\Delta E_{\substack{\{q_{1:r}\}\Delta \{q'_{1:r}\}\\\{\{q_{1:r}\}\Delta \{q'_{1:r}\},m\}}}|\le (\frac{ r}{2\mathcal{N}_r})^2J\epsilon ^{2n_d-2},
\end{equation}
with $\mathcal{N}_r^2=\binom{n_g}{r}$. This inequality has important consequences; Note that it is inversely proportional to $\mathcal{N}_r^2$, which scales exponentially in $n_g$, on the other hand, the number of permutations of momenta increase (at most) exponentially with $r\approx \frac{n_g}{2}$ as well. All of these poles are inside the a neighborhood of the origin which is of order $J \epsilon^{2n_d-1}$. Hence adding a spin to region $\mathcal{R}_d$, pushes all of these poles closer to the origin by $\epsilon^2$. Finding the local overlaps is a similar procedure; we find the minimal set of perturbation operators, which include all of the operators in $\mathcal{R}_d$, following the similar path taken in Appendix \ref{unfold}, and using (\ref{2401}), (\ref{314}), we obtain:

\begin{equation}\label{CCC}
    |\braket{\overline{ {l_{\uparrow q_{1:r}}}}}{\overline{l_{\uparrow q'_{1:r}}}}|\le (\frac{ r}{2\mathcal{N}_r})\epsilon ^{2(n_d+1)}.
\end{equation}
Using this scaling, we can upper-bound the amplitude of side poles induced by the rare Griffiths regions, as described in (\ref{380}). Based on (\ref{bbb}) and (\ref{CCC}), the amplitude of side poles must be upper bounded by (noting that in (\ref{380}), there exist two local overlaps, each upper bounded by (\ref{CCC})):
\begin{equation}\label{enee}
    \bigg(\frac{r}{2\mathcal{N}_r} \epsilon ^{2(n_d+1)}\bigg)^2 \epsilon^2.
\end{equation}
The energy carried by individual side poles is quadratic in (\ref{enee}), scaling as \( \epsilon^{8n_d+12} \) in \( \epsilon \) and as \( (\frac{r}{2\mathcal{N}_r})^4 \) in \( n_g \) (noting that \( \mathcal{N}_r = \binom{n_g}{r} \) and \( r \approx n_g/2 \)). However, the number of side poles grows exponentially with \( n_g \).

Here, we consider the effect induced by the entire Hilbert space underlying the rare region, which consists of approximately \( 2^{n_g} \) eigenmodes, rather than just the \( r \approx n_g/2 \)-magnon sector. Notably, the side poles induced by the \( r \approx n_g/2 \)-magnon sector are significantly more energetic than those induced by other sectors, due to their overlap with the initial Néel order of the system. This feature allows us to use the energy scaling (\ref{enee}) and multiply it by the size of the Hilbert space, to obtain the upper bound for the cumulative energy of the side-poles. (Importantly, in the proof above, we showed that the side-poles are in one-to-one correspondence to the rare-regions eigenstates) Consequently, the cumulative energy carried by the side poles scales as:
\begin{equation}
    2^{n_g} \bigg(\frac{r}{2\mathcal{N}_r}\bigg)^4 \quad \text{in } n_g.
\end{equation}

This scaling, combined with the dependence on \( n_d \) given above, should compete with the energy of the second harmonic, which scales as \( \epsilon^8 \) (see Appendix \ref{lowo}). Given that \( \frac{r}{\mathcal{N}_r} \ll 1 \), and rewriting \( 2^{n_g} = \epsilon^{n_g \log_\epsilon 2} \), we compare the following scalings:
\begin{equation}\label{scaaaaa}
    \epsilon^{n_g \log_\epsilon 2+8n_d+10} \hspace{0.5cm} \text{versus} \hspace{0.5cm} \epsilon^8.
\end{equation}

Note that \( \log_\epsilon 2 < 0 \) for \( 0 < \epsilon < 1 \). When \( \epsilon \to 1^- \), the left-hand side dominates since \( \log_\epsilon 2 \to -\infty \). However, we are interested in the \( \epsilon \to 0^+ \) limit, which corresponds to a regime of high disorder. In this limit, not only does the mean value for \( n_g \) tend to zero while \( n_d \gg n_g \) (see subsection \ref{stattt}), but also \( \log_\epsilon 2 \to 0 \). 

This results in a significantly faster decay of the energy scaling on the left-hand side of (\ref{scaaaaa}), which corresponds to the side-pole energies, compared to the right-hand side, which corresponds to the second-harmonic poles. Therefore, even in the presence of rare regions, side poles induced by Griffiths effects cannot dominate the second-harmonic poles.

\section{Proof of the upper bounds in (\ref{h11}) } \label{pertepsi}
\par Here, we present the time independent perturbation scheme used to calculate energy corrections and inner products in Section \ref{unfold}. Unperturbed eigenstates are product states of local $\mathbf{S}_z$ operators, corresponding to the first term in Hamiltonian (\ref{Hamil}). Structure of the perturbation Hamiltonian $\hat{\mathbf{J}}_k$, defined in Section \ref{higho}, allows simplifying the perturbation series. In general, $k$'s order of perturbation connects $\ket{0}$ to states of the form $\ket{\bar{l_{i_{1:k}}}}=\hat{\mathbf{J}}_{i_{1:k}}\ket{0}$ (see Section \ref{higho} for definition). To obtain the amplitude of this interference, we start with the case where none of the perturbation operators are repeated and they can be applied interchangeably. Using basic formulation of time-independent perturbation theory, the overlap between the states $\ket{0}$ and $\ket{\bar{l_{i_{1:k}}}}$, up to the leading order perturbation, is
\begin{equation}
\begin{split}
    |\frac{(J/2)^k}{(E^{(0)}_0-E^{(0)}_{l_{i_1}}) (E^{(0)}_0-E^{(0)}_{l_{i_{1}:i_2}})\cdots (E^{(0)}_0-E^{(0)}_{l_{i_1:i_k}})}|
    =    |\frac{(J/2)^k}{\prod_{j=1:k} (\sum _{l=1:j} (-1)^{i_{l}}( h_{i_l}-h_{i_{l+1}}))}|
    < \prod_{j=1:k}\frac{\epsilon}{j}<\frac{\epsilon^k}{k!},
\end{split}
\end{equation}
where the definition of $\epsilon$- ideally disordered regions in (\ref{500c}) is used. On the other hand, there are $k!$ permutations of how these $k$ operators can be applied, which gives the upper-bound of $\epsilon^k$ for the overlap between $\ket{l_{i_{1:k}}}$ with $\ket{0}$, up to the leading order of perturbation. If some of the operators are acting on neighboring sites, they can not be permuted and the order is important. Let us demonstrate this case with a simple example of two operators, $ \hat{\mathbf{J}}_{i+1}\hat{\mathbf{J}}_{i}\ket{0}$. The amplitude of interference of this state with $\ket{0}$ is 
\begin{equation}
\begin{split}
      &|\frac{(J/2)^k}{(E^{(0)}_0-E^{(0)}_{l_{i}}) (E^{(0)}_0-E^{(0)}_{l_{i,i+1}})}|= |\frac{(J/2)^k}{ (  h_{i}-h_{i+1})(  h_{i}-\cancel{h_{i+1}}+\cancel{h_{i+1}}-h_{i+2})}|<\epsilon^2.
\end{split}
\end{equation}
Extending this feature to all non-repeated operators, by induction we have:
\begin{equation}\label{2401}
    |\braket{\overline{ {l_{i_{1:k}}}}}{0}|\le{\epsilon
^k}.
\end{equation}

\bibliography{apssamp}

\begin{thebibliography}{63}
\bibitem [{exp({\natexlab{a}})}]{exp6}%
  \BibitemOpen
  \href@noop {} {\bibinfo {title} {This is because when $t$ is finite, the basis fourier functions $e^{-ik_1t}$ and $e^{-ik_2t}; t=[0,t]$ are not orthogonal, i.e., $\int_0^t {e^{i(k_1-k_2)t}}\neq \delta_{k_1,k_2}$. therefore, the inner products are not valid and there is always some leaked amplitude for all frequencies in the fourier spectrum.}} ({\natexlab{a}})\BibitemShut {NoStop}%
\bibitem [{exp({\natexlab{b}})}]{exp7}%
  \BibitemOpen
  \href@noop {} {\bibinfo {title} {If we measure time in the unit $\frac{\hbar}{J}$, then for $w \gg j$, the ratio $\frac{W}{J}$ measures the dominance of the disordered hamiltonian over the interaction hamiltonian. therefore, the time-scale order $o(\frac{\hbar w}{J^2})$ is when the local interactions dominate the dynamics.}} ({\natexlab{b}})\BibitemShut {NoStop}%
\bibitem [{exp({\natexlab{c}})}]{exp2}%
  \BibitemOpen
  \href@noop {} {} ({\natexlab{c}}),\ \bibinfo {note} {given the fact that the entanglement measure is second order in the density matrix, the side poles of the entanglement measure are obtained by convolution between poles of the density matrix. Therefore, the side poles of the dominant frequencies, i.e., first harmonic, in the entanglement measure are the results of two different convolutions between poles of the density matrix: First, the convolution between the zero frequency side poles and the spectrum around the first harmonic, second, the convolution between the side poles of the first harmonic pole with the spectrum around the zero frequency. This requires knowledge of the spectrum near the zero frequency. In contrast, the side poles of the entanglement's second harmonic are the result of convolution between the frequency spectrum near the first harmonics of the certain elements in the density matrix, which is more tractable, and does not require knowledge of the near-zero frequency
  spectrum.}\BibitemShut {Stop}%
\bibitem [{exp({\natexlab{d}})}]{exp1}%
  \BibitemOpen
  \href@noop {} {\bibinfo {title} {The fact that we are able to denote the product state for the full system by $\ket{l}$, which one might try to used a more general notation like $\ket{a \otimes l}$, is because the structure of the interactions prohibits the coupling of both $\ket{\uparrow l}$ and $\ket{\downarrow l}$, with the initial state ($\ket{ 0}$) during the perturbation process (consecutively applying a set of ordered perturbation hamiltonians $\{\mathbf{{j}}_k\}$). {I}n other words, a products state for subsystem $\mathcal{R}$ can only appear in the perturbation series either with $\ket{\uparrow}$, or $\ket{\downarrow}$, but not both. {T}herefore, notation $\ket{a \otimes l}$ is redundant. importantly, given the reasoning above, this property implies that the diagonal part of the {Q}{C}{T}{F}, $\mathcal{K}_d$ in (\ref{main}), vanishes for {H}eisenberg chains with this definition of perturbation operators.}} ({\natexlab{d}})\BibitemShut {NoStop}%
\end{thebibliography}

\end{document}